\documentclass[pdftex,11pt,a4paper,parskip=half,headsepline,bibtotocnumbered]{scrartcl}

\usepackage[english]{babel}
\usepackage[a4paper,left=3.5cm,right=3.5cm,bottom=3.5cm,top=3cm]{geometry}
\usepackage[automark]{scrlayer-scrpage}
\usepackage[utf8]{inputenc}
\usepackage{latexsym}
\usepackage[T1]{fontenc}
\usepackage{amsmath}
\usepackage{amssymb}
\usepackage{amsfonts}
\usepackage{amstext}
\usepackage{pdfsync}
\usepackage{amsthm}
\usepackage{extarrows}
\usepackage{graphicx}
\usepackage{xcolor}
\usepackage{longtable}
\usepackage[longtable]{multirow} 
\usepackage{booktabs} 
\usepackage{listings}
\usepackage{natbib}
\usepackage{url}
\usepackage{hyperref}
\usepackage{todonotes}
\usepackage{placeins}
\usepackage{enumerate}
\usepackage{enumitem}
\usepackage{float}
\extrafloats{100}
\usepackage{algorithm}
\usepackage{algpseudocode}

\definecolor{TUGreen}{rgb}{0.517,0.721,0.094}
 \definecolor{middlegray}{rgb}{0.5,0.5,0.5}
 \definecolor{lightgray}{rgb}{0.8,0.8,0.8}
 \definecolor{orange}{HTML}{e88909}
 \definecolor{yac}{rgb}{0.6,0.6,0.1} 
 \definecolor{bblue}{rgb}{0,0.4470,0.7410}
 \definecolor{rred}{rgb}{0.8500,0.3250,0.0980}
 \definecolor{turquoise}{HTML}{47C2D6}
 \definecolor{ggreen}{HTML}{1bd38a}
 \definecolor{violet}{HTML}{af93e7}
 \definecolor{red}{HTML}{ea7360}

\author{Marléne Baumeister$^{*,\dagger,1,2}$, 
Merle Munko$^{*,3}$, 
Kai-Philipp Gladow$^{4}$, \\
Marc Ditzhaus$^{\ddagger, 3}$, 
Nayden Chakarov$^{4,5}$, 
Markus Pauly$^{1,2}$}

\title{Early and Late Buzzards: Comparing Different Approaches for Quantile-based Multiple Testing in Heavy-Tailed Wildlife Research Data}
\makeatletter

\clearscrheadfoot
\ihead{\headmark}
\automark{section}
\ohead{Early and Late Buzzards}
\newtheorem{theorem}{Theorem}
\newtheorem{lemma}[theorem]{Lemma}
\newtheorem{ass}[theorem]{Assumption}

\newcommand{\diag}{\text{Diag}}

\newcommand{\R}{\mathbb{R}}

\newcommand{\new}[1]{\textcolor{black}{#1}}
\newcommand{\newnew}[1]{\textcolor{black}{#1}}

\begin{document}
\thispagestyle{empty}
\maketitle 
\renewcommand*{\thefootnote}{\fnsymbol{footnote}}
\footnotetext[1]{These authors contributed equally to this work.}
\footnotetext[2]{Corresponding author: \textsf{e-mail: baumeister@statistik.tu-dortmund.de}}
\footnotetext[3]{We would like to thank the late Marc Ditzhaus, a bright mind and wonderful collaboration partner, for his contributions to that paper and for bringing us together which has made us the scientists we are today.}
\renewcommand*{\thefootnote}{\arabic{footnote}}
\footnotetext[1]{Department of Statistics, TU Dortmund University, Germany}
\footnotetext[2]{Research Center Trustworthy Data Science and Security, UA Ruhr, Germany}
\footnotetext[3]{Department of Mathematics, Otto-von-Guericke University Magdeburg, Germany}
\footnotetext[4]{Department of Animal Behaviour, Bielefeld University, Germany}
\footnotetext[5]{Joint Institute for Individualisation in a Changing Environment (JICE), Bielefeld University and University of Münster, Germany} 

\section*{Abstract}
In medical, ecological and psychological research, there is a need for methods to handle multiple testing, for example to consider group comparisons with more than two groups.
Typical approaches that deal with multiple testing are mean or variance based which can be less effective in the context of heavy-tailed and skewed data.
Here, the median is the preferred measure of location and the interquartile range (IQR) is an adequate alternative to the variance.
Therefore, it may be fruitful to formulate research questions of interest in terms of the median or the IQR.
For this reason, we compare different inference approaches for two-sided and non-inferiority hypotheses formulated in terms of medians or IQRs in an extensive simulation study. 
We consider multiple contrast testing procedures combined with a bootstrap method as well as testing procedures with Bonferroni correction.
As an example of a multiple testing problem based on heavy-tailed data we analyse an ecological trait variation in early and late breeding in a medium-sized bird of prey.

\textbf{Keywords:} Bonferroni Adjustment; Factorial Designs; Quantiles; Multiple Contrast Tests; Resampling.

\section{Introduction}\label{sec:intro}
\new{In this paper, we systematically compare possibilities to handle quantile-based multiple testing procedures in general factorial designs.
This comparison is motivated by a testing problem involving the hatching dates of a population of common buzzards (\textit{Buteo buteo}), cf. Figure \ref{fig:buzzards}.}
In the context of species protection it is necessary to analyse the behaviour of animals to understand how animals deal with environmental change and to develop strategies to protect them effectively \citep{halupka_effect_2017}.
It is well known that the hatching dates are influenced by the weather in general \citep{lehikoinen_reproduction_2009}, but in context of increasing temperatures and weather extreme events due to climate change it is interesting to \new{understand} which aspects of weather influence the hatching dates in detail.
Hence, we want to identify years between 2006 and 2022 with an earlier and a later hatching phenology, which could be used in future studies to compare weather conditions and population characteristics between years with early and late breedings.
In the end, this ecological question leads to a non-inferiority multiple testing procedure.
\new{There is the often observed phenomenon in the context of human or animal behaviour that data is skewed and heavy-tailed and therefore substantially deviates from normality.}
Established multiple testing procedures are mean-based and reach their limitations in case of skewed or heavy-tailed data, because they are sensitive to outliers.
\citet{bonett_statistical_2002} pointed out:
"Every student of introductory statistics is taught that the population median may be more meaningful than the population mean when the distribution is skewed."
That is why it can be fruitful to consider quantile-based statistical concepts such as the median or the interquartile range (IQR) instead of mean- or variance-based approaches. 
Another issue is the consideration of multiplicity, because it is natural to formulate further hypotheses regarding post-hoc comparisons after rejecting a global hypothesis \citep{ruxton_time_2008}.
In factorial designs like ours, multiple hypotheses are of potential interest and inferring all of them lead to the problem of the type I error cumulation.

There are some quantile-based methods for statistical inference.
A quantile-based regression was already introduced by \citet{koenker_regression_1978} and these methods are available in the \textsf{R}-Package \texttt{quantreg} \citet{koenker_quantreg_2024}.
Quantile-based testing of global hypotheses in factorial designs has been successfully developed by \citet{chung_exact_2013, ditzhaus_qanova_2021} (univariate), \citet{chung_multivariate_2016} and  \citet{baumeister_quantile-based_2024} (multivariate).
Moreover, \cite{segbehoe_simultaneous_2022} tackle the multiple testing problem regarding quantiles with the development of quantile-based multiple contrast testing procedures (MCTPs).
In general, MCTPs are useful in many situations because they overcome the multiple testing problem by redefining the rejection of the global hypothesis: it is simultaneously rejected if any of the individual comparisons are rejected.
The general concept of MCTPs was introduced for example in \citet{mukerjee_comparison_1987}.
Furthermore, MCTPs are known to be often more powerful than methods with classical p-value adjustment like the Bonferroni procedure \citep{bretz_multiple_2011, konietschke_are_2013}.
Because of these advantages, there are many different adaptations of MCTPs \citep[e.g.][]{bretz_numerical_2001, hasler_multiple_2008, konietschke_are_2013, hasler_multiple_2014, umlauft_wild_2019, noguchi_nonparametric_2020, rubarth_estimation_2022}.

For our approach, the method of \citet{segbehoe_simultaneous_2022} appears to be most suitable.
Among others, they introduced MCTPs regarding differences of quantiles for two-sided statistical hypotheses, but not yet for the non-inferiority testing problem.
Their method is based on an asymptotic approach, which means that the test statistic is compared with a theoretical quantile of a multivariate distribution, and a bootstrap approach, where the test statistic is compared to an empirical bootstrap quantile.
The method is included in the \textsf{R}-Package \texttt{mratios} \citep{djira_mratios_2020}.
\citet{segbehoe_simultaneous_2022} compare the type I error performance of these two approaches in a Monte-Carlo simulation study, which takes into account only scenarios with balanced, predominantly large samples and, most importantly, only three groups.
In a multiple pairwise comparison with three groups, however it follows from the closing testing procedure \new{\citep{marcus_closed_1976}} that there is no need to adjust the levels for the local hypotheses of pairwise comparisons to control the family-wise error rate (FWER) if the global hypothesis can be rejected.
Because of the simplicity of the closing test for three groups it is to be expected that every testing procedure that controls the FWER will perform relatively well in this setting.
The reason for this is that the closing testing procedure works in principle and set-theoretically for every type of test.
\new{See \citet{goeman_comparing_2022} for a discussion of the closing test procedure especially for three groups.}
Furthermore, the simulation of \citet{segbehoe_simultaneous_2022} does not include a comparison with other multiple testing procedures e.g. with Bonferroni-adjusted multiple tests \citep{dunn_multiple_1961}.
It is therefore impossible to get a broader overview about the performance of the tests.
Because our ecological problem regarding the hatch data contains a much larger number of comparisons the simulation study of \citet{segbehoe_simultaneous_2022} can not help us to decide if this method is adequate for our problem.
\new{More generally, it is not possible to decide for one multiple testing procedure to handle multiple testing problems in skewed and heavy-tailed data. }
This is our motivation to consider a more comprehensive and competitive simulation study.

In particular, our aim is to compare the performance of different statistical testing procedures that deal with quantile-based multiple hypotheses.
Beyond the multiple testing problem, statistical questions do not only arise with two-sided hypotheses, as we have seen in our example with regard to buzzards.
To give a broader overview of the methodological possibilities and capabilities of different multiple testing procedures, we study three commonly important versions of hypotheses: non-inferiority, two-sided and equivalence hypotheses.
An intuitive way to deal with the multiple testing problem is to define permutation tests in the framework of the QANOVA by \citet{ditzhaus_qanova_2021} and to adjust them with the well-known Bonferroni correction \citep{dunn_multiple_1961}.
\new{It is a new approach to define tests in this framework that can be applied in one- and two-sided testing problems.}
We also extend the method of \citet{segbehoe_simultaneous_2022} to non-inferiority testing.
Similar to their work we consider two ways of deriving critical values: from the asymptotic distribution or via group-wise bootstrapping.
Besides, we explain that the considered methods are theoretically valid and their inference works without any restrictive distributional assumption and allows for potential heteroscedasticity.
We compare all these quantile-based multiple testing procedures for one- and two-sided hypotheses through an extensive simulation study regarding type I error and power.
In particular, we consider various testing problems (Dunnett, Tukey, Grand Mean), varying sample sizes, distributions, as well as homo- and heteroscedastic settings.
Through our comparison we come to the result that the MCTP methods are not in general superior to the QANOVA permutation approaches with Bonferroni correction regarding empirical family-wise error rate control and power.

The paper is structured as follows.
\new{We state models and hypotheses in Section \ref{sec:setup}.
Afterwards we introduce different statistical testing procedures (Section \ref{sec:methods}) including explanations of the asymptotic Bonferroni-adjusted QANOVA tests  and their permutational version (Section \ref{ssec:bonfasymptotic}, details of the permutation approach in Section \ref{sec:BPerm} in the Appendix), as well as quantile-based MCTPs and a bootstrap version (Section \ref{ssec:asymptoticMCTP}).
An extensive simulation study (Section \ref{sec:sim}) gives an overview of the performances of all methods for various scenarios.
Section \ref{sec:analysis} analyses the data example with these methods.
There, we also explain our motivational data example on buzzards and how it fits to our statistical model.
Section \ref{sec:end} concludes the results.}

\section{Model and Hypotheses}\label{sec:setup}
Suppose we have $k\in\mathbb{N}$ mutually independent samples
$X_{i1},...,X_{in_i} \sim F_i, i\in\{1,...,k\},$ 
where $F_i$ are distribution functions.
Here, $n_i$ represent the sample sizes per group and $n := \sum_{i=1}^k n_i$ denotes the total sample size.
To define the quantity of interest, let $0 < p_1 < ... < p_m < 1$ denote $m\in\mathbb{N}$ probabilities of interest with corresponding quantiles
$$ q_{ij} := F^{-1}_i(p_j) \new{= \inf\{u \in\mathbb R \mid F_i(u) \geq p_j\}},\quad i\in\{1,...,k\}, j\in\{1,...,m\}. $$
The pooled quantile vector is denoted by $\mathbf{q} := (q_{11},...,q_{1m},q_{21},...,q_{km})^{\prime}$. 
For our asymptotic derivations, we need the following assumption throughout this paper.
\begin{ass}\label{Assumptions}
We assume that $F_i$ is continuously differentiable at $q_{i1}, ..., q_{im}$ with positive derivatives $f_i(q_{ij}) > 0$ for all $i\in\{1,...,k\}, j\in\{1,...,m\}.$
Moreover, we assume $n_i/n \to \kappa_i > 0$ as $n\to\infty$ for all $i\in\{1,...,k\}$.
\end{ass}
\new{In practice, the assumption of a continuous derivative at $q_{i1}, ..., q_{im}$ can not really be checked because usually neither $F_i$ nor $q_{i1}, ..., q_{im}$ are known.
However, if there are (many) ties in the data, this is at least an indicator that $F_i$ is not continuous and, thus, not differentiable at the tie points making the previous assumption less plausible.}
Let $\mathbf{H} = [\mathbf{h}_1,...,\mathbf{h}_r]^{\prime}\in\R^{r\times km}$ denote a matrix of vectors $\mathbf{h}_\ell\new{=(h_{\ell 11},...,h_{\ell 1m}, h_{\ell 21},...,h_{\ell km})'}\in\R^{km},\;\ell\in\{1,\dots,r\}$ with the contrast property $\sum_{i=1}^kh_{\ell ij}=0$ for all $j\in\{1,\dots,m\}$. 
This contrast property \new{means that only contrasts over the different groups may be considered and} is actually also needed in \citet{ditzhaus_qanova_2021}.
\new{The property can easily be checked for a known matrix $\mathbf{H}$ and all examples given below fulfil the contrast property.}
Moreover, let $\boldsymbol{\epsilon}=(\epsilon_1,\dots,\epsilon_r)'\in\new{\R}^r$ denote a vector of constants.
Then, we aim to infer the multiple testing problem
\begin{align}\label{eq:TwosidedTestingProc}
    \mathcal{H}_{0,\ell}: \mathbf{h}_{\ell}^{\prime}\mathbf{q} = \epsilon_{\ell} \text{ vs. }
    \mathcal{H}_{1,\ell}: \mathbf{h}_{\ell}^{\prime}\mathbf{q} \neq \epsilon_{\ell},\quad\text{for }
    \ell\in\{1,...,r\}.
\end{align}
\new{These hypotheses follow the usual definition of hypotheses that can be answered through multiple contrast tests, see e.g. \citet{hothorn_simultaneous_2008, konietschke_are_2013}.}
Additionally, we consider a multiple one-sided non-inferiority problem:
\begin{align}\label{eq:OnesidesTestingProc}
    \mathcal{H}^I_{0,\ell}: \mathbf{h}_{\ell}^{\prime}\mathbf{q}\le {\epsilon}_{\ell} \text{ vs. }
    \mathcal{H}^I_{1,\ell}: \mathbf{h}_{\ell}^{\prime}\mathbf{q} > {\epsilon}_{\ell},\quad\text{for }
    \ell\in\{1,...,r\}.
\end{align}
The corresponding global hypotheses are given by $\mathcal{H}_0:\mathbf{Hq}=\boldsymbol{\epsilon}$ and $\mathcal{H}^I_0:\mathbf{Hq}\le\boldsymbol{\epsilon}$, respectively.
Our motivation to consider both types of hypotheses is that they have widely different interpretations despite the methodological similarity.
Interpreting non-inferiority tests is grounded in another research question than the approach of two-sided tests.
Testing non-inferiority means that someone has the aim to show that one treatment/group is not unacceptably worse compared to \new{one other group} \citep{schumi_through_2011}.
What \textit{unacceptably worse} means, is characterized in the vector of constants $\boldsymbol{\epsilon}$.
If the directed deviation in $\mathcal{H}_{0,\ell}^I$ is smaller than $\epsilon_\ell$ something seems to be unacceptably worse and $\mathcal{H}_{0,\ell}^I$ is not rejected.
The constant can be identified with the maximal directed deviation in which a significant difference or improvement is not indicated.
Regarding the testing problem and the question of interest, differences smaller than the constant $\epsilon_{\ell}$ are not indicated as differences.
We refer to \citet{scott_noninferiority_2009} and \citet{schumi_through_2011} for the idea of non-inferiority tests and the interpretation and meaning of $\boldsymbol{\epsilon}$.
Moreover, within this framework it is possible to infer equivalence hypotheses.
To this end, we can adapt the equivalence testing approach of \citet{hauck_new_1984} for quantiles.
Let $[-\delta_{\ell},\delta_{\ell}]$ be equivalence intervals for every $\ell\in \{1,\dots,r\}$.
Then the multiple equivalence hypotheses problem has the form:
\begin{align}\label{eq:equivalence}
    \mathcal{H}^E_{0,\ell}: \vert\mathbf{h}_{\ell}^{\prime}\mathbf{q}\vert\ge {\delta}_{\ell} \text{ vs. }
    \mathcal{H}^E_{1,\ell}: \vert\mathbf{h}_{\ell}^{\prime}\mathbf{q}\vert < {\delta}_{\ell},\quad\text{for }
    \ell\in\{1,...,r\}.
\end{align}
This hypotheses lead to a TOST procedure \citep{schuirmann_comparison_1987} for quantiles, where the statistical question is answered by two one-sided tests with the halved level of significance.
Thus the methodological treatment of \eqref{eq:equivalence} follows from that in \eqref{eq:OnesidesTestingProc}.
We want to point out that it is possible to consider far different statistical questions with similar methodology, but in the following we focus on the two-sided and the non-inferiority hypotheses only. 
Below we give some concrete examples of covered multiple testing problems.

\textbf{Examples of covered hypotheses.} 
The hypotheses $\mathcal{H}_0$ and $\mathcal{H}_0^I$ cover various local and multiple testing problems of interest.
For a single quantile $\mathbf{q}=(q_1,\dots,q_k)^{\prime},\, m=1$\new{,} we can formulate hypotheses that are well known for vectors of means \citep[cf.][]{bretz_multiple_2011, konietschke_are_2013} in terms of medians, quantiles or more general quantile contrasts.
This explicitly includes 
\begin{enumerate}
\item \textbf{All-pairs comparisons for medians.} 
Choosing $p_1 = 0.5$, $m=1$ and the Tukey-type \citep{tukey_problem_1994} matrix as contrast matrix $\mathbf{H}$ leads to the one- and two-sided hypotheses $\mathcal{H}_{0,\ell_1\ell_2}: m_{\ell_1} - m_{\ell_2} = \epsilon_{\ell_1\ell_2}$ and $\mathcal{H}^I_{0,\ell_1\ell_2}: m_{\ell_1} - m_{\ell_2} \leq \epsilon_{\ell_1\ell_2}$, where $\ell_1,\ell_2\in\{1,\dots,k\}$, $\ell_1 > \ell_2$ of all-pairs comparisons for medians $m_i := F_i^{-1}(0.5), i\in\{1,\dots,k\},$ in one-way layouts.
\item \textbf{Many-to-one comparisons for medians.} Similarly, choosing the Dunnett-type 
\citep{dunnett_multiple_1955} matrix gives the one- and two-sided hypotheses $\mathcal{H}_{0,\ell}: m_{\ell} - m_{1} = \epsilon_{\ell}$ and $\mathcal{H}^I_{0,\ell}: m_{\ell} - m_{1} \leq \epsilon_{\ell}, \ell\in\{2,\dots,k\},$ of many-to-one comparisons for medians.
\item  \textbf{Grand-mean comparisons.} Choosing the Grand-mean-type matrix \citep{djira_detecting_2009} instead leads to the one- and two-sided hypotheses $\mathcal{H}_{0,\ell}: m_{\ell} - \bar{m} = \epsilon_{\ell}$ and $\mathcal{H}^I_{0,\ell}: m_{\ell} -\bar{m} \leq \epsilon_{\ell}, \ell\in\{1,\dots,k\},$ of median comparisons to the mean $\bar{m} := k^{-1} \sum_{i=1}^k m_i$ of all group-wise medians in one-way layouts.
\item \textbf{Multiple testing problems in general quantiles or IQR.} In the above hypotheses, the medians $m_1,\dots,m_k$ can be substituted by any quantile or linear contrast of interest. Thus, we can even infer multiple hypotheses about interquartile ranges (IQRs) $IQR_i := F_i^{-1}(0.75) - F_i^{-1}(0.25)$ leading to hypotheses of the form $\mathcal{H}_{0,\ell_1\ell_2}: IQR_{\ell_1} - IQR_{\ell_2} = \epsilon_{\ell_1\ell_2}$ and $\mathcal{H}^I_{0,\ell_1\ell_2}: IQR_{\ell_1} - IQR_{\ell_2} \leq \epsilon_{\ell_1\ell_2}, \ell_1,\ell_2\in\{1,\dots,k\}, \ell_1 > \ell_2$ in the all-pairs comparison setting and similar for the Dunnett- or the Grand-mean-type matrix.
\item \textbf{Simultaneous inference for medians and IQRs.} 
Our test scenario is even more general and also allows for the simultaneous treatment of more than one effect parameter of interest. 
For example, it would be possible to compare the medians and interquartile ranges simultaneously across the groups by setting $p_1 = 0.25, p_2 = 0.5, p_3 = 0.75, m=3$ and choosing a hypothesis matrix 
$$\mathbf{H} \otimes \begin{bmatrix}
0 & 1 & 0\\
-1 & 0 & 1
\end{bmatrix}$$
with $\mathbf{H}$ being one of the Tukey-type, Dunnett-type, or Grand-mean-type matrix, respectively, and $\otimes$ denoting the Kronecker product. 
Here,  the Tukey-type matrix leads to all-pairs comparisons of the medians and IQRs, respectively, with local null hypotheses $\mathcal{H}_{0,\ell_1\ell_2,med}: m_{\ell_1} - m_{\ell_2} = \epsilon_{\ell_1\ell_2,med}, \mathcal{H}_{0,\ell_1\ell_2,IQR}: IQR_{\ell_1} - IQR_{\ell_2} = \epsilon_{\ell_1\ell_2,IQR}, \ell_1,\ell_2\in\{1,\dots,k\}, \ell_1 > \ell_2$ for the two-sided testing problem. 
If $\boldsymbol\epsilon$ is the zero vector, the global null hypothesis that all medians and IQRs are equal is $\mathcal{H}_{0}: m_{1} = \dots = m_{k}, IQR_{1} = \dots = IQR_{k}$. 
Analogously, the hypotheses can be formulated for the one-sided testing problem as well as for the Dunnett-type matrix for many-to-one comparisons and the Grand-mean-type matrix for comparisons of the medians and IQRs to the mean of medians and IQRs, respectively.
\end{enumerate}
Even multiple hypotheses on quantiles in more general factorial designs are covered by splitting up indices as in classical ANOVA (\citealp{paulyETAL2015}).

\section{Statistical Methods}\label{sec:methods}
\new{In the following section, we present four testing procedures that all correspond to the hypotheses in Equation \ref{eq:TwosidedTestingProc} respectively \ref{eq:OnesidesTestingProc} and be compared in Section \ref{sec:sim} by using of simulations.}
An estimator for the quantile $q_{ij}$ is given by the empirical quantile, that is
\begin{align*}
 \widehat{q}_{ij} := \hat{F}_i^{-1}(p_j),
\end{align*}
where $\hat{F}_i$ denotes the empirical distribution function.
Under Assumption~\ref{Assumptions}, \cite{serfling_approximation_1980}
 proved convergence in distribution
\begin{align}\label{eq:conv}
    \sqrt{n} \left(\widehat{q}_{ij} - q_{ij}\right)_{j\in\{1,...,m\}} \xrightarrow{d} \mathbf{Z}_i \sim\mathcal{N}\left(\mathbf{0}, \boldsymbol\Sigma^{(i)}\right)
\end{align}
as $n\to\infty$ for all $i\in\{1,...,k\}$, where 
\begin{align}\label{eq:Sigma}
    \boldsymbol\Sigma^{(i)}_{ab} := \kappa_i^{-1} \frac{1}{f_i(q_{ia})f_i(q_{ib})} \left(\min\{p_a,p_b\} - p_ap_b\right)
\end{align} for all $a,b\in\{1,...m\}.$
Let $\boldsymbol\Sigma := \oplus_{i=1}^k \boldsymbol\Sigma^{(i)} $ denote the direct sum (i.e. block diagonal matrix) of the covariance matrices.
Since we are also interested in directional hypotheses, we consider the family of test statistics
\begin{align}\label{eq:teststat}
    T_n(\mathbf{h}_{\ell},\epsilon_{\ell}) :=\sqrt{n}\frac{\mathbf{h}_{\ell}^{\prime}\widehat{\mathbf{q}} - {\epsilon}_{\ell}}{\sqrt{\mathbf{h}_{\ell}^{\prime} \widehat{\boldsymbol\Sigma} \mathbf{h}_{\ell}}}, \quad \new{\ell}\in\{1,...,r\},
\end{align} instead of the two-sided QANOVA Wald-type test statistic 
that was discussed in \citet{ditzhaus_qanova_2021}. 
We note that for a single contrast $\mathbf{h}_{\ell}$, we obtain the QANOVA Wald-type test statistic of \citet{ditzhaus_qanova_2021} as $T_n^2(\mathbf{h}_{\ell},0)$.

\new{
For appropriate critical values $\widetilde{q}_\ell$, we receive the following test decisions for the \textbf{two-sided multiple testing problem}:
\begin{enumerate}
    \item for each $\ell\in\{1,...,r\}$, $\mathcal{H}_{0,\ell}$ is rejected if and only if $\left|T_n(\mathbf{h}_{\ell},\epsilon_{\ell})\right| > \widetilde{q}_\ell $,
    \item the global hypothesis $\mathcal{H}_0 = \bigcap\limits_{\ell=1}^r \mathcal{H}_{0,\ell}$ is rejected if and only if at least one $\mathcal{H}_{0,\ell}$ is rejected, i.e. if $\max\limits_{\ell\in\{1,...,r\}}\left|T_n(\mathbf{h}_{\ell},\epsilon_{\ell})\right| > \widetilde{q}_\ell $.
\end{enumerate}
Corresponding simultaneous two-sided confidence intervals for $\mathbf{h}_{\ell}'\mathbf{q}, \ell\in\{1,...,r\},$ can be obtained as
\begin{align*}
\left[  \mathbf{h}_{\ell}'\widehat{\mathbf{q}} - \sqrt{\mathbf{h}_{\ell}'\widehat{\boldsymbol{\Sigma}}\mathbf{h}_{\ell}}\frac{\widetilde{q}_\ell}{\sqrt{n}}, \mathbf{h}_{\ell}'\widehat{\mathbf{q}} + \sqrt{\mathbf{h}_{\ell}'\widehat{\boldsymbol{\Sigma}}\mathbf{h}_{\ell}}\frac{\widetilde{q}_\ell}{\sqrt{n}} \right]  , \quad\ell\in\{1,...,r\}.
\end{align*}
Alternatively, there is the ability to formulate simultaneous tests $\mathbf{1}\{\left|T_n(\mathbf{h}_{\ell},\epsilon_{\ell})\right| > \widetilde{q}_\ell\}$ for every Hypothesis $\mathcal{H}_{0,\ell},\,\ell\in\{1,...,r\}$ and a test $\mathbf{1}\{\max_{\ell\in\{1,...,r\}}\left|T_n(\mathbf{h}_{\ell},\epsilon_{\ell})\right| > \widetilde{q}_\ell\}$ for the global Hypothesis $\mathcal{H}_0$.
}
\new{
Analogously the test decisions for the \textbf{non-inferiority multiple testing problem} are
\begin{enumerate}
    \item for each $\ell\in\{1,...,r\}$, $\mathcal{H}^{I}_{0,\ell}$ is rejected if and only if $T_n(\mathbf{h}_{\ell},\epsilon_{\ell}) > q_\ell $,
    \item the global hypothesis $\mathcal{H}^{I}_0 = \bigcap\limits_{\ell=1}^r \mathcal{H}_{0,\ell}$ is rejected if and only if at least one $\mathcal{H}^{I}_{0,\ell}$ is rejected, i.e. if $\max\limits_{\ell\in\{1,...,r\}}\left(T_n(\mathbf{h}_{\ell},\epsilon_{\ell})\right) > q_\ell $
\end{enumerate}
with appropriate critical values $q_\ell$
and the corresponding simultaneous one-sided confidence intervals for $\mathbf{h}_{\ell}'\mathbf{q}, \ell\in\{1,...,r\},$ are given by
\begin{align}\label{eq:oneCI}
\left[  \mathbf{h}_{\ell}'\widehat{\mathbf{q}} - \sqrt{\mathbf{h}_{\ell}'\widehat{\boldsymbol{\Sigma}}\mathbf{h}_{\ell}}\frac{q_\ell}{\sqrt{n}} , \infty \right), \quad\ell\in\{1,...,r\} .
\end{align}
This testing problem can also be formulated in short test notation as $\mathbf{1}\{T_n(\mathbf{h}_{\ell},\epsilon_{\ell}) > q_\ell \}$ for the simultaneous hypotheses $\mathcal{H}^{I}_{0,\ell},\,\ell\in\{1,...,r\}$, and for the global Hypothesis $\mathcal{H}^{I}_0$ as $\mathbf{1}\{\max_{\ell\in\{1,...,r\}}\left(T_n(\mathbf{h}_{\ell},\epsilon_{\ell})\right) > q_\ell \}$.
Note, that both testing problems comply with the union-intersection principle introduced by \citet{roy_heuristic_1953} and that they are in fact quantile-based versions  of so-called max-t tests \citep{bretz_numerical_2001}. 
}

\new{Of note, an application of a stepwise procedure as the closed-testing procedure \citep{gabriel_simultaneous_1969}, the well-known Holm procedure \citep{holm_simple_1979} or Shaffer's method \citep{shaffer_modified_1986} may increase the power of the proposed multiple tests but lacks the obtainment of corresponding simultaneous confidence regions.
However, we focus on multiple testing procedures that come along with corresponding simultaneous confidence intervals in the following.
See \cite{pigeot_basic_2000} for a methodologically overview about multiple testing and \cite{gabriel_simultaneous_1969} for the foundation of simultaneous testing procedures.
}

\new{In order to determine appropriate critical values, we firstly need to investigate the joint asymptotic behavior of the test statistics.}
Due to (\ref{eq:conv}), it follows that we have convergence in distribution
\begin{align}\label{eq:conv2}
    \left( T_n(\mathbf{h}_1,\epsilon_1), ..., T_n(\mathbf{h}_r,\epsilon_r) \right)^{\prime} \xrightarrow{d} 
    \mathcal{N}\left(\mathbf{0}, \mathbf{D}\mathbf{H}\boldsymbol\Sigma \mathbf{H}^{\prime} \mathbf{D} \right)
\end{align}
as $n\to\infty$ under the null hypotheses in (\ref{eq:TwosidedTestingProc}), where
\begin{align}\label{eq:multilimit}
\mathbf{D} := \text{diag}\left( \left( \mathbf{h}_{1}^{\prime} {\boldsymbol\Sigma} \mathbf{h}_{1} \right)^{-1/2} ,..., \left( \mathbf{h}_{r}^{\prime} {\boldsymbol\Sigma} \mathbf{h}_{r} \right)^{-1/2}  \right).
\end{align} 
Note that the covariance matrix in (\ref{eq:multilimit}) in the limit is a correlation matrix, i.e. has a diagonal of ones, and, thus, each test statistic $T_n(\mathbf{h}_{\ell},\epsilon_{\ell})$ is asymptotically standard normally distributed.
Since $\boldsymbol\Sigma$ is usually unknown, the joint limiting distribution is unknown.
To get a consistent estimator $\widehat{\boldsymbol\Sigma}$ for ${\boldsymbol\Sigma}$, we use three different approaches as discussed in \citet{ditzhaus_qanova_2021}:
a kernel-estimator, a bootstrap-estimator and an interval-based approach.
In \citet{ditzhaus_qanova_2021}, there was no clear recommendation for one of them.
We thus analyze all of them.
The concrete forms are given in Section~\ref{sec:covEst} in the Appendix. 
It should be noted that further technical assumptions are needed for the consistency of the covariance estimator $\widehat{\boldsymbol{\Sigma}}$ for ${\boldsymbol{\Sigma}}$, see \citet{ditzhaus_qanova_2021} for details.
With each of the three consistent estimators, we are able to obtain an approximation for the critical values.
\new{
In the following subsections, we elaborate on different asymptotic- and resampling-based choices of $\widetilde{q}_\ell$ and $q_\ell$.
}

\subsection{Bonferroni-adjusted QANOVA}\label{ssec:bonfasymptotic}
Let $\alpha\in (0,1)$ represent the level of significance.
An intuitive and well-known method to deal with multiple testing problems is 
the Bonferroni correction \citep{dunn_multiple_1961}, where each individual hypothesis is tested at a smaller local level of $\alpha/r$.
To realize this for our statistical question, recall that $T_n(\mathbf{h}_{\ell},\epsilon_{\ell})$ is asymptotically standard normal distributed. 
This motivates to consider standard normal quantiles as critical values. 
Let $z_{\beta}$ denote the $\beta$-quantile of the standard normal distribution.
\new{Then, choosing $\widetilde{q}_\ell = z_{1-\alpha/(2r)}$ for the two-sided multiple testing problem or $q_\ell = z_{1-\alpha/r}$ for the non-inferiority multiple testing problem, respectively, yield the \textbf{Bonferroni-adjusted asymptotic testing procedures}.}

Regarding \eqref{eq:conv2}, this method is expected to work well for large sample sizes.
However, resampling methods have proven useful in several different statistical fields if the sample sizes are small \citep{paulyETAL2015, doblerpauly2018, dobler_nonparametric_2020, sattler_testing_2022, ditzhaus2023studentized, munko2023rmstbased, baumeister_quantile-based_2024}.
This particularly holds for permutation tests that even are finitely exact under exchangeability \citep{hemerik2018exact,lehmann_testing_2022}.
\citet{ditzhaus_qanova_2021} already proposed permutation tests for the QANOVA.
\new{In our model, exchangeability means that the distribution functions are equal across the groups, i.e., $F_1 = ... = F_k$.} 
\new{The idea of the permutation approach is to draw the permuted samples $X_{i1}^{\pi},...,X_{in_i}^{\pi}, i\in\{1,...,k\},$ without replacement from the pooled sample $X_{11},...,X_{1n_1},X_{21},...,X_{kn_k}$.
Statistics and estimators based on the permuted data $X_{i1}^{\pi},...,X_{in_i}^{\pi}, i\in\{1,...,k\},$ are denoted here and throughout with a $\pi$ in the superscript.
The permutation QANOVA approach is derived by using permutation-based critical values instead of the standard normal quantiles.
Therefore, let $q^{\pi}_{\ell,\beta}$ and $\widetilde q^{\pi}_{\ell,\beta}$ denote the $\beta$-quantiles of the conditional distribution of the permutation test statistics given the data for all $\ell\in\{1,...,r\}$. 
By Equation (\ref{eq:convPerm}) in the Appendix \ref{sec:BPerm}, the quantiles are converging in probability to quantiles of the standard normal distribution or its absolute value, respectively.
That is why we set  $\widetilde{q}_\ell=\tilde{q}^{\pi}_{\ell, 1-\alpha/2r}$ and $q_\ell=q^{\pi}_{\ell, 1-\alpha/r}$, respectively, for the \textbf{Bonferroni-adjusted permutation testing procedure}.}
The concrete computation of these critical values \new{and necessary assumptions for the asymptotic validity can be found} in Section~\ref{sec:BPerm} in the Appendix, see also \citet{ditzhaus_qanova_2021}.
\new{Note that if exchangeability is given, i.e., if $F_1 = ... = F_k$ holds, the permutation test is finitely exact.
However, we do not need the exchangeability assumption for proving the asymptotic validity of the permutation test.
Hence, the permutation approach also works asymptotically under non-exchangeable data.} 

\subsection{Multiple Contrast Test Procedures}\label{ssec:asymptoticMCTP}

In this Section, we \new{firstly} extend the asymptotic approach of \citet{segbehoe_simultaneous_2022} to inference settings with more than one quantile of interest and to allow for one-sided testing problems.
For the \new{a}symptotic Multiple Contrast Test Procedure (MCTP), the main ideas are to replace $\boldsymbol\Sigma$ by $\widehat{\boldsymbol\Sigma}$ in the limit distribution in (\ref{eq:conv2}) \new{to consider the asymptotic multivariate distribution of the test statistics.}
Since the local test statistics $T_n(\mathbf{h}_1,\epsilon_1), ..., T_n(\mathbf{h}_r,\epsilon_r)$ all have the same marginal limit distribution, we may choose the same critical value for all local hypotheses. 
Then, rejecting the global null hypothesis whenever a local hypothesis is rejected translates into comparing the maximum of the test statistics to the critical value.
Hence, in order to determine the critical value for the asymptotic approach, let $(Y_1,...,Y_r)^{\prime} \sim \mathcal{N}(\mathbf{0}, \widehat{\mathbf{D}}\mathbf{H}\widehat{\boldsymbol\Sigma} \mathbf{H}^{\prime}\widehat{\mathbf{D}}  )$ given the data with
$$ \widehat{\mathbf{D}} :=  \text{diag}\left( \left( \mathbf{h}_{1}^{\prime} \widehat{\boldsymbol\Sigma} \mathbf{h}_{1} \right)^{-1/2} ,..., \left( \mathbf{h}_{r}^{\prime} \widehat{\boldsymbol\Sigma} \mathbf{h}_{r} \right)^{-1/2}  \right).$$
Moreover, denote by $q_{1-\alpha}$ the $(1-\alpha)$-quantile of the conditional distribution of $ \max_{\ell\in\{1,...,r\}} Y_\ell$
and by $\widetilde{q}_{1-\alpha}$ the $(1-\alpha)$-quantile of the conditional distribution of $ \max_{\ell\in\{1,...,r\}} |Y_\ell|$ given the data.
Due to the consistency of the covariance estimators, ${q}_{1-\alpha}$ and $\widetilde{q}_{1-\alpha}$ are converging in probability to the $(1-\alpha)$-quantiles of $ \max_{\ell\in\{1,...,r\}} Z_\ell$ and $ \max_{\ell\in\{1,...,r\}} |Z_\ell|$, respectively, \new{under Assumption~\ref{Assumptions},} see Section~\ref{sec:Proofs} in the Appendix for details.
This ensures the asymptotic control of the family-wise error rate (FWER) \new{under Assumption~\ref{Assumptions}} by using \new{$\widetilde{q}_\ell = \widetilde{q}_{1 - \alpha}$} for the \textbf{\new{Asymptotic} MCTP for the two-sided problem} \new{and $q_\ell = {q}_{1-\alpha}$ for the \textbf{Asymptotic MCTP for the non-inferiority testing problem}, respectively.}

For a better small sample performance {in the MCTP approach}, we \new{also} consider a groupwise bootstrap similarly to the bootstrap proposed in \citet{segbehoe_simultaneous_2022} to approximate the limiting distribution.
This approach is identical to the bootstrap approach in \citet{baumeister_quantile-based_2024}.
To realize this, we draw a nonparametric bootstrap sample $X_{i1}^*,...,X_{in_i}^*$ with replacement from the original $i$th sample $X_{i1},...,X_{in_i}$ as in Section~2.4 of \cite{segbehoe_simultaneous_2022}. 
\new{In detail, $X_{i1}^*,...,X_{in_i}^* \sim \hat{F}_i$ are independent identically distributed given the data $X_{i1},...,X_{in_i}$.} 
\new{Note, that this is simply the adoption of Efron's in the context of MCTPs.} 
In the following, the estimators based on the bootstrap samples are denoted with a superscript $*$, respectively.
Then, we define the groupwise bootstrap counterpart of the test statistics by
\begin{align*}
T_n^*(\mathbf{h}_{\ell}) := \sqrt{n}\frac{\mathbf{h}_{\ell}^{\prime}(\widehat{\mathbf{q}}^* - \widehat{\mathbf{q}})}{\sqrt{\mathbf{h}_{\ell}^{\prime}\widehat{\boldsymbol\Sigma}^{*} \mathbf{h}_{\ell}}}, \quad \ell\in\{1,...,r\}.
\end{align*}
Note that in comparison to \cite{segbehoe_simultaneous_2022}, we consider the counterpart of our studentized test statistics (\ref{eq:teststat}). 
Let $q^*_{1-\alpha}$ and $\tilde q_{1-\alpha}^*$ denote the $(1-\alpha)$-quantiles of the conditional distribution of the max-test statistics $\max_{\ell\in\{1,...,r\}} T_n^*(\mathbf{h}_{\ell}) $ and $\max_{\ell\in\{1,...,r\}} |T_n^*(\mathbf{h}_{\ell})| $, respectively, given the data. 
In Section~\ref{sec:bootstrapdetails} in the Appendix, we prove that \new{choosing $\widetilde{q}_\ell = \tilde q^*_{1-\alpha}$ and $q_\ell = q^*_{1-\alpha}$}  results in asymptotically valid \new{\textbf{groupwise bootstrap MCTPs}} \new{under Assumption~\ref{Assumptions}} whenever the kernel or interval-based covariance estimator is used.
\new{Explicit algorithms for the bootstrap MCTP can be found in Section~\ref{ssec:Algorithms} in the appendix.}

\section{Simulations}\label{sec:sim}
Having discussed some asymptotic properties of the different multiple testing approaches, we now evaluate their finite sample performance in various settings. 
To this end, we did an intensive simulation study using the statistical software \textsf{R} version 4.2.1 \citep{r_core_team_r_2024}.
\new{The complete material of the simulation study can be found in the Supporting Information.}

\subsection{\newnew{Simulation for Small Sample Sizes}}\label{ssec:smallsim}
\newnew{In this section, w}e consider $k=4$ groups and compare the medians, i.e. $p_1 = 0.5, m = 1$. Therefore, we use the Dunnett-type, Tukey-type and Grand-mean-type hypothesis matrix as $\mathbf{H}$, respectively, and $\epsilon_1 = ... = \epsilon_r = 0$ for the two-sided and non-inferiority hypotheses. 
Further simulations that focused on the comparison of medians and interquartile ranges simultaneously can be found in the Supplement and are summarized at the end of this section. 
 For the data generation, we consider the same setup as in \citet{ditzhaus_qanova_2021}, i.e we simulate groupwise data from the model
\begin{align}\label{eq:simmodel}
X_{is} = \sigma_i (\eta_{is} - m_i) + \mu_i \sim F_i, \quad i\in\{1,...,k\}, s\in\{1,...,n_i\}.
\end{align} 
Here, we consider different variance settings given by
$\boldsymbol{\sigma}_1 = (\sigma_1,\sigma_2,\sigma_3,\sigma_4) = (1,1,1,1)$, $\boldsymbol{\sigma}_2 = (1,1.25, $ $1.5, 1.75)$, $\boldsymbol{\sigma}_3 = (1.75,1.5,1.25,1)$ and two different sample size allocations given by $\mathbf{n}_1$ $ = $ $(n_1,n_2,n_3, $ $n_4) $ $= (15,15,15,15)$, $\mathbf{n}_2 = (10,10,20,20)$. 
This leads to balanced ($\mathbf{n}_1$) and unbalanced ($\mathbf{n}_2$) homo-  ($\boldsymbol{\sigma}_1$) and heteroscedastic ($\boldsymbol{\sigma}_2$ and $\boldsymbol{\sigma}_3$) scenarios. 
In the case of $\mathbf{n}_2$, these can be further divided into heteroscedastic settings with positive ($\boldsymbol{\sigma}_2$) and negative ($\boldsymbol{\sigma}_3$) pairing similar to \citet{paulyETAL2015}.
The random variables \new{$\eta_{11},...,\eta_{1n_1}, \eta_{21}, ..., \eta_{kn_k}$}  are drawn \new{independently} from five different distributions: $\mathcal{N}(0,1),\, \mathcal{LN}(0,1),\, \chi^2_3,\, t_2,\, t_3$. 
\new{Here, $\mathcal{LN}(0,1)$ denotes the log-normal distribution with parameters $0$ and $1$, $\chi^2_3$ denotes the $\chi^2$-distribution, and $t_m$ denotes the t-distribution with $m$ degrees of freedom.}
The constants $m_i$ in equation \eqref{eq:simmodel} represent the medians of the corresponding distribution.
We set $\mu_1 = ... = \mu_k = 0$ under the null hypothesis. 
For power simulations, a shift parameter \new{$\delta$} is added to the fourth group as in \citet{ditzhaus_qanova_2021}, that is $\mu_4 \new{= \delta} \in\{ 0.5,1,1.5,2\}$.
We run $N_{sim} = 5000$ simulation runs for each setting and use $B=2000$ resampling (permutation resp. bootstrap) iterations. 
The global level of significance was set to $\alpha = 5\%$. 
Furthermore, the three different covariance matrix estimators as described in \citet{ditzhaus_qanova_2021} are considered for all approaches.
\new{For the kernel estimator, we used the gaussian kernel and determined the bandwidth by using \newnew{the following \texttt{nrd0} method implemented in the \textsf{R} \citep{r_core_team_r_2024} function \texttt{bw.nrd0}, which is a version of} Silverman's rule-of-thumb \new{\citep[\newnew{p. 48}]{silverman_density_1998}}\newnew{:
The bandwidth is chosen as $0.9 n^{-1/5} \min\{SD, IQR/1.34\}$, where $SD$ denotes the standard deviation, $IQR$ the interquartile range, and $n$ the sample size, if $IQR > 0$.}
This ensures that the densities are well estimated which in turn ensures that the kernel estimator for the covariance works well.}
The multiple testing procedures that we compare are the asymptotic MCTP, 
the bootstrap MCTP, 
and the Bonferroni-adjusted (abbreviated as B.) asymptotic and permutation-based QANOVA tests of \citet{ditzhaus_qanova_2021}, as explained in Section~\ref{sec:methods}.
This leaves us with twelve different methods which are compared in 120 simulation scenarios. We first discuss their performance in terms of family-wise error rate (FWER) control.

\begin{figure}[t]
\includegraphics[width=\textwidth]{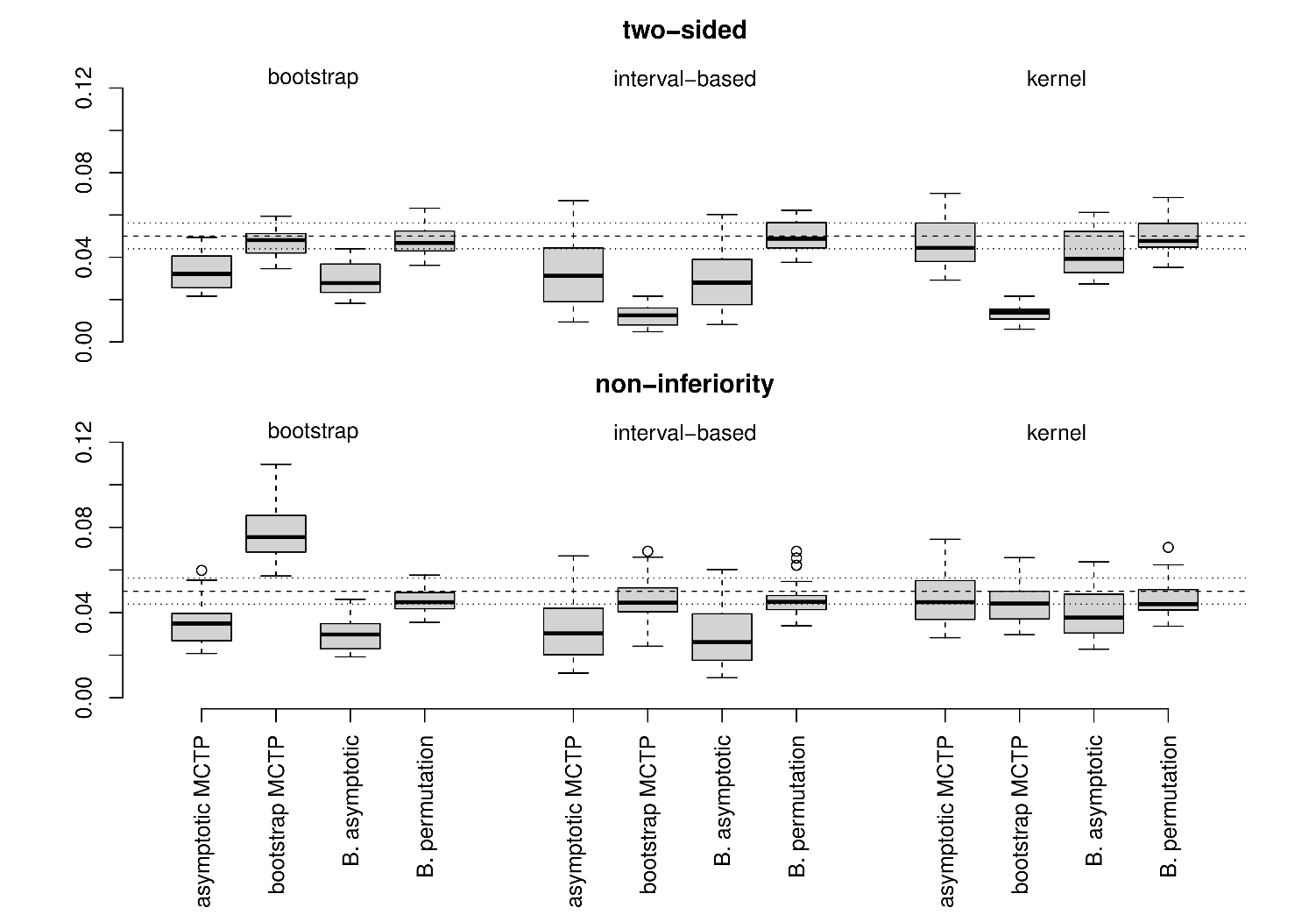} 
\caption{Empirical FWERs for \textit{Dunnett-type} contrasts with different hypotheses (top: two-sided and bottom: non-inferiority) and variance estimators (from left to right: bootstrap, interval-based or kernel). The dashed line represents the desired level of $\alpha=5\%$ and the dotted lines represent the Binomial interval $[0.044,0.0562]$ for $N_{sim} = 5000$ repetitions. }
\label{fig:Dunnett}
\end{figure}

\begin{figure}[t]
\includegraphics[width=\textwidth]{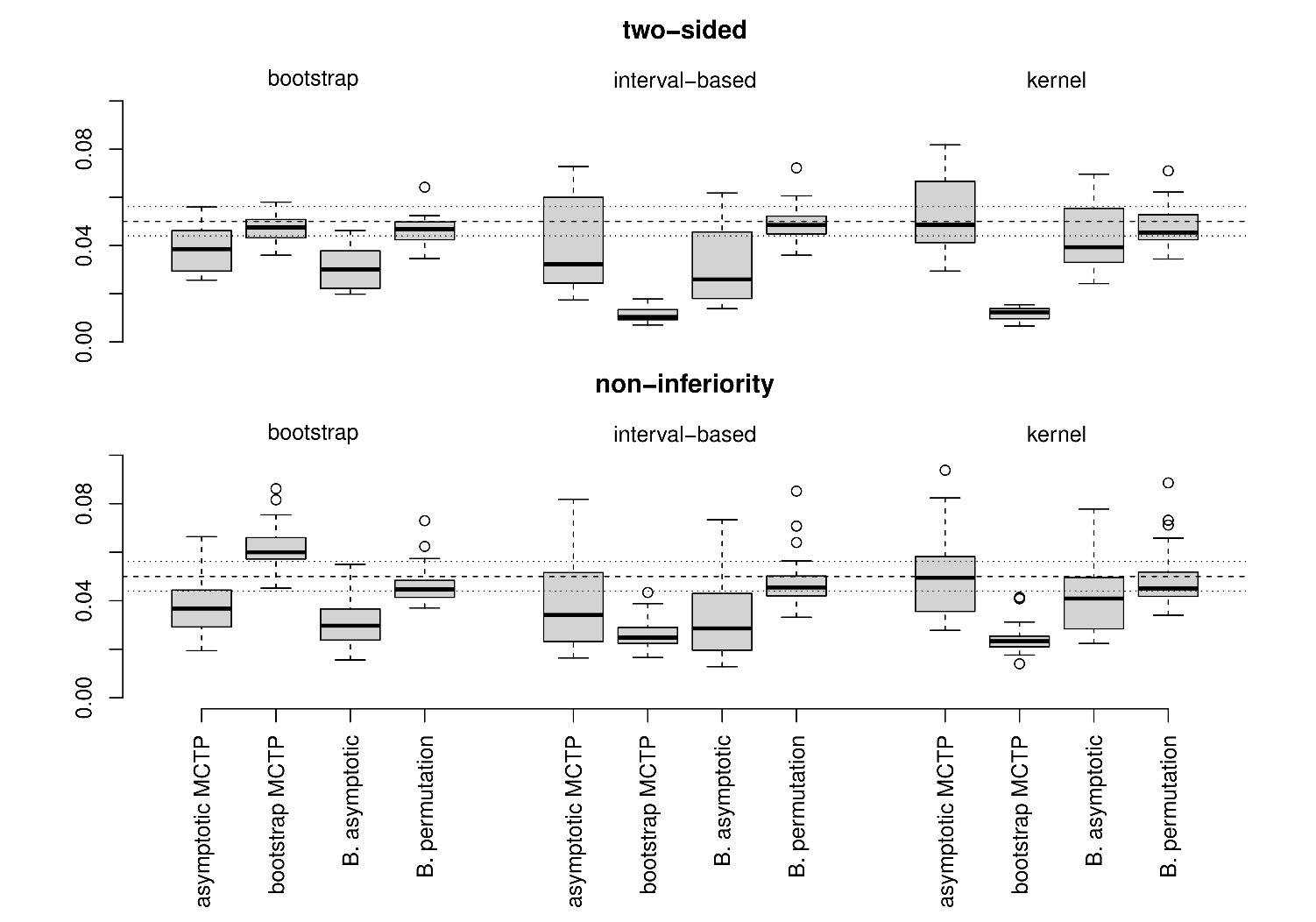} 
\caption{Empirical FWERs for \textit{Tukey-type} contrasts with different hypotheses (top: two-sided and bottom: non-inferiority) and variance estimators (from left to right: bootstrap, interval-based or kernel). 
The dashed line represents the desired level of significance of $\alpha=5\%$ and the dotted lines represent the Binomial interval $[0.044,0.0562]$ for $N_{sim} = 5000$ repetitions.  }
\label{fig:Tukey}
\end{figure}

\begin{figure}[t]
\includegraphics[width=\textwidth]{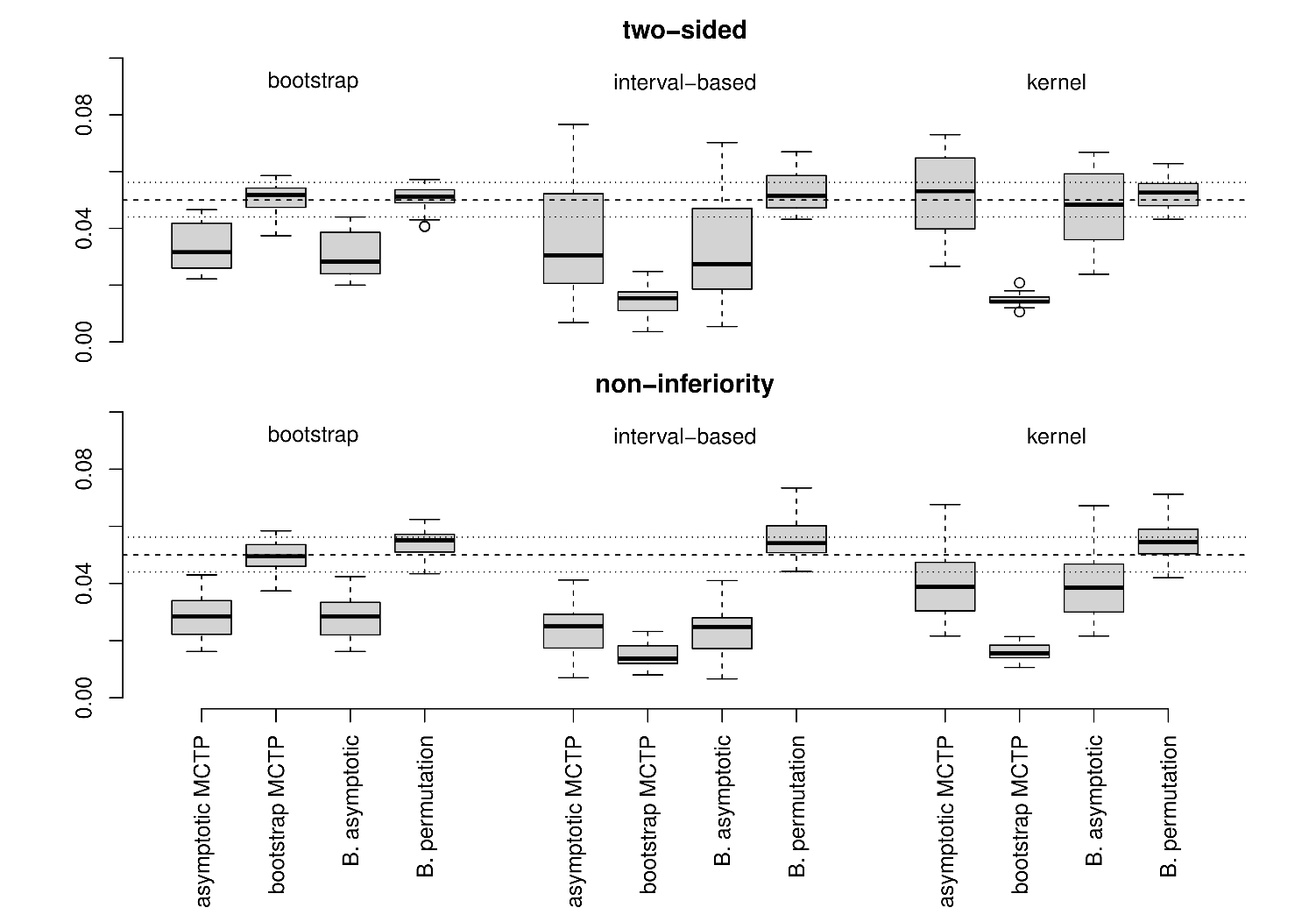} 
\caption{Empirical FWERs for \textit{Grand-mean-type} contrasts with different hypotheses (top: two-sided and bottom: non-inferiority) and variance estimators (from left to right: bootstrap, interval-based or kernel). 
The dashed line represents the desired level of significance of $\alpha=5\%$ and the dotted lines represent the Binomial interval $[0.044,0.0562]$ for $N_{sim} = 5000$ repetitions.  }
\label{fig:GM}
\end{figure}

{\bf Control of the family-wise error rate.} In Figures~\ref{fig:Dunnett}--\ref{fig:GM}, the empirical FWERs across all different scenarios are illustrated. 
The empirical FWERs for the asymptotic MCTP and the asymptotic Bonferroni-adjusted test vary more across the different settings.
These tests tend to be too conservative for the bootstrap and interval-based variance estimator, where the Bonferroni adjustment leads to slightly more conservative results than the asymptotic MCTP of Section~\ref{ssec:asymptoticMCTP}. 
Such a conservative behaviour can also be observed in many scenarios for the bootstrap MCTP with interval-based or kernel variance estimator. 
However, by using the bootstrap MCTP in combination with the bootstrap variance estimator, the type I error of the tests seem to increase and exceeds the desired level of 5\% in most of the scenarios for the Dunnett- and Tukey-type contrasts.
In contrast, the Bonferroni-adjusted permutation test has a most accurate FWER control across all scenarios. 
It only exhibits a slight liberality in case of the non-inferiority testing for the Grand mean multiple testing family.

\begin{figure}[t]
\includegraphics[width=\textwidth]{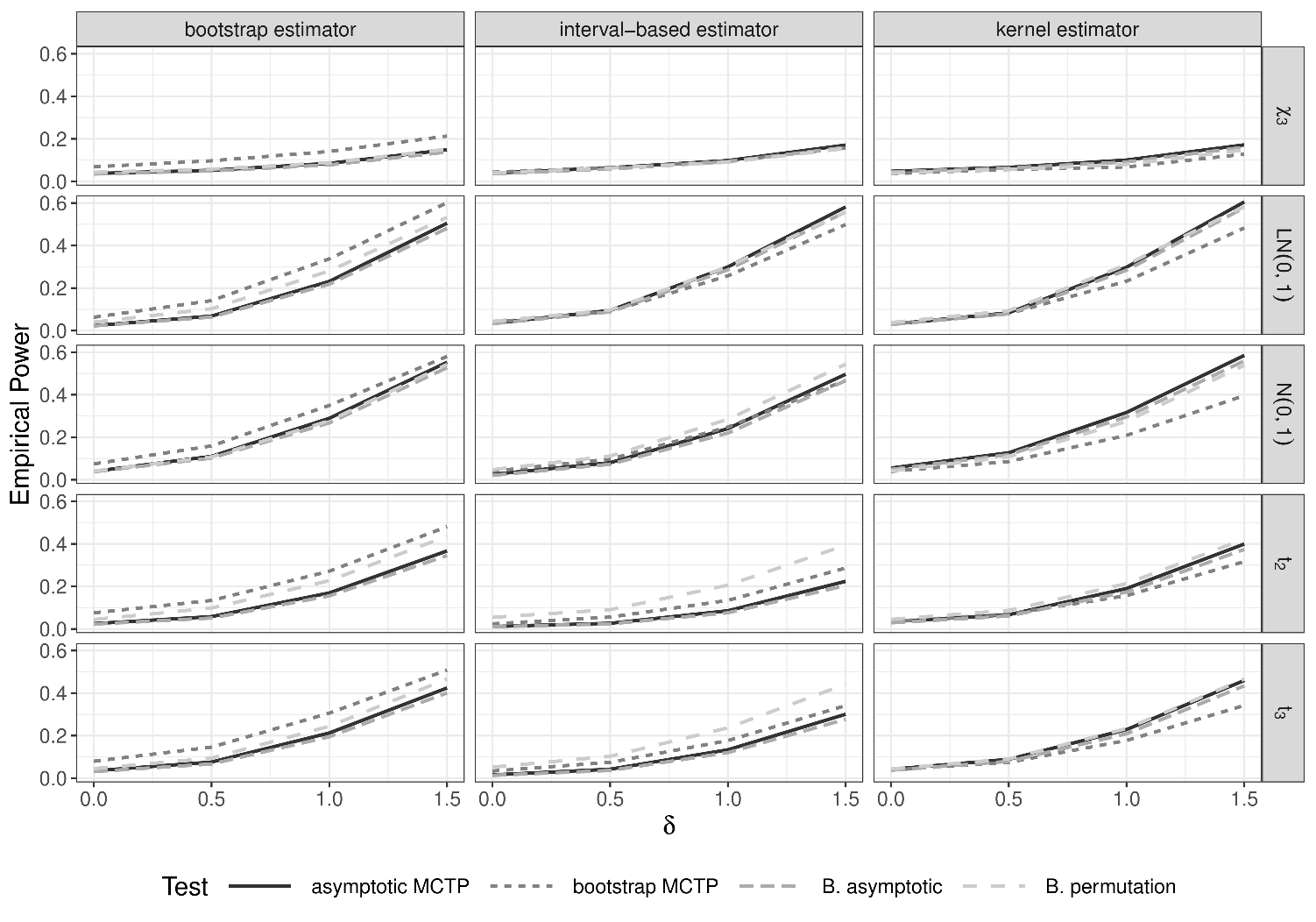} 
\caption{Empirical power for non-inferiority Dunnett-type contrast tests in the unbalanced ($\mathbf{n}_2$) heteroscedastic ($\boldsymbol{\sigma}_2$) design with positive pairing under different distributions and variance estimators (from left to right: bootstrap, interval-based or kernel). }
\label{fig:Power}
\end{figure}

\textbf{Power results.} The simulation results for the empirical global and local power can mainly be found in the Supplement (Figures~2-10). 
Here, the empirical global power denotes the rejection rate for a false global hypothesis, while the empirical local power is the rejection rate for a false local hypothesis.
It is observable that tests that performed too liberal in terms of type I error control generally also lead to a higher empirical global and local power (as expected). 
Moreover, the empirical global and local power is always comparable between the asymptotic MCTP and the Bonferroni-adjusted asymptotic and permutation test. 
In Figure~\ref{fig:Power}, exemplary empirical global power curves are shown for non-inferiority Dunnett-type contrast tests in the unbalanced heteroscedastic design with positive pairing ($\mathbf{n}_2$ and $\boldsymbol\sigma_2$). 
It can be seen that the bootstrap MCTP with the bootstrap covariance estimator has generally the highest empirical global power. However, the procedure is also often too liberal as we have seen before. 
By considering the other methods, we observe that the Bonferroni-adjusted permutation test is usually one of the methods with the highest global and local power or at least with a comparable power to the method with the highest global and local power, respectively. 
This is also the case for the other variance estimators. 
Especially for the interval-based estimator and the standard normal and t-distributions, the Bonferroni-adjusted permutation test clearly outperforms the other methods in terms of empirical global power.
Furthermore, it is observable that the Bonferroni-adjusted asymptotic test is slighly less powerful than the asymptotic MCTP in all scenarios.
All in all, regarding the local and global power one cannot make a clear recommendation, but the power of the Bonferroni-adjusted methods is in general not worse than the MCTP methods.

\textbf{Other effect parameters.} The results of the additional simulation study in the Supplement, where medians and IQRs are inferred simultaneously, are similar:
The Bonferroni-adjusted permutation test performs quite accurate in terms of FWER control while the asymptotic approaches and the bootstrap MCTP tend to be conservative in most scenarios.
Regarding the empirical power, the Bonferroni-adjusted permutation test is comparable and in some scenarios even more powerful than the other approaches.

\subsection{\newnew{Simulation Motivated by the Data Example}}\label{ssec:largesim}
 We also conducted an additional simulation study with $r = 16$ tests and larger sample sizes of 58-549 individuals per group as in the data example in Section~\ref{sec:analysis}. 
\newnew{We consider a modification of the simulation study in Section~\ref{ssec:smallsim} to analyze the performance of the methods in a framework that is closer to the considered data example in Section~\ref{sec:analysis}.
Therefore, we considered $k=17$ groups and used the Dunnett-type contrast matrix as $\mathbf{H}$ with group $17$ as base.
Furthermore, the constants are set to $\epsilon_1 = ... = \epsilon_{16} = 0$. 
 For the data generation, we used the model as in Section~\ref{ssec:smallsim}.
The sample sizes are set to $n =(59, 175,  98,  78, 280, 176, 351, 128, 368, 403, 240, 376, 278, 549, 428, $ $379, 250)$, which is similarly heterogeneous as the number of individuals in the 17 groups of the hatch data in Section~\ref{sec:analysis}, see Figure~\ref{fig:hatchdates}.
The variance setting is motivated by the data example as the parameters $\sigma_1,...,\sigma_{17}$ are chosen such that the variances of $X_{is}$ match the empirical variances in group $i$ for the hatch data of Section~\ref{sec:analysis}.
This  yields a heteroscedastic variance setting.
The random variables {$\eta_{11},...,\eta_{1n_1}, \eta_{21}, ..., \eta_{kn_k}$}  are drawn {independently} from four different distributions: $\mathcal{N}(0,1),\, \mathcal{LN}(0,1),\, \chi^2_3,\, t_3$. 
The reason why we exclude the $t_2$-distribution is that the variances of $X_{is}$ would not exist in this case.
Hence, it would not be possible to choose the parameters $\sigma_1,...,\sigma_{17}$ such that the variances of $X_{is}$ equal the empirical variances. 
The constants $m_i$ represent the medians of the corresponding distribution.
We set $\mu_1 = ... = \mu_{17} = 0$ under the null hypothesis. 
For power simulations, we set $\mu_i$ to the empirical median of group $i$ for the hatch data of Section~\ref{sec:analysis} for all $i\in\{1,...,16\}$ and $\mu_{17}$ to the empirical median of group $17$ for the hatch data minus $7$ (which is the constant $\epsilon_\ell$ in the data analysis).
All other parameters are set as in Section~\ref{ssec:smallsim}.
}

\newnew{\textbf{Control of the family-wise error rate.} The empirical FWERs under the null hypothesis across all scenarios are illustrated in Figure~\ref{fig:newFWER}.}
Here, we see that the results are not as surprising as for smaller sample sizes and less groups.
Particularly for the bootstrap variance estimator, the empirical FWERs of the MCTPs are quite close to the desired level $\alpha = 0.05$, while the interval and kernel estimators still show an observable deviance from $\alpha = 0.05$.
\newnew{
The Bonferroni-adjusted tests tend to be too conservative.
This might be explained by the large number of tests, that is 16.
The asmptotic MCTP with interval-based estimator performs slightly too liberal in the considered simulation settings.
\begin{figure}[bt]\centering
\includegraphics[width=\textwidth]{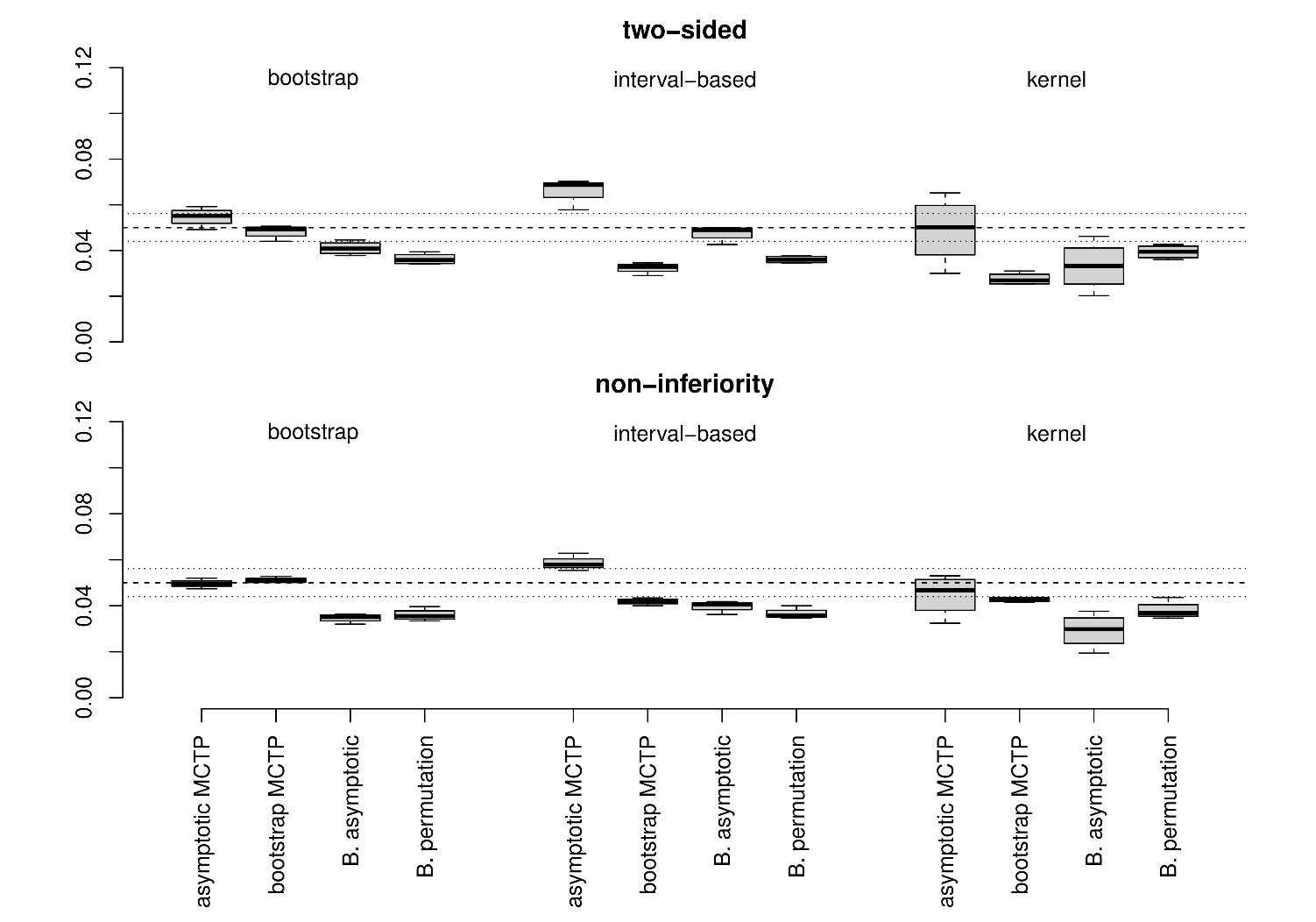} 
\caption{\newnew{Empirical FWERs for Dunnett-type contrasts with different hypotheses (top: two-sided and bottom: non-inferiority) and variance estimators (from left to right: bootstrap, interval-based or kernel).}}
\label{fig:newFWER}
\end{figure}
}

\newnew{\textbf{Power results.} The empirical global power, which is the rejection rate of the global null hypothesis under the alternative, has been exactly 1 for all scenarios.
This means that the global null hypothesis could be rejected in all simulation runs for all settings under the alternative.
}

\subsection{\newnew{Discussion of the results}} 
 The simulation results are quite surprising in several ways.
There are two well-known and often discussed problems with the Bonferroni adjustment in general: a loss of power \citep[e.g.][]{holm_simple_1979, olejnik_multiple_1997} and a rather conservative behaviour \citep[e.g.][]{westfall_p_1989, gordon_control_2007, chen_general_2017}.
From the method's definition it is clear that the conservative behaviour occurs if a large number of hypotheses is simultaneously tested or the hypotheses are highly correlated.
The situation that is described in other articles is vice-versa for MCTPs.
\citet{hasler_multiple_2008} and many others \citep[e.g.][]{bretz_numerical_2001, hasler_multiple_2008, konietschke_are_2013, hasler_multiple_2014, umlauft_wild_2019, noguchi_nonparametric_2020, rubarth_estimation_2022}
 showed in simulation studies that MCTPs hold their level of significance quite satisfactorily.
Furthermore, \cite{konietschke_are_2013} showed that the power of the global test decision of some mean-based MCTPs is comparable to the power of an ANOVA-F-test.
\newnew{In fact, this is exactly what we could observe in the simulation study of Section~\ref{ssec:largesim} with larger sample sizes and many hypotheses.}
From these observations one would assume that the MCTPs are the preferred method compared to Bonferroni-adjusted procedures.
\newnew{H}owever, the simulation results \newnew{of Section~\ref{ssec:smallsim} with smaller sample sizes} are in favour of the Bonferroni-adjusted permutation approach.
We point out that the good behaviour of the Bonferroni adjustment can only be observed \new{for small sample sizes} in combination with the permutation approach, the standard asymptotic version was often observed to be too conservative.
\new{Moreover, it is important to note that the Bonferroni correction can not really be improved by a MCTP for large negative correlations between the test statistics.
However, this situation mainly occurs for the non-inferiority tests with Tukey- and Grand-mean-type matrix in our simulations, cf. Section~\ref{ssec:Correlation} for details.
For highly positive correlated tests it is well known that the Bonferroni correction performs too conservative, which is simply a consequence of the Bonferroni inequality.
In our simulation settings we can observe the highest positive correlation in Dunnett-type tests with non-inferiority hypothesis (median correlation $0.524$ with bootstrap covariance estimator, cf. Section~\ref{ssec:Correlation}), but can not observe that the Bonferroni-correction Permutation approach behaves very conservative. }
It should also be emphasized that most MCTPs and corresponding simulations or analyses use means and not quantiles as an estimand.
Furthermore, \citet{vanderweele_desirable_2019} stated that there are still many testing problems where the behaviour is not or only a little bit conservative and the tests are still rejecting even if they are less powerful than other tests that control the FWER.
Moreover, the Bonferroni-adjusted QANOVA  permutation approach and the MCTPs are not directly comparable as they use \new{different} techniques to derive critical values. 
In particular, the estimation of the covariance matrix is more crucial for the MCTPs than for the QANOVA as the latter may have a balancing effect through the studentized permutation approach. 
As the estimation of the underlying covariance structures is much more complex in the case of quantiles compared to classical mean-based approaches, this could be one reason for our results.
In context of the simulation study of \citet{segbehoe_simultaneous_2022}, our simulation results lead to the conclusion, that MCTPs regarding quantiles do not perform well for \new{smaller sample sizes and} more than three groups, especially less well than the Bonferroni-adjusted permutation test.
 
\textbf{Recommendation.} To conclude, we recommend to use the Bonferroni-adjusted permutation test \new{for small sample sizes and few hypotheses} due to a quite accurate FWER control and comparable power to other methods.
The simulations indicate that the choice of the variance estimator has no big impact on the permutation tests decision.
\new{All methods are expected to perform similarly well for larger sample sizes regarding the FWER control.
However, if the number of tests increase, the Bonferroni adjustment may lead to conservative test results.
This observation is not suprising and refers to the well-known disadvantages of the Bonferroni-correction. 
In order to ensure more powerful test decisions, we recommend to use the bootstrap MCTPs with bootstrap variance estimator in the case of many hypotheses and large sample sizes.
}

\section{Data Example: Early and Late Buzzards}\label{sec:analysis}
Birds living in temperate climates have to cope with changing seasons during the year; they have to adapt to different weather conditions, temperatures and length of daytimes \citep{begon_ecology_2021}.
Parental care is probably one of the most important activities of birds regulated by the seasons due to the strong connection to reproduction and fitness \citep{caro_antipredator_2005}.
For this energy-demanding task, most birds must rely on sufficient resources to feed their young and thus are dependent on a small time frame during the year, when enough of these resources are available \citep{verhulst_timing_2007}.
To do so, most birds rely on hints from temperature or length of daylight \citep{verhulst_timing_2007} to time hatching in the best possible way. 
Since human-induced climate change alters weather conditions as well as temperature developments through the year way faster than during earlier decades and centuries \citep{sippel_climate_2020}, birds relying on these influences to time their reproduction were shown to change their reaction accordingly \citep{halupka_effect_2017}.
At first glance this might seem positive as climate change leads in general to warmer temperatures and hence the reproductive period during the year should potentially increase \citep{mcdermott_long-term_2016}.
However, not all organisms react in the same way and at the same pace to these changes, leading to potential mismatches in the food web \citep{drever_spring_2007}.

\begin{figure}\centering
\includegraphics[height=4.4cm]{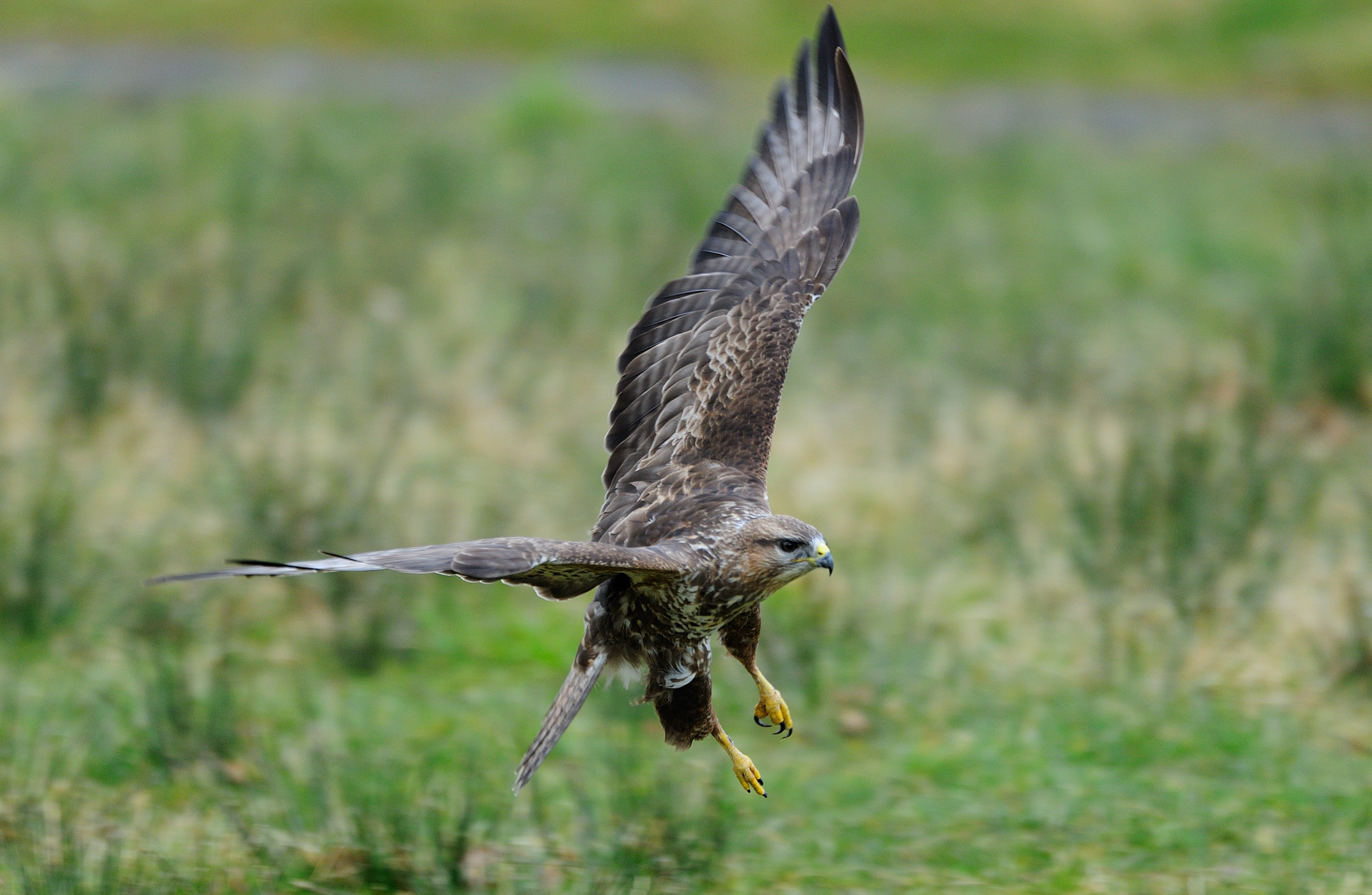}
\includegraphics[height=4.4cm]{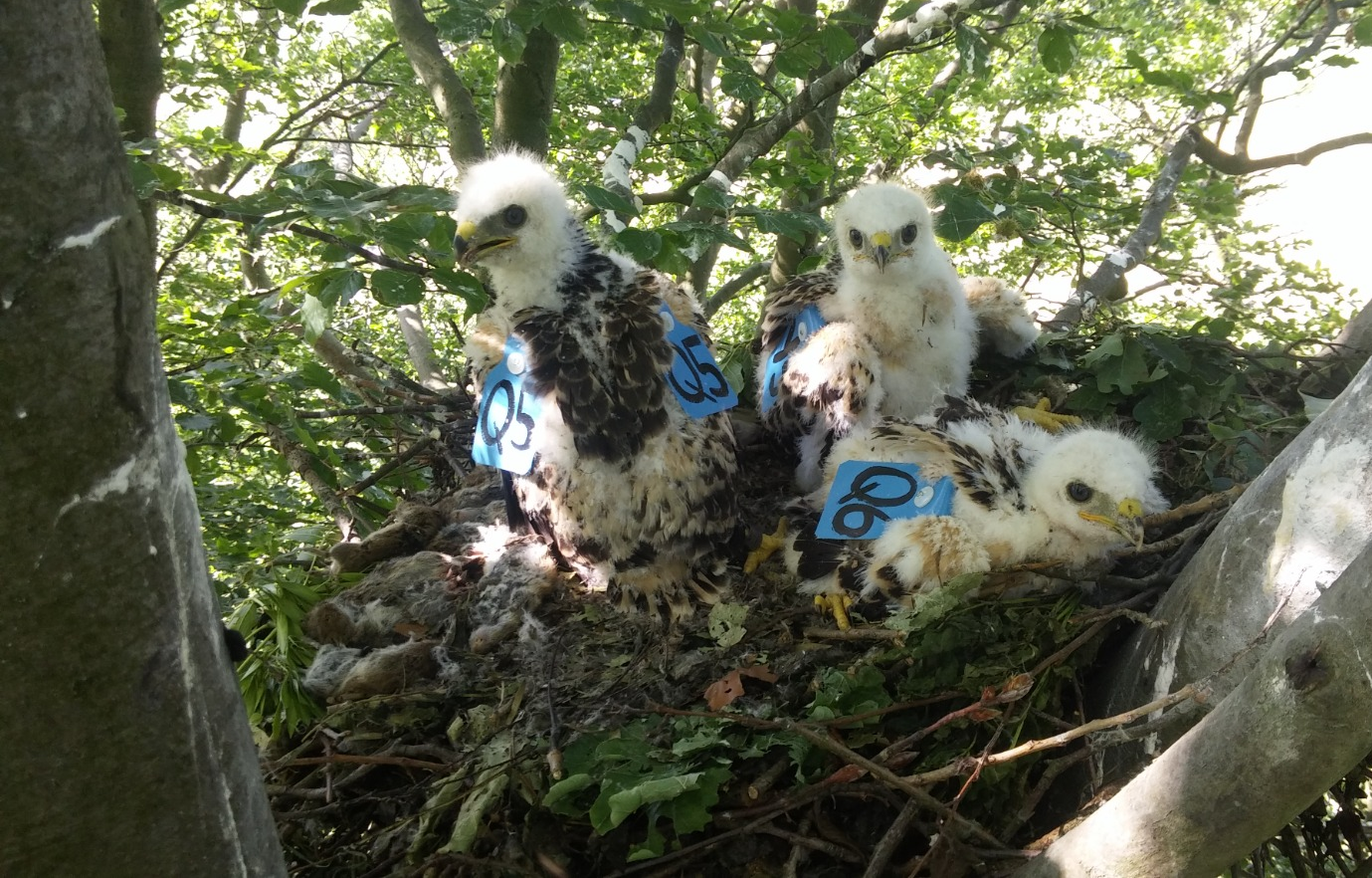}
\caption{Adult common buzzard during flight (left) and buzzard nestlings in the nest (right). $\copyright$ O. Krüger, N. Chakarov.}
\label{fig:buzzards}
\end{figure}

Common buzzards (\textit{Buteo buteo}, Figure \ref{fig:buzzards}) are medium-sized birds of prey and feed mostly on small mammals and birds \citep{walls_common_2020}.
As being predators, they are dependent on the performance of many other organisms, not only their prey, but also their prey’s food resources \citep{mittelbach_community_2019}.
It is known that buzzards have higher breeding success under certain weather conditions \citep{kostrzewa_relationship_1990, kruger_dissecting_2002, kruger_importance_2004}.
Their main prey, field voles (\textit{Microtus arvalis}), often shows fluctuating population densities between years \citep{frank_causality_1957} caused by different factors like predation pressure or snow level during winter \citep{boyce_population_1988}, also influencing breeding success in buzzards \citep{lehikoinen_reproduction_2009}.
In their study, \cite{lehikoinen_reproduction_2009} showed as well that common buzzards in Finland started breeding earlier and shifted their range more towards the north because of the warmer climate.

\begin{figure}\centering
\includegraphics[width=0.8\textwidth]{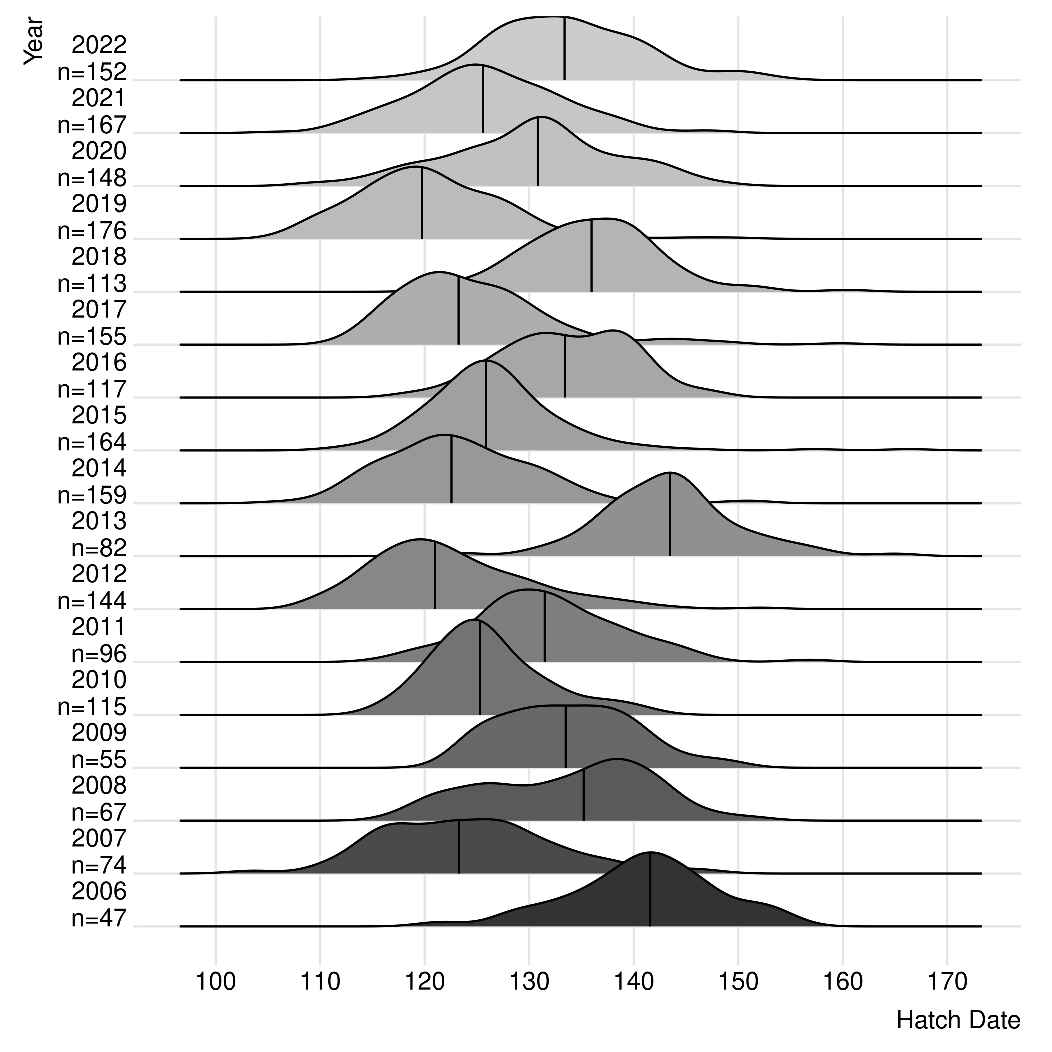} 
\caption{Kernel density estimators of hatching dates \new{(in days)} for the years $2006-2022$ with the sample sizes for every year. The black vertical line marks the median.}
\label{fig:hatchdates}
\end{figure}

Data was collected from 2006 to 2022 by the Department of Animal Behaviour in a study area in north-west of Germany (see \cite{chakarov_variation_2013} for a description of the study area and sampling procedure).
\new{We only consider the age of the first-hatched nestling of each brood for this analysis to avoid dependencies between siblings.}
With the relationship between the age of the chicks and their wing length observed by \citet{bijlmsa_1999_sex}, we are able to calculate the hatch dates of the chicks.
\new{
We use \textsf{R} 4.4.2 \citep{r_core_team_r_2024} and  the implementation in \textsf{R} by \citet{ottensmann_r-package_2022} for the calculation.
Here, we use as a scaling the \textit{day of the year}, where $32$ means 1st February and $120$ means 30th April in non-leap years.
From this, the hatch dates are calculated by the day of observation minus the age in days.
Therefore, the considered hatch dates are the result of a polynomial model and accordingly metric.}
This data is shown in Figure \ref{fig:hatchdates} as kernel density estimators with gaussian kernels and a bandwi\new{d}th determined \newnew{by the \texttt{nrd0} method as explained in Section~\ref{ssec:smallsim}}.
\new{The complete material of this analysis as well as the data can be found in the Supporting Information.}
 
In Figure \ref{fig:hatchdates}, there is a high variability between years regarding the hatch dates of common buzzard nestlings.
Biologists who study these animals often have the impression of particular \textit{early} and \textit{late} years, especially if the behaviour of the buzzards differs from the years before.
As this is also observable in the kernel density estimators in Fig. \ref{fig:hatchdates}, it is a motivation to search for possible reasons.
In the years from 2019 onwards this pattern seemed to be changing as the years 2019, 2020 and 2021 tended to be earlier in contrast to the year 2022.
This is our motivation to take 2022 as a reference year for \textit{late} years.
There are potentially several reasons for this phenomenon which are difficult to measure, but the division into two groups (early and late years) is a simplification that makes it possible to get a more accurate view.
For that, we do a multiple testing procedure that identifies similar years.
In context of a directed scenario, the multiple testing procedure that identifies similar years can be understood as a multiple non-inferiority testing problem as described in (\ref{eq:OnesidesTestingProc}).
Regarding the structure of the data, heavy-tailed distributions appear quite frequently (Fig. \ref{fig:hatchdates}).
The simulation studies of \citet{ditzhaus_qanova_2021} show that \new{median-based tests have a higher power than mean-based tests in the context of heavy-tailed data.}
So if one wants to identify similar years to investigate possible reasons, it is in this case most suitable to use a median-based approach.
\new{The sample sizes of the $k=17$ groups are shown in Figure \ref{fig:hatchdates}.
From this it can be seen that the groups are highly unbalanced.
This is no problem because from the simulation study we observed that all methods can deal with high variation in sample sizes between samples.
As this behaviour can also be observed in the simulations of \citet{ditzhaus_qanova_2021} and \citet{baumeister_quantile-based_2024} this seems to be a useful property of testing regarding quantiles.}
Consider Figure~1 in the Supplement for an analysis of that property in our simulation study.
\new{Since the simulation setup of Section~\ref{sec:sim} does not perfectly fit to the data example, we have conducted a further simulation motivated by the data example.
The detailed description of the scenarios and the results can be found in Section~5 in the Supplement.}

We use the median $m_\ell$ of the year $\ell\in\{06,07,08,\dots,19,20,21\}$ ($m=1,\,p_1=0.5$, $k=17$ groups) of the hatch dates and $\epsilon_\ell=7$ for all $\ell$ regarding to the intuitive observations of the ecologists.
Hatches one week \new{(seven days)} later do not lead to the conclusion that a year is late.
This situation leads to a Dunnett's test or many-to-one procedure and has the concrete form:
\newnew{
\begin{align}\label{eq:birdhypothesis}
 \mathcal{H}_{0,\ell}^I:\,m_{22}-m_{\ell}\ge7\quad\text{vs.}\quad\mathcal{H}_{1,\ell}^I:\,m_{22}-m_{\ell}<7,\,\ell\in\{06,07,08,\dots,19,20,21\}.
 \end{align}
 }
To realize these hypotheses of interest we use the Dunnett-type hypothesis matrix
\new{
\begin{align*}
\mathbf{H}=[\diag(\mathbf{1}_{16}),\,-\mathbf{1}_{16}].
\end{align*}
}
Note, that this framework assumes independent groups, which means in terms of content that the years are assumed to be independent.
This is a plausible assumption because of the high fluctuation of the buzzards in the data sample.
As the data collection is based on the defined area and not on the individual buzzard, the sample size differs through the years \new{(see Figure \ref{fig:hatchdates})}, every year birds migrate to the area, some leave it and other change their nest within the area. 
Therefore, the data is not collected as a paired sample.

\new{In the context of the not entirely clear simulation results, we consider the four presented testing methods (cf. Section \ref{sec:methods}) with the bootstrap covariance estimator as the simulation results for larger sample sizes indicates the best performance for that covariance estimator in the bootstrap MCTP while this does not seems to be relevant in other situations and for other methods.
We used $1999$ iterations for both resampling methods.}
\new{The results for the selected test decisions are given as $p$-values and as confidence intervals in Table \ref{tab:testresult}, cf. Section \ref{sec:methods}.}
\new{Here, all testing methods indicate that the null hypotheses $\mathcal{H}^I_{0,06}$, $\mathcal{H}^I_{0,08}$, $\mathcal{H}^I_{0,09}$, $\mathcal{H}^I_{0,11}$, $\mathcal{H}^I_{0,13}$, $\mathcal{H}^I_{0,16}$, $\mathcal{H}^I_{0,18}$ and $\mathcal{H}^I_{0,20}$ are rejected to a significance level of $\alpha=0.05$.}
\new{In line with the non-inferiority multiple testing problem, the global hypothesis $\mathcal{H}^{I}_0 = \bigcap_{\ell=1}^r \mathcal{H}_{0,\ell}$ is also rejected.} 
This means that the hatch days in the years 2006, 2008, 2009, 2011, 2013, 2016, 2018, and 2020 are identified as \new{\newnew{at least as late} as the year 2022 because the difference of the median hatch days between these} years is not significantly bigger as one week.
\newnew{These \textit{late} years} can be used for further investigation of reasons for different hatching dates.
Here, the most interesting result is that the year 2020 is still categorized as an late year, although the density plots in Figure \ref{fig:hatchdates} suggest a seemingly high similarity of the years 2019, 2020 and 2021.
\new{However, the tests indicate \newnew{that 2020 is at least as late as 2022}, \newnew{while 2019 and 2021 could not be identified as at least late as 2022}, which is an important information to look for possible reasons.}
This is a good motivation for using the median as estimand because it is not sensitive to the heavy-tailed data.
\new{In line with the simulation results for bigger sample sizes the testing procedures perform similarly.
In the sense of the simulation study, special attention should be paid to the MCTP with bootstrap.
}
\begin{table}[t]
\centering
\scriptsize
\newnew{
\begin{tabular}{lrrrrrrrr}
\toprule
 &\multicolumn{2}{c}{asymp. MCTP} & \multicolumn{2}{c}{boot. MCTP} & \multicolumn{2}{c}{B. asymp.} & \multicolumn{2}{c}{B. perm.}\\
\cmidrule(lr){2-3}\cmidrule(lr){4-5}\cmidrule(lr){6-7}\cmidrule(lr){8-9}
Test & $p$-value & SCI & $p$-value & SCI & $p$-value & SCI & $p$-value & SCI\\
 \midrule
$\hat m_{22}-\hat{m}_{06} =-\,08.19$ & $<0.0001$ & -5.01 & $<0.0005$  & -4.93 & $<0.0001$ & -4.83 & $<0.0006$  & -4.94 \\ 
$\hat m_{22}-\hat{m}_{07}=\quad10.11$ & $1.0000$ & 14.15 & 1.0000 & 14.25 & 1.0000 & 14.37 & 1.0000 & 14.04 \\ 
$\hat m_{22}-\hat{m}_{08}=-\,01.83$ & $<0.0001$ & 2.71 & $<0.0006$  & 2.83 & $<0.0001$ & 2.97 & $<0.0006$  & 2.87 \\ 
$\hat m_{22}-\hat{m}_{09}=-\,00.11$ & $<0.0001$ & 3.77 & $<0.0006$  & 3.87 & $<0.0001$ & 3.99 & $<0.0006$  & 4.35 \\ 
$\hat m_{22}-\hat{m}_{10}=\quad08.08$ & $0.9998$ & 10.11 & 1.0000 & 10.17 & 1.0000 & 10.23 & 1.0000 & 10.05 \\ 
$\hat m_{22}-\hat{m}_{11}=\quad01.90$ & $<0.0001$ & 4.82 & $<0.0006$  & 4.90 & $<0.0001$ & 4.98 & $<0.0006$  & 5.08 \\ 
$\hat m_{22}-\hat{m}_{12}=\quad12.39$ & 1.0000 & 14.94 & 1.0000 & 15.01 & 1.0000 & 15.09 & 1.0000 & 15.04 \\ 
$\hat m_{22}-\hat{m}_{13}=-\,10.09$ & $<0.0001$ & -7.51 & $<0.0006$ & -7.44 & $<0.0001$ & 7.36 & $<0.0006$  & -7.31 \\ 
$\hat m_{22}-\hat{m}_{14}=\quad10.82$ & 1.0000 & 13.44 & 1.0000 & 13.51 & 1.0000 & 13.59 & 1.0000 & 13.91 \\ 
$\hat m_{22}-\hat{m}_{15}=\quad07.51$ & 0.9930 & 9.97 & 0.9990 & 10.03 & 1.0000 & 10.10 & 1.0000 & 10.41 \\ 
$\hat m_{22}-\hat{m}_{16}=-\,00.05$ & $<0.0001$ & 2.95 & $<0.0006$  & 3.02 & $<0.0001$ & 3.12 & $<0.0006$  & 3.23 \\ 
$\hat m_{22}-\hat{m}_{17}=\quad10.14$ & 1.0000 & 13.01 & 1.0000 & 13.08 & 1.0000 & 13.17 & 1.0000 & 13.10 \\ 
$\hat m_{22}-\hat{m}_{18}=-\,02.58$ & $<0.0001$ & 0.16 & $<0.0006$  & 0.23 & $<0.0001$ & 0.31 & $<0.0006$  & 0.21 \\ 
$\hat m_{22}-\hat{m}_{19}=\quad13.64$ & 1.0000 & 16.01 & 1.0000 & 16.07 & 1.0000 & 16.15 & 1.0000 & 16.12 \\ 
$\hat m_{22}-\hat{m}_{20}=\quad02.56$ & $<0.0001$ & 4.91 & $<0.0006$  & 4.97 & $<0.0001$ & 5.05 & $<0.0006$  & 5.09 \\ 
$\hat m_{22}-\hat{m}_{21}=\quad07.80$ & 0.9967 & 10.43 & 0.9990 & 10.50 & 1.0000 & 10.59 & 1.0000 & 10.66 \\ 
\bottomrule
\end{tabular}}
\caption{\new{Test results of the four testing procedures (from left to the right: \textit{asymptotic MCTP, bootstrap MCTP, Bonferroni asymptotic} and \textit{Bonferroni permutation}, cf. Section \ref{sec:methods}) with bootstrap covariance estimator regarding the testing problem given in Equation \ref{eq:birdhypothesis}. The test results are given as $p$-value and as the right value of the one-sided confidence interval $(-\infty,\,\cdot\,]$ for \newnew{$m_{22} - m_\ell$}, cf. Equation \ref{eq:oneCI}.  Interpretation: \newnew{Rejecting local null hypotheses $\mathcal{H}_{0\ell}^I$ means that we can rule out that the median hatch date of year $\ell$ is at least 7 days later than for year 2022.} The $p$-values for the bootstrap MCTP and the Bonferroni permutation consider the number of resampling iterations and are calculated by Equation 17.7 in \citet{lehmann_testing_2022}. This testing problem uses Dunnett-type contrasts, the median hatch date of the year 2022 is compared with the median hatch dates from the years $2006$ to $2021$. The first column contains the differences of the empirical medians between the year 2022 and the respective other years.}}
\label{tab:testresult}
\end{table}

\section{Conclusion and Outlook}\label{sec:end}
We have compared different approaches to solve one-sided and two-sided multiple testing problems regarding one or more quantiles simultaneously.
To this end we have presented and extended two Bonferroni-adjusted methods, an asymptotic and a permutation approach, and an asymptotic and a bootstrap multiple contrast testing procedure in a comparable multiple testing framework for two-sided and non-inferiority hypotheses.
As a motivation for this kind of testing problems we gave a non-inferiority example from ecology, which deals with hatch dates in context of climate change.
To investigate the behaviour of the methods we have conducted an intensive simulation study.
Here, our main motivation was to compare Bonferroni adjustement and MCTPs in context of testing regarding quantiles.
In-line with \cite{vanderweele_desirable_2019} we have found out that the Bonferroni adjustment can be conservative, but when combined with a permutation approach \new{in the situation of small sample sizes} it performs better than its reputation.
The often-read claim that the Bonferroni method is \new{in general} too conservative \citep{gordon_control_2007}, cannot be confirmed when inferring quantiles.
We also wanted to ask the question whether the MCTPs are less conservative and have more power than Bonferroni-adjusted approaches. 
Our clear answer in this quantile-based setting \new{with small samples} is: no.
This is because of the behaviour of the Bonferroni-adjusted permutation approach, which is very stable.
Independently from the considered distributions, the covariance structure or the sample sizes, its empirical FWER-control was quite accurate and there was almost no power loss compared to the MCTPs.
\new{In contrast to the asymptotic and resampling-based MCTP-approaches the permutation-based method does not seem to need bigger sample sizes to work well.
For both small and large samples, the resampling-based methods show a clear improvement in the test performance.}

We also want to point out that hypotheses formulated in terms of quantiles can be useful in lots of situations.
This is particularly important in the context of data that refers to animal and human behaviour, as this situations are known to be rather skewed and can be rarely modelled as homoscedastic and normally distributed \citep[e.g.][]{gardiner_fitting_2014}.
As multiple testing problems occur very often in this field of science \citep{farcomeni_review_2008}, our analyses can be helpful in the selection of the appropriate method.

For future research, it remains to create an implementation in \textsf{R} for the presented methods as well as for other quantile-based methods for factorial designs, e.g., \citet{ditzhaus_qanova_2021, baumeister_quantile-based_2024}.
Additionally, it can be investigated how multiple testing regarding multivariate quantiles can be realized by extending the QMANOVA of \citet{baumeister_quantile-based_2024}.
Furthermore, it would be interesting to have more systematic comparisons between MCTPs and other multiple testing procedures like the Bonferroni adjustment for other estimands of interest.
\new{Especially for mean-based methods it would be interesting to investigate if a similar simulation-based comparison comes to the same conclusion as our simulation does.
Then a general statement could be made about whether this relationship between the behaviour of the Bonferroni-correction and sample size occurs systematically.}
\new{As \citet[Sec. 6.3]{besag_bayesian_1995} introduced quantile-based simultaneous credible regions, there are also bayesian approaches which could be compared with the methods presented in this paper in further comparisons.}
From this we hope to gain a better overview of the behaviour of MCTPs in relation to FWER-control, power and further concepts.

\vspace{12 pt}
\noindent\textbf{Author Contributions (CRediT)}

M.B.: Data curation, Formal analysis, Methodology, Project administration, Software, Validation, Visualization, Writing - original draft;

M.M.: Formal analysis, Methodology, Software, Validation, Visualization, Writing - original draft;

K.P.G.: Data curation, Investigation, Resources, Writing - original draft;

M.D.: Conceptualization, Funding acquisition, Methodology, Supervision, Writing - review \& editing;

N.C.: Investigation, Resources, Writing - review \& editing;

M.P.: Conceptualization, Funding acquisition, Methodology,  Supervision, Writing - review \& editing.

\vspace{12 pt}
\noindent {\textbf{Conflict of Interest}}
\noindent {\textit{The authors have declared no conflict of interest.}}
\vspace{12 pt}

\noindent{\textbf{Data Availability Statement}}
The data that support the findings of this study e.g. the data example, simulation scripts and results are openly available in TUDOdata at \url{http://doi.org/10.17877/TUDODATA-2025-M6TDKFDE}.

\noindent{\textbf{Acknowledgement}}
M.M. and M.D. gratefully acknowledge funding by the Deutsche Forschungsgemeinschaft - 314838170, GRK 2297 MathCoRe.
The work of M.B. and M.P. has been partly supported by the Research Center Trustworthy Data Science and Security (\url{https://rc-trust.ai}), one of the Research Alliance Centers within the \href{https://uaruhr.de}{UA Ruhr}.
The work of K.P.G. has been partly supported by the Friedrich-Ebert-Stiftung (FES) and the work of K.P.G and N.C. has been partly supported by the SFB TRR 212
(NC³) (Project Number 396780709) of the Deutsche Forschungsgemeinschaft (DFG, German Research Foundation).
We thank Meinolf Ottensmann for his help with the calculation of the hatch dates.
Furthermore, we thank Paavo Sattler for many helpful advices on multiple testing.

\section{Appendix}

\subsection{Covariance Estimators}\label{sec:covEst}
Let $\boldsymbol\Sigma^{(i)} = (\boldsymbol\Sigma^{(i)}_{ab})_{a,b\in\{1,\dots,m\}}$ denote the covariance matrix of group $i\in\{1,\dots,k\}$.
In our analysis, we consider the following three different covariance estimators for $\boldsymbol\Sigma^{(i)}_{ab}, a,b \in\{1,\dots,m\},$ as discussed in \citet{ditzhaus_qanova_2021}:
\begin{enumerate}
\item {\bf Kernel estimator.}  
\new{The main idea of the kernel estimator is to replace the unknown densities in \eqref{eq:Sigma} by kernel density estimators. 
Therefore, l}et $K_i:\R\to [0,\infty)$ with $\int_{\R} K_i(x)\;\mathrm{d}x = 1$ denote a Lebesgue density, $h_{ni}\to 0$ as $n\to\infty$ a bandwidth, and $$\widehat{f}_{K_i,i} (x) = (n_ih_{ni})^{-1} \sum_{i=1}^{n_i} K_i\left( \frac{x - X_{ij}}{h_{ni}} \right)$$ the kernel density estimator for $f_i$ for all $i\in\{1,\dots,k\}$. 
Then, the kernel estimator for $\boldsymbol\Sigma^{(i)}_{ab}$ is given by
$$ \widehat{\boldsymbol\Sigma}^{(i),K}_{ab} = \frac{n}{n_i}\frac{\min\{p_a,p_b\} - p_ap_b}{\widehat{f}_{K_i,i}(\widehat{q}_{ia})\widehat{f}_{K_i,i}(\widehat{q}_{ib})}. $$
\item {\bf Bootstrap estimator.} 
\new{For the bootstrap estimator, we use the fact that the mean squared error of the bootstrapped sample quantile, that can be calculated as}
$$ \widehat{\sigma}_i^*(p_r) := \left( n_i \sum_{j=1}^{n_i} (X^{(i)}_{j:n_i} - \widehat{q}_{ir})^2 P_{ijr} \right)^{1/2}\quad\text{for all } r\in\{1,\dots,m\}, $$
\new{converges in probability to the asymptotic standard deviation of the corresponding sample quantile $\sqrt{\kappa_i\boldsymbol{\Sigma}^{(i)}_{rr}} = \sqrt{p_r - p_r^2}/f_i(q_{ir})$ \citep{ditzhaus_qanova_2021},}
where $X^{(i)}_{j:n_i}$ denote the $j$th smallest element of the ordered $i$th sample and $$P_{ijr} := B_{n_i,(j-1)/n_i}\left( (-\infty, \lceil n_ip_r\rceil -1] \right) - B_{n_i,j/n_i}\left( (-\infty, \lceil n_ip_r\rceil -1] \right)$$
for $B_{n_i,p}$ denoting the binomial distribution with size parameter $n_i$ and success probability $p$.
\new{Further, equation (8) in \citet{ditzhaus_qanova_2021} shows that $\boldsymbol{\Sigma}^{(i)}_{ab}$ only depends on $\boldsymbol{\Sigma}^{(i)}_{aa}$, $\boldsymbol{\Sigma}^{(i)}_{bb}$, $p_a$ and $p_b$ through
\begin{align}\label{eq:equation8}
{\boldsymbol\Sigma}^{(i)}_{ab} = \sqrt{\boldsymbol{\Sigma}^{(i)}_{aa}\boldsymbol{\Sigma}^{(i)}_{bb}}\frac{\min\{p_a,p_b\} - p_ap_b}{\sqrt{(p_a-p_a^2)(p_b-p_b^2)}}. 
\end{align}
Thus,}
 the bootstrap estimator for $\boldsymbol\Sigma^{(i)}_{ab}$ is given by
$$ \widehat{\boldsymbol\Sigma}^{(i),B}_{ab} = \frac{n}{n_i}\widehat{\sigma}_i^*(p_a)\widehat{\sigma}_i^*(p_b)\frac{\min\{p_a,p_b\} - p_ap_b}{\sqrt{(p_a-p_a^2)(p_b-p_b^2)}}. $$
\item {\bf Interval-based estimator.} 
\new{
For the interval-based estimator, we use the extended estimator for the standard deviation of the $p$th sample quantile \citep{ditzhaus_qanova_2021}, which is motivated by an estimator of \citet{mckean_comparison_1984} based on a standardized confidence interval. 
The extended estimator of \citet{ditzhaus_qanova_2021} is given by}
$$ \widehat{\sigma}_i^{PB}(p) := n_i^{1/2}\frac{X^{(i)}_{u_i(p):n_i}-X^{(i)}_{l_i(p):n_i}}{2z_{1-\alpha^*_{n_i}(p)/2} + 2n_i^{-1/2}} \quad\text{for all } p\in (0,1), $$
where $l_i(p) := \max\{1, \lfloor n_ip - z_{1-\alpha/2}\sqrt{n_ip(1-p)} \rfloor\}$, $u_i(p):=\min\{n_i, \lfloor n_ip + z_{1-\alpha/2}\sqrt{n_ip(1-p)} \rfloor\}$, $z_{1-\alpha/2}$ denotes the $(1-\alpha/2)$-quantile of the standard normal distribution and 
$$\alpha^*_{n_i}(p):=  1 - \sum_{j=l_i(p)+1}^{u_i(p) -1} \begin{pmatrix}n_i\\j\end{pmatrix}p^j (1-p)^{n_i-j}. $$ 
\new{In \citet{ditzhaus_qanova_2021}, it is shown that $\widehat{\sigma}_i^{PB}(p_a)$ is consistent for the asymptotic standard deviation of the corresponding sample quantile $\sqrt{\kappa_i\boldsymbol{\Sigma}^{(i)}_{aa}}$.
By using \eqref{eq:equation8}, we obtain}
the interval-based estimator for $\boldsymbol\Sigma^{(i)}_{ab}$
$$ \widehat{\boldsymbol\Sigma}^{(i),PB}_{ab} = \frac{n}{n_i}\widehat{\sigma}_i^{PB}(p_a)\widehat{\sigma}_i^{PB}(p_b)\frac{\min\{p_a,p_b\} - p_ap_b}{\sqrt{(p_a-p_a^2)(p_b-p_b^2)}}. $$
\end{enumerate}

\subsection{Details on the Bonferroni-adjusted Permutation QANOVA}\label{sec:BPerm}
\new{Here, we want to explain the details of the QANOVA permutation approach.
Draw the permuted samples $X_{i1}^{\pi},...,X_{in_i}^{\pi}, i\in\{1,...,k\},$ without replacement from the pooled sample $X_{11},...,X_{1n_1},X_{21},...,X_{kn_k}$. }
As in \cite{ditzhaus_qanova_2021}, let $F := \sum_{i=1}^k \kappa_i F_i$ denote the pooled cumulative distribution function and assume the following.
\begin{ass}\label{PermAssumptions}
	We assume \new{that $F$ is differentiable with uniformly continuous derivative $f$ and that} $f(F^{-1}(p_j)) $ $ > 0$ for all $j\in\{1,...,m\}$.
	\new{Moreover, we assume $|n_i/n - \kappa_i| = O(n^{-1/2})$ as $n\to\infty$ for all $i\in\{1,...,k\}$.}
\end{ass}
\new{Note that the latter assumption equals Assumption~4 in \citet{ditzhaus_qanova_2021}.
Similarly as Assumption~\ref{Assumptions}, this assumption can not really be checked in practice because usually $F$ is not known.
However, ties in the pooled data indicate that $F$ can not be continuous and, thus, not differentiable.
The assumption $|n_i/n - \kappa_i| = O(n^{-1/2})$ guarantees that the group fractions  converge sufficiently fast to their limits.}
Then, the permutation counterpart of the test statistics are defined as
\begin{align*}
T_n^{\pi} (\mathbf{h}_{\ell}) := \sqrt{n}\frac{\mathbf{h}_{\ell}^{\prime} \widehat{\mathbf{q}}^{\pi}}{\sqrt{\mathbf{h}_{\ell}^{\prime}\widehat{\boldsymbol\Sigma}^{\pi}\mathbf{h}_{\ell}}}, \quad \ell\in\{1,...,r\}.
\end{align*}
By Lemma~S1, S2 and S3 in the Supplement of \cite{ditzhaus_qanova_2021}, we have
\begin{align}\label{eq:convPerm}
T_n^{\pi} (\mathbf{h}_{\ell}) \xrightarrow{d^*} \mathcal{N}(0,1)
\end{align}
 as $n\to\infty$ for each $\ell\in\{1,...,r\}$ under $|n_i/n - \kappa_i| = O(n^{-1/2})$ for all $i\in\{1,...,k\}$ whenever the kernel or interval-based covariance estimator is used, where here and throughout $\xrightarrow{d^*}$ denote conditional convergence in distribution in probability given the data $X_{11}, X_{12},\dots, X_{21}, \dots, X_{k1}, \dots$.
For the bootstrap estimator, we need the stronger assumption that $|n_i/n - \kappa_i| = \new{o}(n^{-1})$ holds\new{, which means that the group fractions converge sufficiently fast to their limits}.
To show the consistency of the bootstrap-estimator in this case, we firstly note that $X_{11}^{\pi},...,X_{kn_k}^{\pi} \sim \sum_{i=1}^k n_i/n F_i$ (unconditionally) independent and identically distributed. 
Hence, we can construct random variables $Y_{11},...,Y_{kn_k} \sim F$ independent and identically distributed with $\mathrm{P}(Y_{ij} \neq X_{ij}^{\pi})\leq |n_i/n - \kappa_i|$ for all $i\in\{1,...,k\}, j\in\{1,...,n_i\}.$ 
Thus, we get
$$\mathrm{P}\left(\exists j\in\{1,...,n_i\} \: : \: Y_{ij} \neq X_{ij}^{\pi}\right) \leq \sum_{j=1}^{n_i} |n_i/n - \kappa_i| \to 0$$ as $n\to\infty$.
for all $i\in\{1,...,k\}$. Since the bootstrap-estimator based on $Y_{11},...,Y_{kn_k}$ is consistent as discussed in \cite{ditzhaus_qanova_2021}, it easily follows that the permutation counterpart of the bootstrap-estimator is consistent as well.
Mathematically, (\ref{eq:convPerm}) means
\begin{align*}
\sup_{x\in\mathbb R}\left|\mathrm{P}\left( T_n^{\pi} (\mathbf{h}_{\ell})\leq x \mid X_{11},...,X_{1n_1},X_{21},...,X_{kn_k} \right) - \Phi(x) \right| \xrightarrow{P} 0
\end{align*} as $n\to\infty$, where $\Phi$ denotes the standard normal distribution function.
Thus, each test statistic $T_n^{\pi} (\mathbf{h}_{\ell})$ can mimic the distribution of $T_n (\mathbf{h}_{\ell}, \epsilon_{\ell})$ asymptotically. However, the joint distribution of the whole vector of test statistics $(T_n^{\pi} (\mathbf{h}_{1}),...,T_n^{\pi} (\mathbf{h}_{r}))$ turns out to converge to a centered normal distribution with different covariance matrix than in (\ref{eq:conv2}) in general. Hence, this approach is not able to mimic the joint distribution asymptotically and a correcting procedure for multiple testing, e.g. a Bonferroni correction, needs to be applied. Therefore, let $q^{\pi}_{\ell,\beta}$ and $\widetilde q^{\pi}_{\ell,\beta}$ denote the $\beta$-quantiles of the conditional distribution of $T_n^{\pi}(\mathbf{h}_{\ell})$ and $|T_n^{\pi}(\mathbf{h}_{\ell})|$, respectively, given the data for all $\ell\in\{1,...,r\}$. 
By (\ref{eq:convPerm}), the quantiles are converging in probability to quantiles of the standard normal distribution or its absolute value, respectively.
\new{
The Bonferroni-adjusted permutation tests can be obtained by setting $q_\ell = q^{\pi}_{\ell,1-\alpha/r}$ and $\widetilde q_\ell = \widetilde q^{\pi}_{\ell,1-\alpha/r}$ in Section~\ref{sec:methods}.
}

\subsection{Details on the Asymptotic MCTPs}\label{sec:Proofs}
In this section, we prove that the critical values in Section~\ref{ssec:asymptoticMCTP} converge in probability to the $(1-\alpha)$-quantiles of $ \max_{\ell\in\{1,...,r\}} Z_\ell$ and $ \max_{\ell\in\{1,...,r\}} |Z_\ell|$, respectively, for $(Z_1,\dots,Z_r)^{\prime}\sim \mathcal{N}\left(\mathbf{0}, {\mathbf{D}}\mathbf{H}{\boldsymbol\Sigma} \mathbf{H}^{\prime}{\mathbf{D}}  \right)$.
Therefore, we firstly state an auxiliary lemma.

\begin{lemma}\label{Auxiliary}
	Let $(Z_1,\dots,Z_r)\sim F$, where $F:\R^r\to[0,1]$ denotes a continuous distribution function, and $(Y_{n1},\dots,Y_{nr})$ be a sequence of random vectors with $(Y_{n1},\dots,Y_{nr})$ $ \xrightarrow{d^*} (Z_1,\dots,Z_r)$ as $n\to\infty$ conditionally on a random variable $\mathbf{X}$. 
Moreover, denote by $G_n$ the conditional distribution function of $\max_{\ell\in\{1,...,r\}} Y_{n\ell}$ or $\max_{\ell\in\{1,...,r\}} |Y_{n\ell}|$ given $\mathbf{X}$ and by $G$ the distribution function of $\max_{\ell\in\{1,...,r\}} Z_\ell$ or $\max_{\ell\in\{1,...,r\}} |Z_\ell|$, respectively.
If $G$ is strictly increasing on $[a,b]\subset\R$ with $G(a) < 1-\alpha < G(b)$ for $\alpha\in(0,1)$, we have $G_n^{-1}(1-\alpha) \xrightarrow{P} G^{-1}(1-\alpha)$.  
\end{lemma}

\begin{proof}
By the conditional convergence in distribution, we get $\max_{\ell\in\{1,...,r\}} Y_{n\ell} \xrightarrow{d^*} \max_{\ell\in\{1,...,r\}} Z_\ell$ and $\max_{\ell\in\{1,...,r\}} |Y_{n\ell}| \xrightarrow{d^*} \max_{\ell\in\{1,...,r\}} |Z_\ell|$ as $n\to\infty$ conditionally on $\mathbf{X}$ by the continuous mapping theorem. Since $F$ is continuous, $G$ is continuous. Hence, it follows $\sup_{t\in\R} |G_n(t)-G(t)| \xrightarrow{P} 0$. By Lemma~S3 in the Supplement of \citet{munko2023rmstbased} together with the subsequence criterion, we obtain $G_n^{-1}(1-\alpha) \xrightarrow{P} G^{-1}(1-\alpha)$.
\end{proof}

In Section~\ref{ssec:asymptoticMCTP}, the conditional convergence in distribution follows from  the consistency of the covariance estimators.
Hence, ${q}_{1-\alpha}$ and $\widetilde{q}_{1-\alpha}$ are converging in probability to the corresponding quantiles of $ \max_{\ell\in\{1,...,r\}} Z_\ell$ and $ \max_{\ell\in\{1,...,r\}} |Z_\ell|$, respectively. 

\subsection{Details on the Bootstrap MCTPs}\label{sec:bootstrapdetails}
\new{For the groupwise bootstrap MCTP}, Theorem~3.6.1 in \cite{vaartWellner_1996} implies $\sqrt{n_i}(\widehat{F}^*_i - \widehat{F}_i) \xrightarrow{d^*} B \circ F_i$ on $D(\mathbb{R})$ as $n\to\infty$ for all $i\in\{1,...,k\}$, where $D(\mathbb R)$ denotes the Skorohod space on $\mathbb R$ equipped with the sup-norm and $B$ denotes a Brownian bridge on $[0, 1]$. 
By the delta method (\citealp{vaartWellner_1996}, Theorem~3.9.11), it follows that we have conditional convergence in distribution $\sqrt{n} \left( \widehat{q}_{ij}^* - \widehat{q}_{ij} \right)_{j\in\{1,...,m\}}\xrightarrow{d^*} \mathbf{Z}_i$ as $n\to\infty$ for all $i\in\{1,...,k\}$ similarly as in the proof of Proposition~1 in the Supplement of \citet{ditzhaus_qanova_2021}.
Moreover, the consistency of the group-wise bootstrap counterpart of the kernel and interval-based covariance estimator follows as in Lemma~S.2 and S.3 in the Supplement of \cite{ditzhaus_qanova_2021} by just replacing $\widehat{F}^{\pi}_i$ by $\widehat{F}^{*}_i$ and $f$ by $f_i$.
Hence, combining everything with Slutsky's lemma and the continuous mapping theorem yields
\begin{align*}
(T_n^*(\mathbf{h}_{1}),...,  T_n^*(\mathbf{h}_{r}))^{\prime} \xrightarrow{d^*} \mathcal{N}(\mathbf{0},\mathbf{DH}\boldsymbol{\Sigma}\mathbf{H}^{\prime}\mathbf{D})
\end{align*}
 as $n\to\infty$ whenever the kernel or interval-based covariance estimator is used.
Hence, even the joint limit distribution in (\ref{eq:conv2}) can be approximated by the group-wise bootstrap.
By Lemma~\ref{Auxiliary}, $q^*_{1-\alpha}$ and $\widetilde q_{1-\alpha}^*$ are converging in probability to the corresponding quantiles of $ \max_{\ell\in\{1,...,r\}} Z_\ell$ and $ \max_{\ell\in\{1,...,r\}} |Z_\ell|$, respectively, whenever the kernel or interval-based covariance estimator is used.

\new{\subsection{Algorithms for the Bootstrap MCTP}\label{ssec:Algorithms}}
\begin{algorithm}
\caption{\new{Bootstrap MCTP algorithm for the two-sided testing problem}}
\new{\begin{algorithmic}[1]
\State \textbf{Input:} Original samples $X_{i1}, \dots, X_{in_i}$ for $i \in \{1, \dots, k\}$, probabilities $p_1, \dots, p_m$, contrasts $\mathbf{h}_1, \dots, \mathbf{h}_r$, constants $\epsilon_1, \dots, \epsilon_r$, global significance level $\alpha$, and number of bootstrap samples $B$.
\State Calculate $\widehat{\mathbf{q}}$ and $\widehat{\boldsymbol\Sigma}$.
\For{$\ell = 1, \dots, r$}
    \State Calculate the original test statistic $T_n(\mathbf{h}_{\ell},\epsilon_{\ell}) := \sqrt{n} \frac{\mathbf{h}_{\ell}^{\prime} \widehat{\mathbf{q}} - \epsilon_{\ell}}{\sqrt{\mathbf{h}_{\ell}^{\prime} \widehat{\boldsymbol\Sigma} \mathbf{h}_{\ell}}}.$
\EndFor
\State \textbf{Bootstrap Procedure:}
\For{$b = 1, \dots, B$}
    \State Draw bootstrap samples $X_{i1}^*, \dots, X_{in_i}^* \sim \hat{F}_i, i\in\{1,...,k\},$ independently conditionally on the data.
    \State Calculate $\widehat{\mathbf{q}}^*$ and $\widehat{\boldsymbol\Sigma}^{*}$ based on $X_{i1}^*, \dots, X_{in_i}^*, i\in\{1,...,k\}.$
    \For{$\ell = 1, \dots, r$}
        \State Calculate the bootstrap test statistic $T_n^{*}(\mathbf{h}_{\ell}) := \sqrt{n}\frac{\mathbf{h}_{\ell}^{\prime}(\widehat{\mathbf{q}}^* - \widehat{\mathbf{q}})}{\sqrt{\mathbf{h}_{\ell}^{\prime}\widehat{\boldsymbol\Sigma}^{*} \mathbf{h}_{\ell}}}.$
    \EndFor
    \State Compute $\widetilde{M}_b := \max_{\ell \in \{1, \dots, r\}} |T_n^{*}(\mathbf{h}_{\ell})|.$
\EndFor
\State Estimate the quantile $\widetilde{q}^*_{1-\alpha}$ as empirical 
$(1-\alpha)$-quantiles of $\widetilde{M}_1,...,\widetilde{M}_B$.
\State \textbf{Test Decisions:}
    \For{$\ell = 1, \dots, r$}
        \State Reject $\mathcal{H}_{0,\ell}$ if and only if $\left|T_n(\mathbf{h}_{\ell},\epsilon_{\ell})\right| > \widetilde{q}^*_{1-\alpha}.$
    \EndFor
    \State Reject the global hypothesis $\mathcal{H}_0 = \bigcap_{\ell=1}^r \mathcal{H}_{0,\ell}$ if and only if $\max_{\ell \in \{1, \dots, r\}}\left|T_n(\mathbf{h}_{\ell},\epsilon_{\ell})\right| > \widetilde{q}^*_{1-\alpha}.$
\State \textbf{Output:} Multiple test decisions of the bootstrap MCTP for the two-sided testing problem.
\end{algorithmic}
}
\end{algorithm}

\begin{algorithm}
\caption{\new{Bootstrap MCTP algorithm for the non-inferiority testing problem}}
\new{\begin{algorithmic}[1]
\State \textbf{Input:} Original samples $X_{i1}, \dots, X_{in_i}$ for $i \in \{1, \dots, k\}$, probabilities $p_1, \dots, p_m$, contrasts $\mathbf{h}_1, \dots, \mathbf{h}_r$, constants $\epsilon_1, \dots, \epsilon_r$, global significance level $\alpha$, and number of bootstrap samples $B$.
\State Calculate $\widehat{\mathbf{q}}$ and $\widehat{\boldsymbol\Sigma}$.
\For{$\ell = 1, \dots, r$}
    \State Calculate the original test statistic $T_n(\mathbf{h}_{\ell},\epsilon_{\ell}) := \sqrt{n} \frac{\mathbf{h}_{\ell}^{\prime} \widehat{\mathbf{q}} - \epsilon_{\ell}}{\sqrt{\mathbf{h}_{\ell}^{\prime} \widehat{\boldsymbol\Sigma} \mathbf{h}_{\ell}}}.$
\EndFor
\State \textbf{Bootstrap Procedure:}
\For{$b = 1, \dots, B$}
    \State Draw bootstrap samples $X_{i1}^*, \dots, X_{in_i}^* \sim \hat{F}_i, i\in\{1,...,k\},$ independently conditionally on the data.
    \State Calculate $\widehat{\mathbf{q}}^*$ and $\widehat{\boldsymbol\Sigma}^{*}$ based on $X_{i1}^*, \dots, X_{in_i}^*, i\in\{1,...,k\}.$
    \For{$\ell = 1, \dots, r$}
        \State Calculate the bootstrap test statistic $T_n^{*}(\mathbf{h}_{\ell}) := \sqrt{n}\frac{\mathbf{h}_{\ell}^{\prime}(\widehat{\mathbf{q}}^* - \widehat{\mathbf{q}})}{\sqrt{\mathbf{h}_{\ell}^{\prime}\widehat{\boldsymbol\Sigma}^{*} \mathbf{h}_{\ell}}}.$
    \EndFor
    \State Compute $M_b := \max_{\ell \in \{1, \dots, r\}} T_n^{*}(\mathbf{h}_{\ell}).$
\EndFor
\State Estimate the quantiles $q^*_{1-\alpha}$ as the empirical $(1-\alpha)$-quantile of $M_1,...,M_B$.
\State \textbf{Test Decisions:}
    \For{$\ell = 1, \dots, r$}
        \State Reject $\mathcal{H}^{I}_{0,\ell}$ if and only if $T_n(\mathbf{h}_{\ell},\epsilon_{\ell}) > q^*_{1-\alpha}.$
    \EndFor
    \State Reject the global hypothesis $\mathcal{H}^{I}_0 = \bigcap_{\ell=1}^r \mathcal{H}^{I}_{0,\ell}$ if and only if $\max_{\ell \in \{1, \dots, r\}} T_n(\mathbf{h}_{\ell},\epsilon_{\ell}) > q^*_{1-\alpha}.$
\State \textbf{Output:} Multiple test decisions of the bootstrap MCTP for the non-inferiority testing problem.
\end{algorithmic}
}
\end{algorithm}
\FloatBarrier

\new{
\subsection{Correlation between the test statistics in our simulations}\label{ssec:Correlation}
For a better understanding and investigation of the behavior of the different test procedures in our simulation study of Section~\ref{sec:sim}, we report summaries of the correlations between the test statistics in this section.
Therefore, we calculated the empirical correlations for the $N_{sim} = 5000$ test statistics resulting from the $5000$ data sets used in the simulation for each setting.
The same is done for the absolute values of the test statistics, which are used for the two-sided multiple testing problem, cf. Section~\ref{sec:methods}.
In order to get a broad overview over the correlations, the minimal (Min.), maximal (Max.), median, and minimal absolute (min abs.) correlations are reported in Table~\ref{tab:corr_table}.}
\begin{table}[h]
    \centering
    \scriptsize
    \new{
\begin{tabular}{lllrrrr}
        \toprule
        Contrast matrix & Testing problem & Variance estimator & Min. & Max. & Median & Min. abs. \\
        \midrule
        Dunnett         & Two-sided       & Bootstrap          & 0.067   & 0.666   & 0.276  & 0.067           \\
        Dunnett         & Two-sided       & Interval-based     & 0.063   & 0.653   & 0.263  & 0.063           \\
        Dunnett         & Two-sided       & Kernel             & 0.068   & 0.695   & 0.272  & 0.068           \\
        Dunnett         & Non-inferiority & Bootstrap          & 0.259   & 0.820   & 0.524  & 0.259           \\
        Dunnett         & Non-inferiority & Interval-based     & 0.258   & 0.814   & 0.520  & 0.258           \\
        Dunnett         & Non-inferiority & Kernel             & 0.256   & 0.823   & 0.513  & 0.256           \\
        Tukey           & Two-sided       & Bootstrap          & -0.029  & 0.666   & 0.203  & 0.000           \\
        Tukey           & Two-sided       & Interval-based     & -0.033  & 0.653   & 0.203  & 0.000           \\
        Tukey           & Two-sided       & Kernel             & -0.030  & 0.695   & 0.205  & 0.001           \\
        Tukey           & Non-inferiority & Bootstrap          & -0.593  & 0.820   & 0.247  & 0.000           \\
        Tukey           & Non-inferiority & Interval-based     & -0.589  & 0.814   & 0.249  & 0.000           \\
        Tukey           & Non-inferiority & Kernel             & -0.589  & 0.823   & 0.241  & 0.000           \\
        Grand-mean      & Two-sided       & Bootstrap          & 0.008   & 0.249   & 0.116  & 0.008           \\
        Grand-mean      & Two-sided       & Interval-based     & 0.006   & 0.256   & 0.122  & 0.006           \\
        Grand-mean      & Two-sided       & Kernel             & 0.007   & 0.308   & 0.133  & 0.007           \\
        Grand-mean      & Non-inferiority & Bootstrap          & -0.525  & 0.130   & -0.321 & 0.097           \\
        Grand-mean      & Non-inferiority & Interval-based     & -0.532  & 0.131   & -0.319 & 0.092           \\
        Grand-mean      & Non-inferiority & Kernel             & -0.524  & 0.137   & -0.322 & 0.096           \\
        \bottomrule
    \end{tabular}
    \normalsize
    \caption{Correlation summary between the different test statistics in our simulation study across all settings under the null hypothesis for the different scenarios.}
    \label{tab:corr_table}}
\end{table}
\new{The Bonferroni correction is known to perform too conservative for a large positive correlation.
The largest negative correlations are realized for the non-inferiority tests for  Tukey- and Grand-mean-type contrast matrix with correlations less than -0.52.
For the Grand-mean-type contrast matrix, we do not observe a visible difference of the asymptotic MCTP compared to the Bonferroni-adjusted asymptotic test for the non-inferiority testing problem in Figure~\ref{fig:GM}.
This is a consequence of the rather negative correlations between the test statistics for this scenario which yield a good performance of the Bonferroni correction.
}
\newpage

\bibliographystyle{apalike}

\newpage
\appendix

\end{document}


\thispagestyle{empty}
\maketitle

\footnotetext[1]{Department of Statistics, TU Dortmund University, Germany}
\footnotetext[2]{Research Center Trustworthy Data Science and Security, UA Ruhr, Germany}
\footnotetext[3]{Department of Mathematics, Otto-von-Guericke University Magdeburg, Germany}
\footnotetext[4]{Department of Animal Behaviour, Bielefeld University, Germany}
\footnotetext[5]{Joint Institute for Individualisation in a Changing Environment (JICE), Bielefeld University and University of Münster, Germany}

\section*{Abstract}
In this project, we compare different  inference approaches for two-sided and non-inferiority hypotheses formulated in terms of medians or IQRs in an extensive simulation study. 
We consider multiple contrast testing procedures combined with a bootstrap method as well as testing procedures with Bonferroni correction.
As an example of a multiple testing problem based on heavy-tailed data we analyse an ecological trait variation in early and late breeding in a medium-sized bird of prey.
This Supplement contains the detailed simulation results.
Further plots and tables on the simulation results of the paper are provided.

This document is licensed under Creative Commons Attribution 4.0 International (CC-BY 4.0).

\tableofcontents
\newpage

\section{Plot on the Impact of (Un)Balanced Designs}
\begin{figure}[H]\centering
\includegraphics[width=0.85\textwidth]{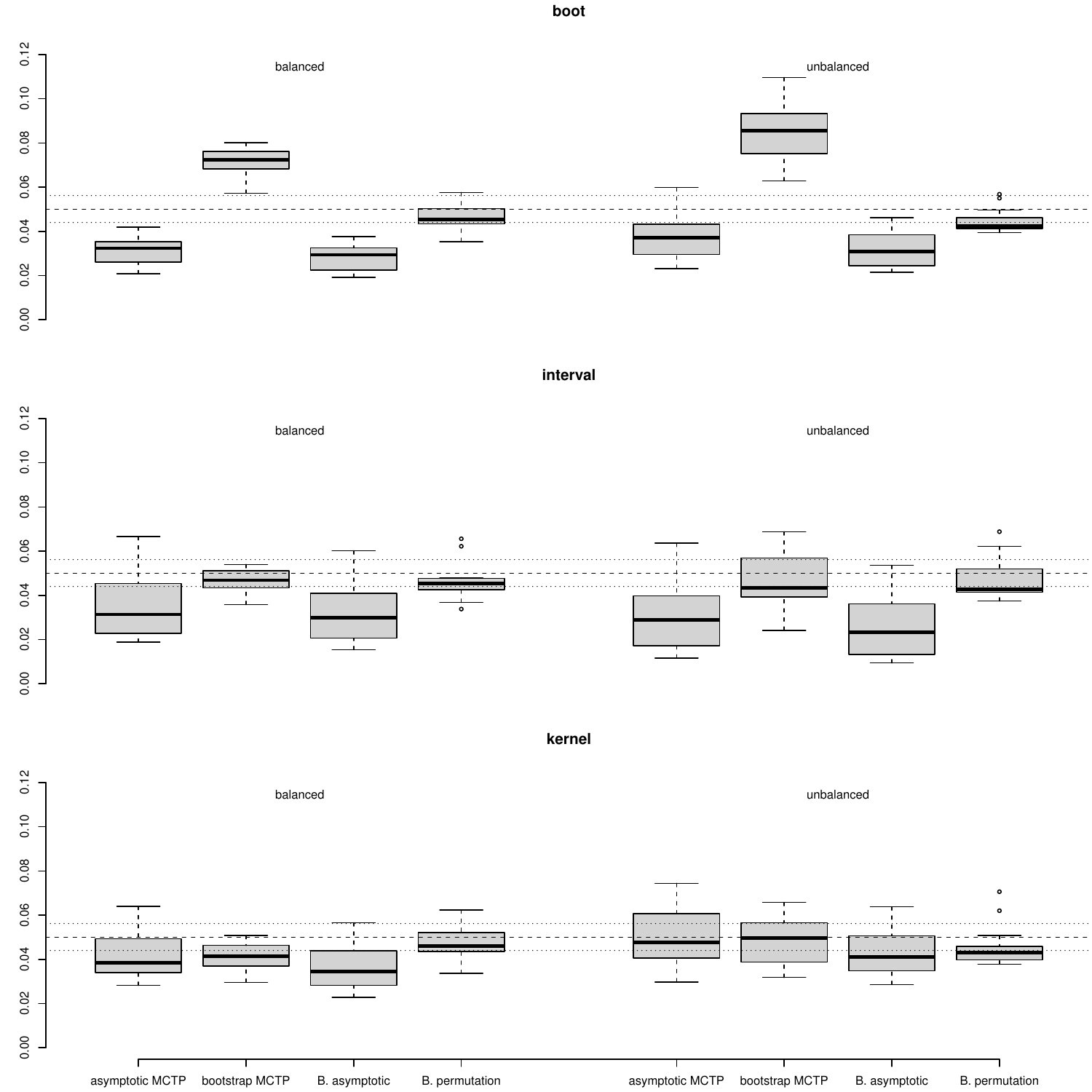} 
\caption{Empirical FWERs for non-inferiority Dunnett-type tests with different designs (left: balanced and right: unbalanced) and variance estimators (from left to right: bootstrap, interval-based or kernel).}
\label{fig:balanced}
\end{figure}

To investigate the impact of balanced and unbalanced designs, Figure~\ref{fig:balanced} shows the rejection rates under the null hypothesis exemplarily for the non-inferiority Dunnett-type tests. 
This setting is chosen for illustration due to the real data example in Section~\new{5} of the paper.
It is observable that the empirical FWERs for balanced and unbalanced designs are comparable for all methods and variance estimators.

\section{Plots and Tables for Empirical Global Power}
\begin{figure}[H]\centering
\includegraphics[height=0.4\textheight]{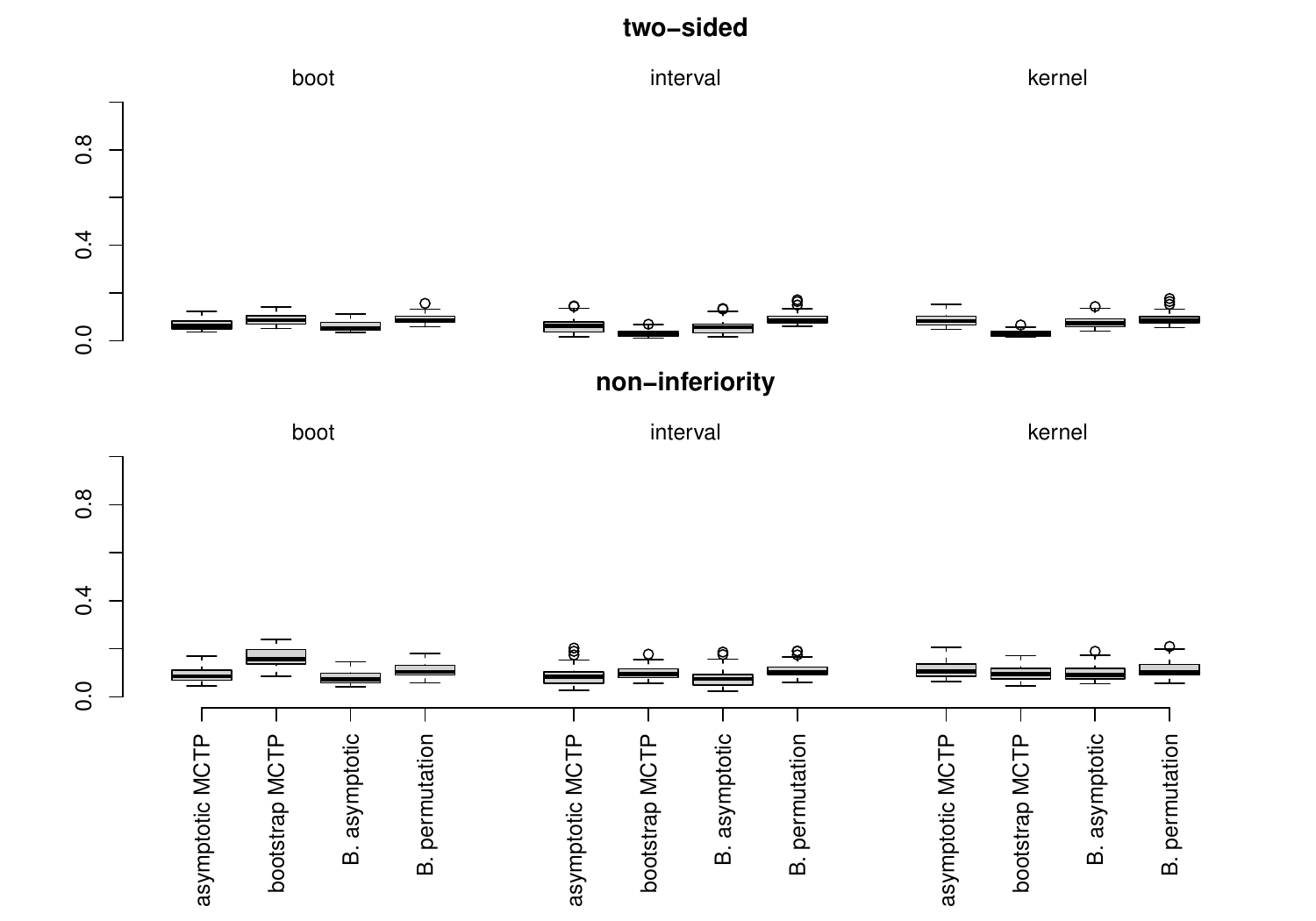} 
\caption{Empirical global power with $\delta = 0.5$ for Dunnett-type contrasts with different hypotheses (top: two-sided, bottom: non-inferiority) and variance estimators. }
\label{fig:DunnettStart}
\end{figure}

\begin{figure}[H]\centering
\includegraphics[height=0.4\textheight]{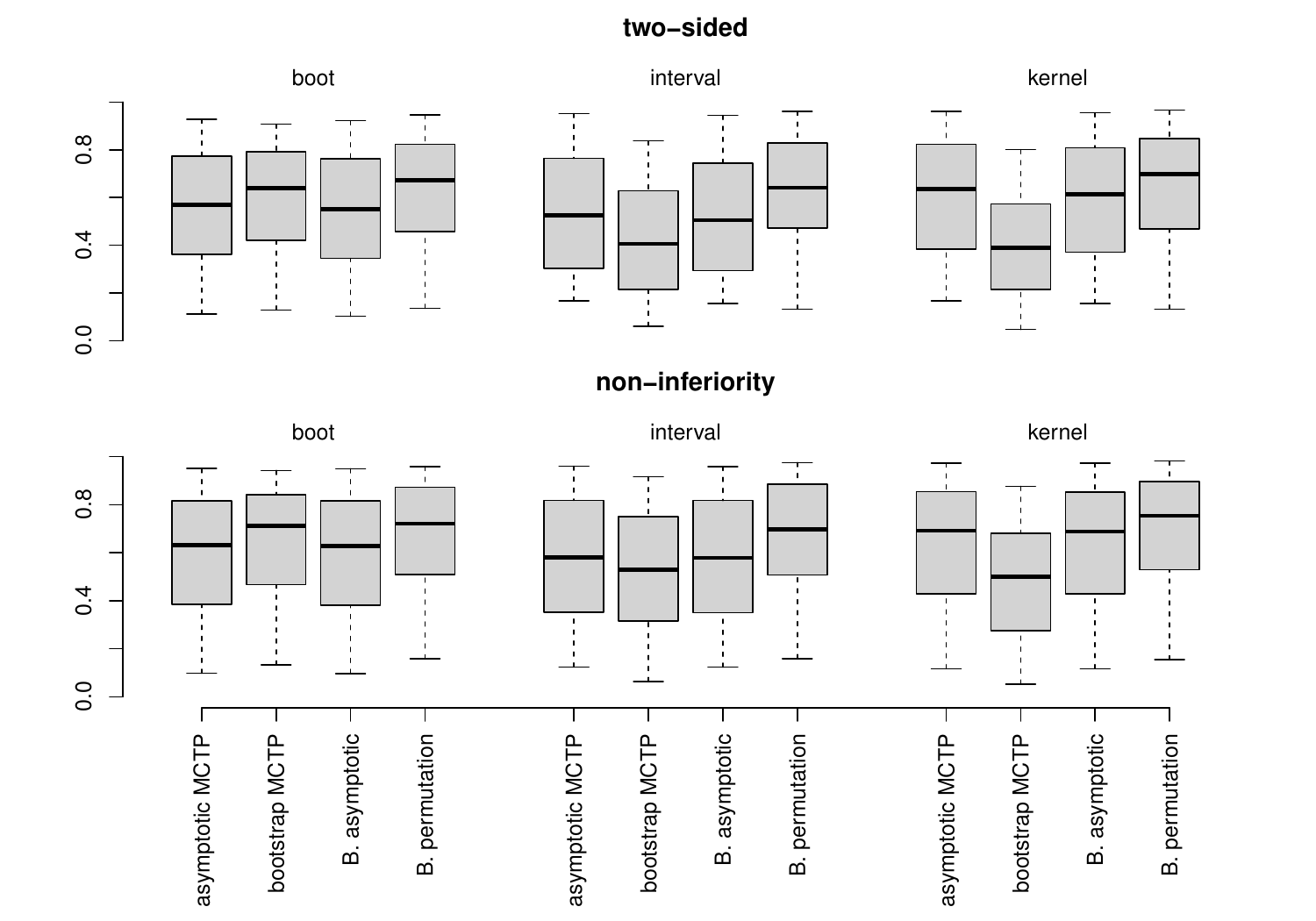} 
\caption{Empirical global power with $\delta = 1.5$ for Grand-mean-type contrasts with different hypotheses (top: two-sided, bottom: non-inferiority) and variance estimators.}
\label{fig:GMEnd}
\end{figure}

Plots for the analyzing the global power are shown in Figures~\ref{fig:DunnettStart}--\ref{fig:GMEnd}. Here and throughout, the following abbreviations are used:  boot, interval, kernel - bootstrap, interval-based, and kernel variance estimator, respectively.
The results of all different scenarios can be found in Tables~\ref{tab:1}--\ref{tab:2}.
The following abbreviations are used in the tables: bal and unb for balanced ($\mathbf{n}_1$) and unbalanced ($\mathbf{n}_2$) designs, hom, pos and neg for homoscedastic scenario ($\boldsymbol{\sigma}_1$), positive ($\boldsymbol{\sigma}_2$) and negative ($\boldsymbol{\sigma}_3$) pairing, asymp., boot. and perm. for asymptotic, bootstrap and permutation, respectively, and B. for Bonferroni-adjusted.

\begin{figure}[H]\centering
\includegraphics[height=0.4\textheight]{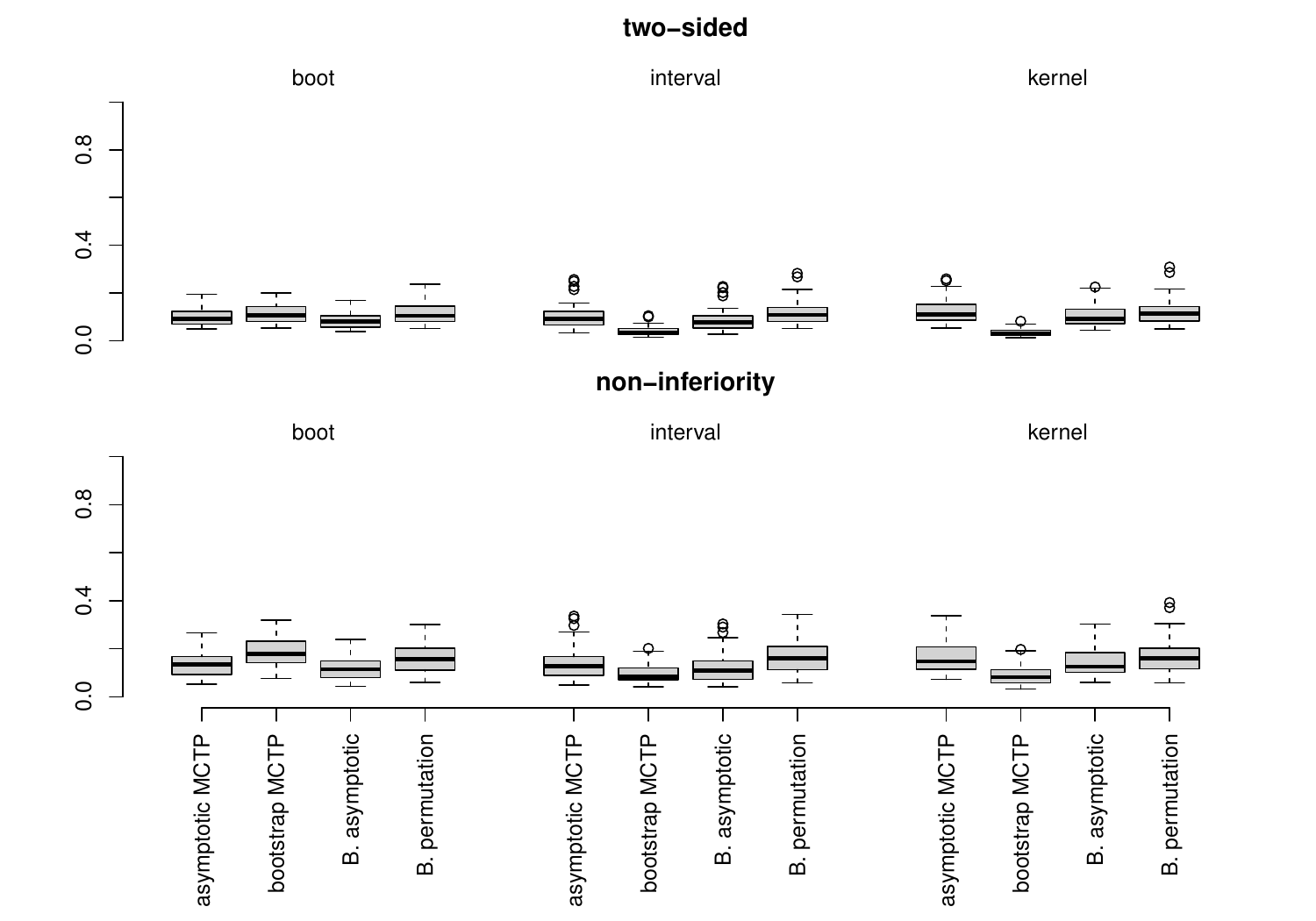} 
\caption{Empirical global power with $\delta = 0.5$ for Tukey-type contrasts with different hypotheses (top: two-sided, bottom: non-inferiority) and variance estimators.}
\end{figure}

\begin{figure}[H]\centering
\includegraphics[height=0.4\textheight]{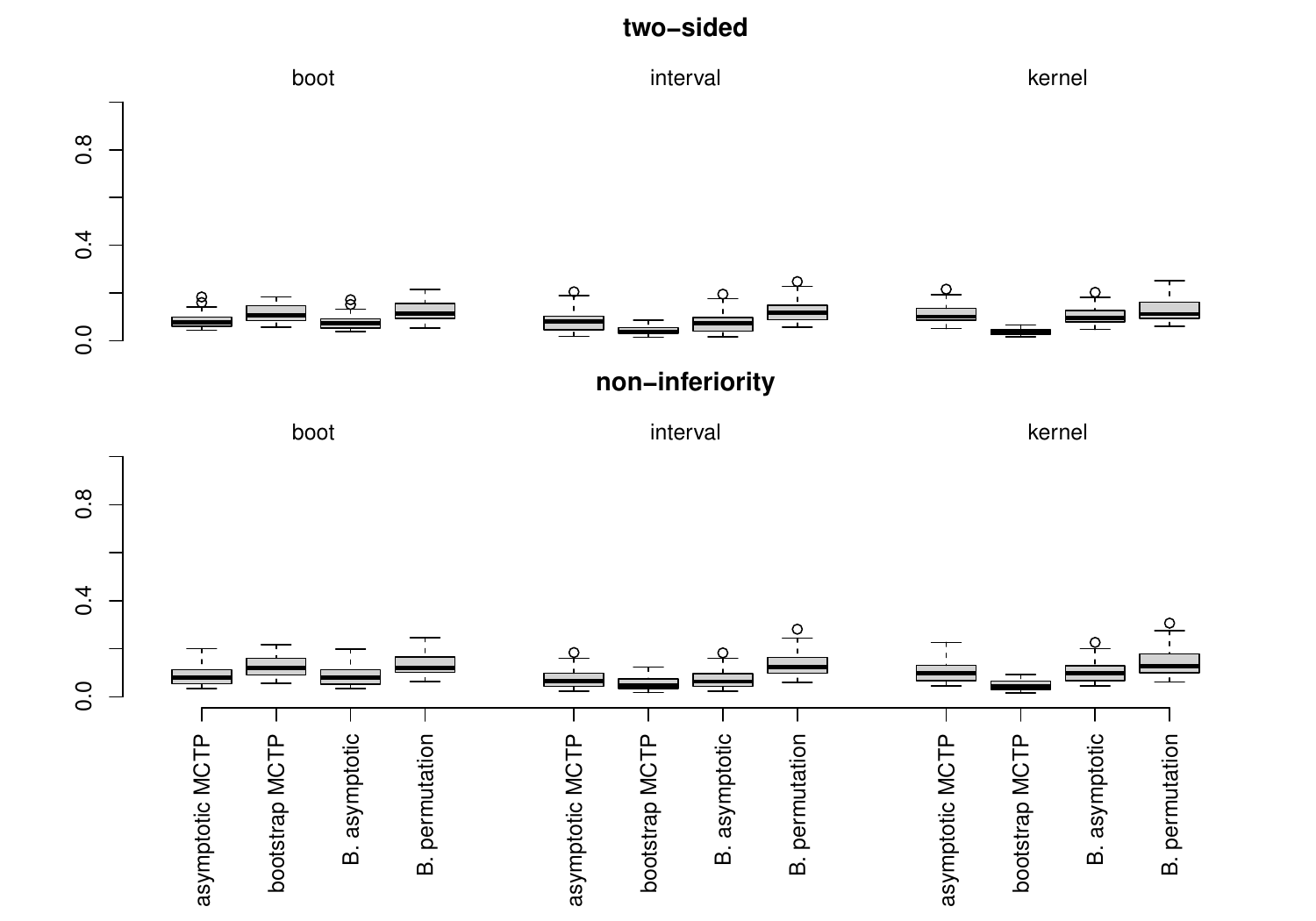} 
\caption{Empirical global power with $\delta = 0.5$ for Grand-mean-type contrasts with different hypotheses (top: two-sided, bottom: non-inferiority) and variance estimators.}
\end{figure}

\begin{figure}[H]\centering
\includegraphics[height=0.4\textheight]{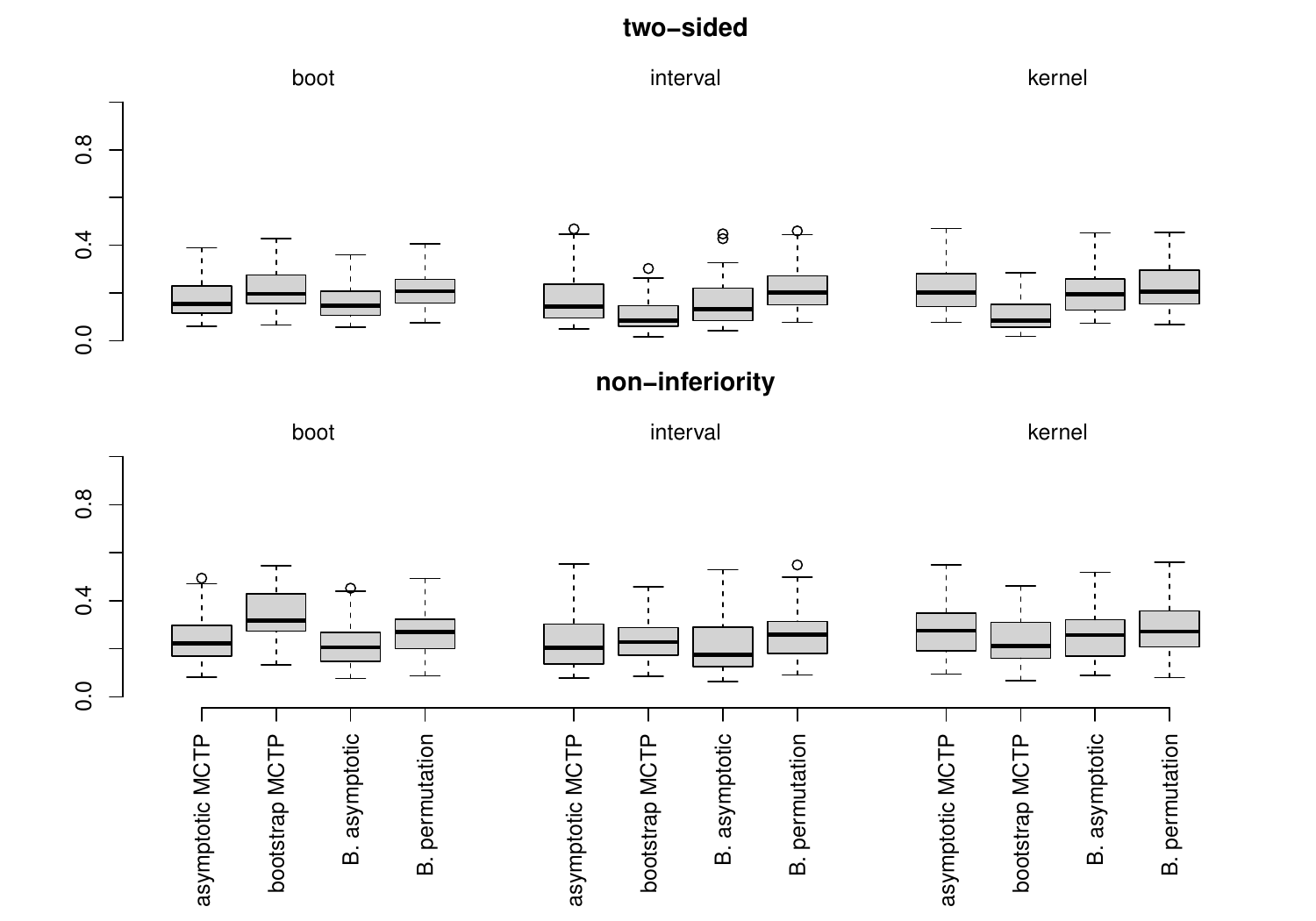} 
\caption{Empirical global power with $\delta = 1.0$ for Dunnett-type contrasts with different hypotheses (top: two-sided, bottom: non-inferiority) and variance estimators.}
\end{figure}

\begin{figure}[H]\centering
\includegraphics[height=0.4\textheight]{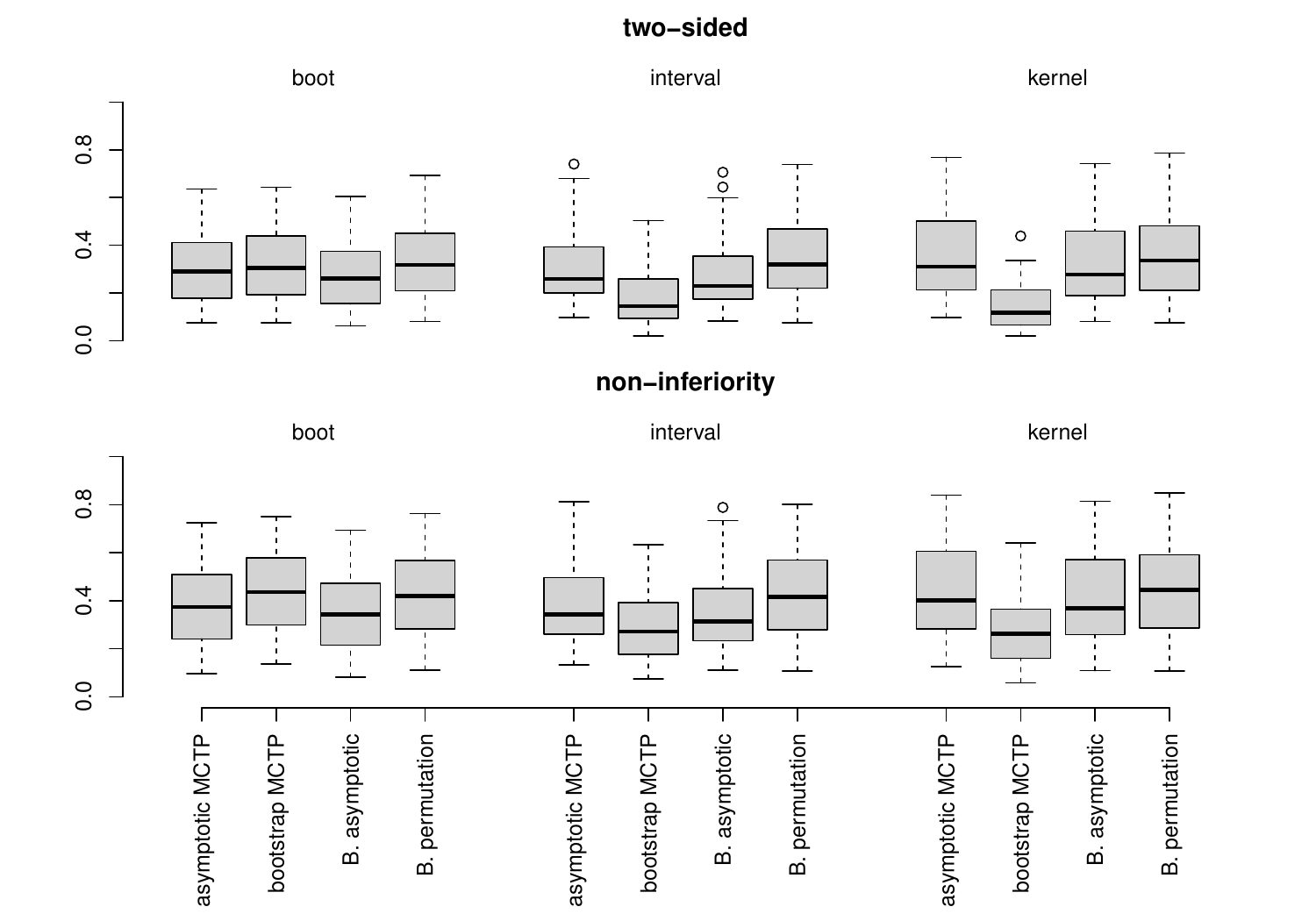} 
\caption{Empirical global power with $\delta = 1.0$ for Tukey-type contrasts with different hypotheses (top: two-sided, bottom: non-inferiority) and variance estimators.}
\end{figure}

\begin{figure}[H]\centering
\includegraphics[height=0.4\textheight]{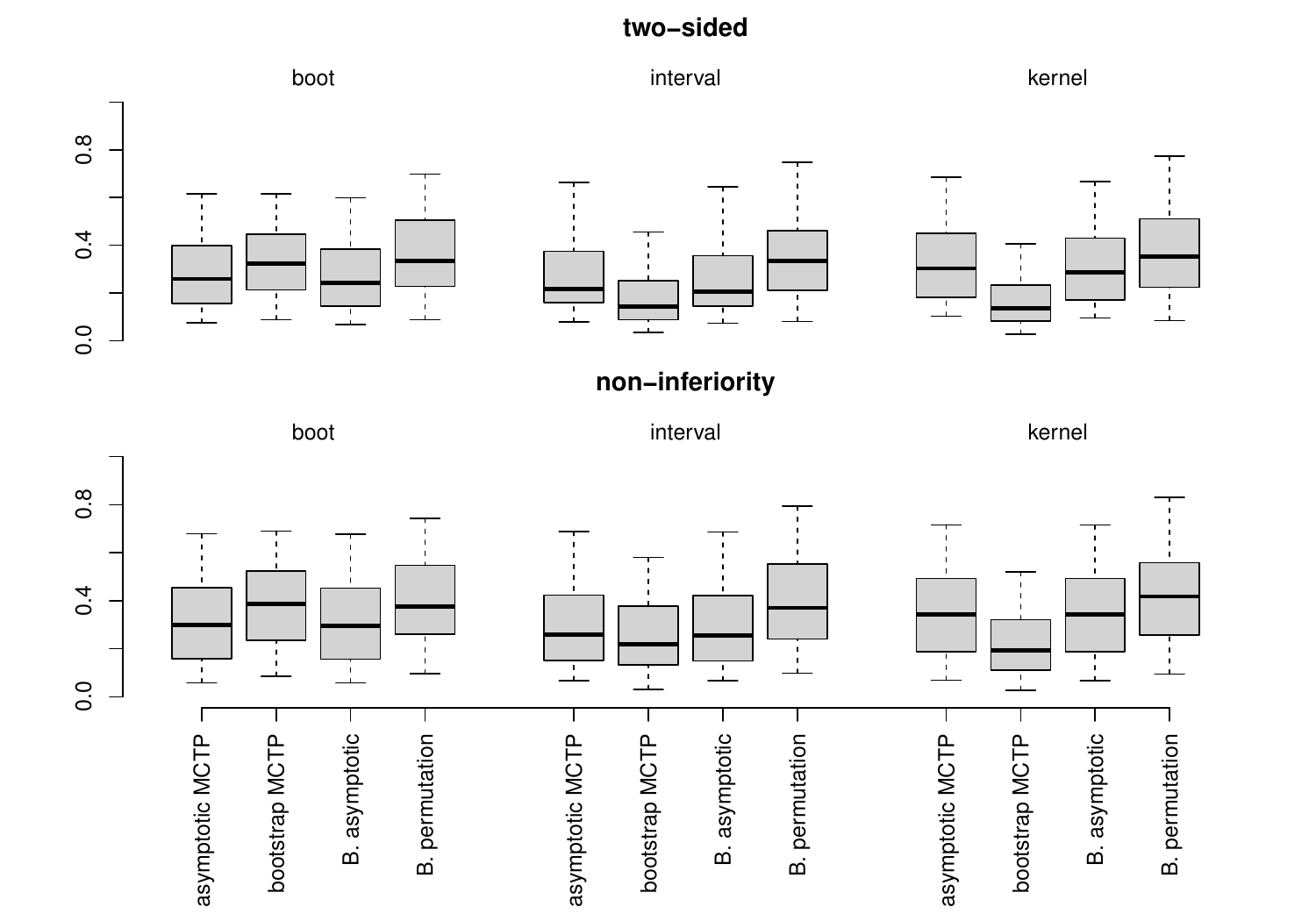} 
\caption{Empirical global power with $\delta = 1.0$ for Grand-mean-type contrasts with different hypotheses (top: two-sided, bottom: non-inferiority) and variance estimators.}
\end{figure}

\begin{figure}[H]\centering
\includegraphics[height=0.4\textheight]{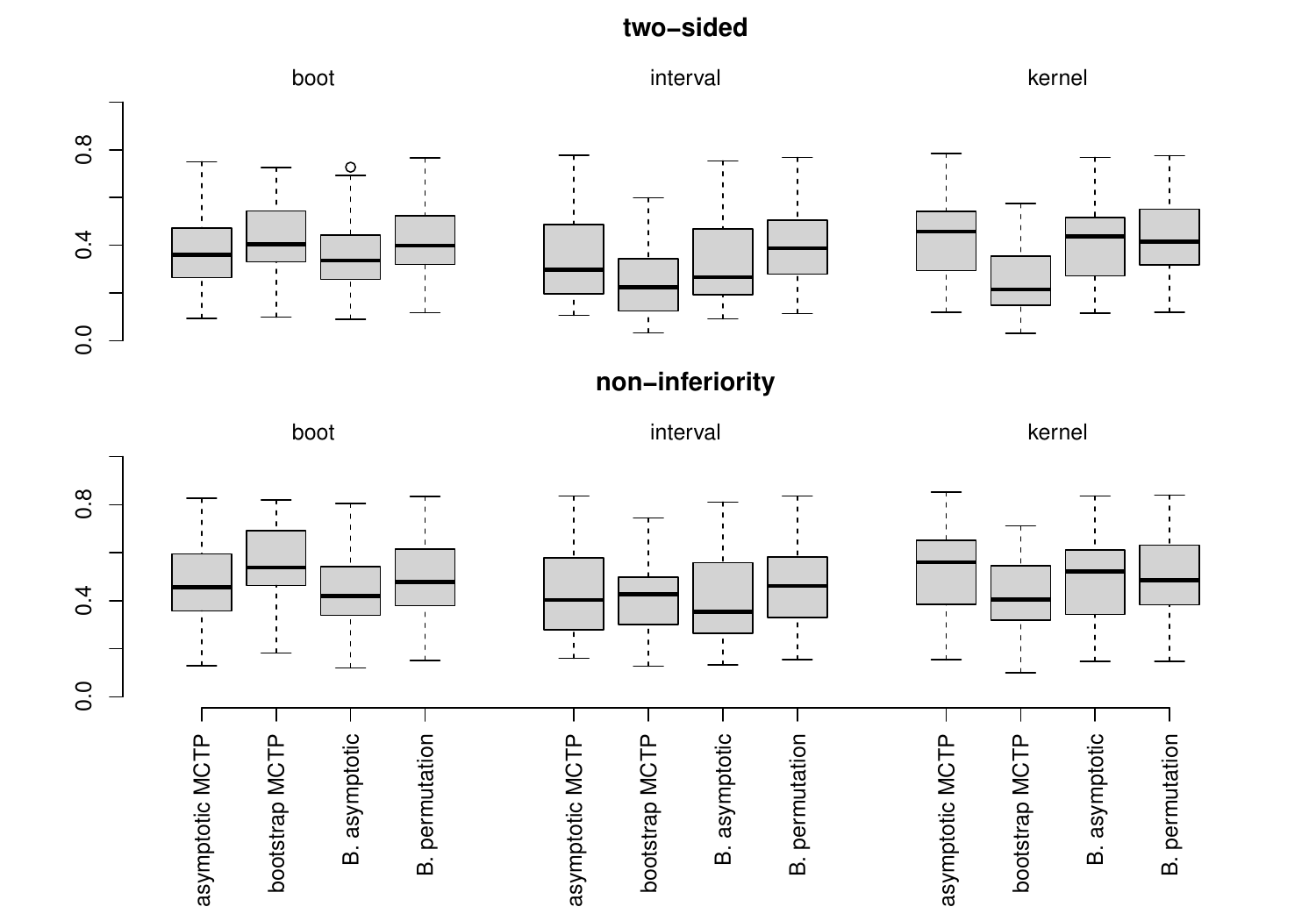} 
\caption{Empirical global power with $\delta = 1.5$ for Dunnett-type contrasts with different hypotheses (top: two-sided, bottom: non-inferiority) and variance estimators.}
\end{figure}

\begin{figure}[H]\centering
\includegraphics[height=0.4\textheight]{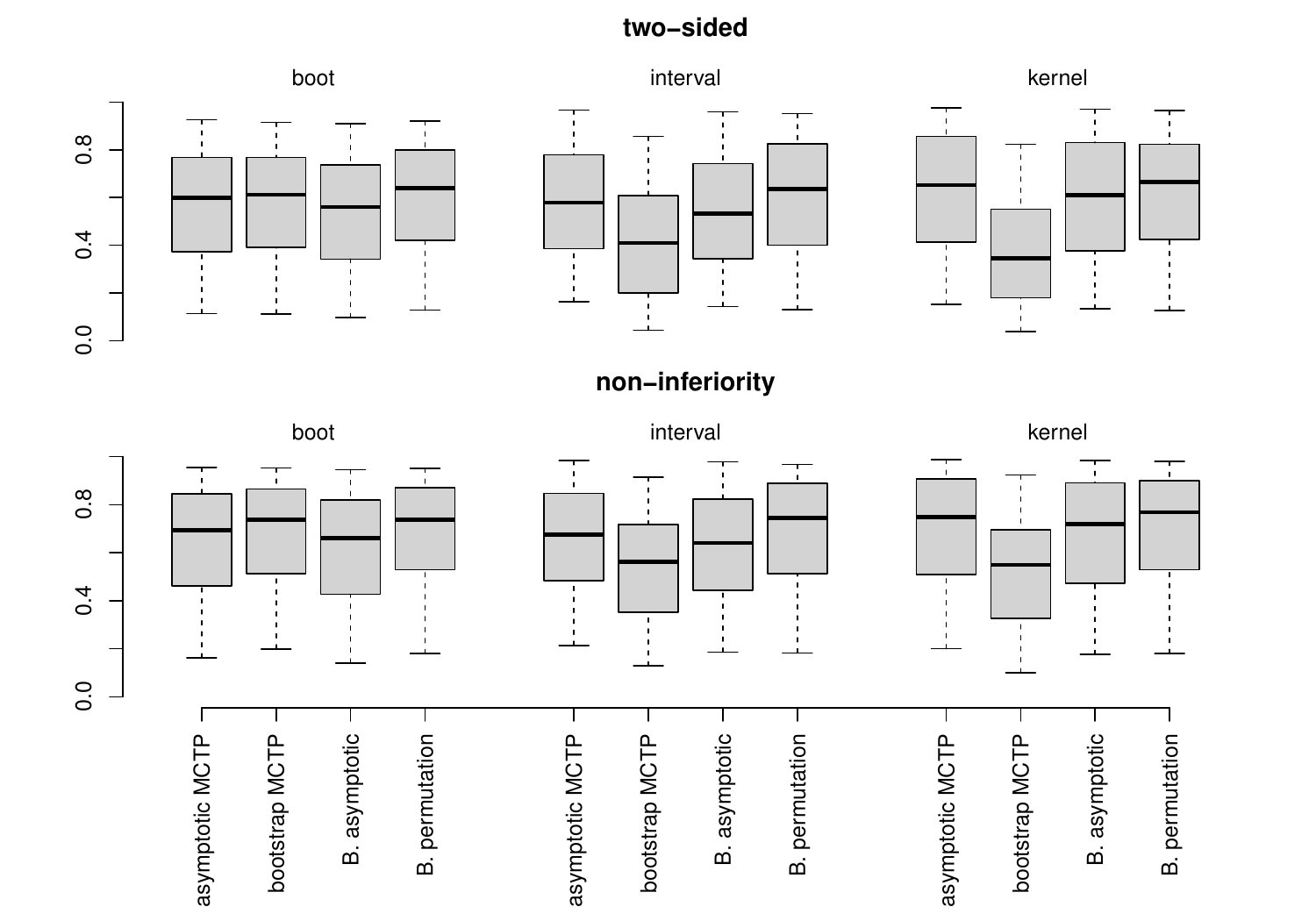} 
\caption{Empirical global power with $\delta = 1.5$ for Tukey-type contrasts with different hypotheses (top: two-sided, bottom: non-inferiority) and variance estimators.}
\end{figure}

\FloatBarrier
\begin{table}[H]
\centering\footnotesize
\begin{tabular}{clllrrrr}
\toprule

\multicolumn{4}{c}{Setting} & \multicolumn{4}{c}{Method}\\

\cmidrule(lr){1-4}\cmidrule(lr){5-8}

Distribution & \multicolumn{2}{c}{Design} & $\delta$ & asymp. MCTP & boot. MCTP & B. asymp. & B. perm. \\
\midrule
$\chi^2_3$&bal&hom&0.0&3.66 & \textbf{4.74} & 3.32 & \textbf{4.68} \\
$\chi^2_3$&bal&neg&0.0&4.30 & 5.72 & 3.68 & \textbf{5.40} \\
$\chi^2_3$&bal&pos&0.0&\textbf{4.60} & \textbf{5.02} & 4.40 & 6.32 \\
$\chi^2_3$&unb&hom&0.0&3.42 & \textbf{4.66} & 3.08 & 3.80 \\
$\chi^2_3$&unb&neg&0.0&\textbf{4.44} & 5.64 & 3.76 & \textbf{5.00} \\
$\chi^2_3$&unb&pos&0.0&4.02 & 4.32 & 3.74 & \textbf{4.66} \\
\bottomrule
\end{tabular}
\caption{Empirical FWER in \% of two-sided Dunnett-type tests with bootstrap estimator}\label{tab:1}
\end{table}
\newpage
\begin{table}[H]
\centering\footnotesize
\begin{tabular}{clllrrrr}
\toprule

\multicolumn{4}{c}{Setting} & \multicolumn{4}{c}{Method}\\

\cmidrule(lr){1-4}\cmidrule(lr){5-8}

Distribution & \multicolumn{2}{c}{Design} & $\delta$ & asymp. MCTP & boot. MCTP & B. asymp. & B. perm. \\
\midrule
$\chi^2_3$&bal&hom&0.0&3.52 & 6.80 & 3.28 & \textbf{4.52} \\
$\chi^2_3$&bal&neg&0.0&4.20 & 7.74 & 3.76 & \textbf{5.24} \\
$\chi^2_3$&bal&pos&0.0&3.52 & 6.48 & 3.32 & \textbf{5.12} \\
$\chi^2_3$&unb&hom&0.0&4.18 & 8.78 & 3.46 & 4.22 \\
$\chi^2_3$&unb&neg&0.0&\textbf{5.52} & 10.26 & \textbf{4.62} & \textbf{5.50} \\
$\chi^2_3$&unb&pos&0.0&3.72 & 6.84 & 3.48 & 4.28 \\
\bottomrule
\end{tabular}
\caption{Empirical FWER in \% of non-inferiority Dunnett-type tests with bootstrap estimator}
\end{table}\begin{table}[H]
\centering\footnotesize
\begin{tabular}{clllrrrr}
\toprule

\multicolumn{4}{c}{Setting} & \multicolumn{4}{c}{Method}\\

\cmidrule(lr){1-4}\cmidrule(lr){5-8}

Distribution & \multicolumn{2}{c}{Design} & $\delta$ & asymp. MCTP & boot. MCTP & B. asymp. & B. perm. \\
\midrule
$\chi^2_3$&bal&hom&0.0&\textbf{5.00} & 1.54 & \textbf{4.54} & 4.32 \\
$\chi^2_3$&bal&neg&0.0&6.68 & 2.16 & 5.64 & \textbf{5.56} \\
$\chi^2_3$&bal&pos&0.0&6.34 & 1.68 & 6.02 & 6.22 \\
$\chi^2_3$&unb&hom&0.0&4.04 & 1.18 & 3.54 & 3.98 \\
$\chi^2_3$&unb&neg&0.0&\textbf{4.44} & 1.64 & 3.90 & \textbf{5.40} \\
$\chi^2_3$&unb&pos&0.0&4.36 & 1.08 & 3.90 & 5.64 \\
\bottomrule
\end{tabular}
\caption{Empirical FWER in \% of two-sided Dunnett-type tests with interval-based estimator}
\end{table}\begin{table}[H]
\centering\footnotesize
\begin{tabular}{clllrrrr}
\toprule

\multicolumn{4}{c}{Setting} & \multicolumn{4}{c}{Method}\\

\cmidrule(lr){1-4}\cmidrule(lr){5-8}

Distribution & \multicolumn{2}{c}{Design} & $\delta$ & asymp. MCTP & boot. MCTP & B. asymp. & B. perm. \\
\midrule
$\chi^2_3$&bal&hom&0.0&\textbf{4.88} & \textbf{4.42} & 4.24 & 4.12 \\
$\chi^2_3$&bal&neg&0.0&6.66 & \textbf{5.30} & 5.86 & 6.22 \\
$\chi^2_3$&bal&pos&0.0&4.20 & 4.26 & 3.94 & 4.14 \\
$\chi^2_3$&unb&hom&0.0&\textbf{5.18} & 6.22 & \textbf{4.60} & 4.28 \\
$\chi^2_3$&unb&neg&0.0&6.04 & 6.88 & \textbf{5.04} & 6.22 \\
$\chi^2_3$&unb&pos&0.0&3.90 & 4.34 & 3.58 & 3.96 \\
\bottomrule
\end{tabular}
\caption{Empirical FWER in \% of non-inferiority Dunnett-type tests with interval-based estimator}
\end{table}\begin{table}[H]
\centering\footnotesize
\begin{tabular}{clllrrrr}
\toprule

\multicolumn{4}{c}{Setting} & \multicolumn{4}{c}{Method}\\

\cmidrule(lr){1-4}\cmidrule(lr){5-8}

Distribution & \multicolumn{2}{c}{Design} & $\delta$ & asymp. MCTP & boot. MCTP & B. asymp. & B. perm. \\
\midrule
$\chi^2_3$&bal&hom&0.0&\textbf{5.08} & 1.40 & \textbf{4.78} & 4.14 \\
$\chi^2_3$&bal&neg&0.0&6.58 & 1.90 & 5.70 & 5.84 \\
$\chi^2_3$&bal&pos&0.0&6.56 & 1.48 & 6.12 & 6.82 \\
$\chi^2_3$&unb&hom&0.0&\textbf{5.50} & 1.42 & \textbf{5.00} & 4.14 \\
$\chi^2_3$&unb&neg&0.0&6.62 & 1.84 & 5.74 & 5.78 \\
$\chi^2_3$&unb&pos&0.0&\textbf{5.62} & 1.26 & \textbf{5.26} & \textbf{4.82} \\
\bottomrule
\end{tabular}
\caption{Empirical FWER in \% of two-sided Dunnett-type tests with kernel estimator}
\end{table}\begin{table}[H]
\centering\footnotesize
\begin{tabular}{clllrrrr}
\toprule

\multicolumn{4}{c}{Setting} & \multicolumn{4}{c}{Method}\\

\cmidrule(lr){1-4}\cmidrule(lr){5-8}

Distribution & \multicolumn{2}{c}{Design} & $\delta$ & asymp. MCTP & boot. MCTP & B. asymp. & B. perm. \\
\midrule
$\chi^2_3$&bal&hom&0.0&\textbf{5.04} & 3.72 & \textbf{4.46} & 4.38 \\
$\chi^2_3$&bal&neg&0.0&6.40 & \textbf{4.58} & 5.66 & 6.02 \\
$\chi^2_3$&bal&pos&0.0&\textbf{4.54} & 3.58 & 4.32 & \textbf{4.78} \\
$\chi^2_3$&unb&hom&0.0&6.00 & \textbf{5.32} & \textbf{5.12} & 4.32 \\
$\chi^2_3$&unb&neg&0.0&7.44 & 6.58 & 6.38 & 6.20 \\
$\chi^2_3$&unb&pos&0.0&\textbf{4.76} & 3.64 & \textbf{4.50} & 3.86 \\
\bottomrule
\end{tabular}
\caption{Empirical FWER in \% of non-inferiority Dunnett-type tests with kernel estimator}
\end{table}\begin{table}[H]
\centering\footnotesize
\begin{tabular}{clllrrrr}
\toprule

\multicolumn{4}{c}{Setting} & \multicolumn{4}{c}{Method}\\

\cmidrule(lr){1-4}\cmidrule(lr){5-8}

Distribution & \multicolumn{2}{c}{Design} & $\delta$ & asymp. MCTP & boot. MCTP & B. asymp. & B. perm. \\
\midrule
$\chi^2_3$&bal&hom&0.0&4.38 & \textbf{4.78} & 3.56 & 4.34 \\
$\chi^2_3$&bal&neg&0.0&\textbf{4.62} & \textbf{4.82} & 3.60 & \textbf{4.84} \\
$\chi^2_3$&bal&pos&0.0&\textbf{4.76} & \textbf{4.78} & 3.82 & \textbf{5.20} \\
$\chi^2_3$&unb&hom&0.0&\textbf{4.56} & \textbf{4.84} & 3.78 & 4.14 \\
$\chi^2_3$&unb&neg&0.0&\textbf{5.40} & \textbf{5.44} & \textbf{4.58} & \textbf{5.24} \\
$\chi^2_3$&unb&pos&0.0&4.14 & 4.24 & 3.32 & 4.24 \\
\bottomrule
\end{tabular}
\caption{Empirical FWER in \% of two-sided Tukey-type tests with bootstrap estimator}
\end{table}\begin{table}[H]
\centering\footnotesize
\begin{tabular}{clllrrrr}
\toprule

\multicolumn{4}{c}{Setting} & \multicolumn{4}{c}{Method}\\

\cmidrule(lr){1-4}\cmidrule(lr){5-8}

Distribution & \multicolumn{2}{c}{Design} & $\delta$ & asymp. MCTP & boot. MCTP & B. asymp. & B. perm. \\
\midrule
$\chi^2_3$&bal&hom&0.0&3.82 & 5.86 & 3.16 & 4.30 \\
$\chi^2_3$&bal&neg&0.0&\textbf{4.44} & 6.34 & 3.66 & \textbf{5.50} \\
$\chi^2_3$&bal&pos&0.0&4.00 & \textbf{5.48} & 3.24 & \textbf{4.52} \\
$\chi^2_3$&unb&hom&0.0&\textbf{4.52} & 7.04 & 3.84 & 4.14 \\
$\chi^2_3$&unb&neg&0.0&6.64 & 8.62 & \textbf{5.50} & 6.24 \\
$\chi^2_3$&unb&pos&0.0&4.08 & 6.20 & 3.46 & 3.96 \\
\bottomrule
\end{tabular}
\caption{Empirical FWER in \% of non-inferiority Tukey-type tests with bootstrap estimator}
\end{table}\begin{table}[H]
\centering\footnotesize
\begin{tabular}{clllrrrr}
\toprule

\multicolumn{4}{c}{Setting} & \multicolumn{4}{c}{Method}\\

\cmidrule(lr){1-4}\cmidrule(lr){5-8}

Distribution & \multicolumn{2}{c}{Design} & $\delta$ & asymp. MCTP & boot. MCTP & B. asymp. & B. perm. \\
\midrule
$\chi^2_3$&bal&hom&0.0&6.00 & 1.34 & \textbf{4.56} & 4.24 \\
$\chi^2_3$&bal&neg&0.0&7.28 & 1.50 & 6.18 & 6.04 \\
$\chi^2_3$&bal&pos&0.0&6.78 & 1.20 & \textbf{5.36} & \textbf{5.04} \\
$\chi^2_3$&unb&hom&0.0&\textbf{5.04} & 1.46 & 4.02 & 4.20 \\
$\chi^2_3$&unb&neg&0.0&6.00 & 1.66 & \textbf{4.86} & 6.06 \\
$\chi^2_3$&unb&pos&0.0&\textbf{4.54} & 0.92 & 3.68 & \textbf{4.78} \\
\bottomrule
\end{tabular}
\caption{Empirical FWER in \% of two-sided Tukey-type tests with interval-based estimator}
\end{table}\begin{table}[H]
\centering\footnotesize
\begin{tabular}{clllrrrr}
\toprule

\multicolumn{4}{c}{Setting} & \multicolumn{4}{c}{Method}\\

\cmidrule(lr){1-4}\cmidrule(lr){5-8}

Distribution & \multicolumn{2}{c}{Design} & $\delta$ & asymp. MCTP & boot. MCTP & B. asymp. & B. perm. \\
\midrule
$\chi^2_3$&bal&hom&0.0&\textbf{5.50} & 3.04 & \textbf{4.52} & 4.06 \\
$\chi^2_3$&bal&neg&0.0&7.56 & 3.08 & 6.42 & 6.40 \\
$\chi^2_3$&bal&pos&0.0&\textbf{4.52} & 2.40 & 3.62 & 3.54 \\
$\chi^2_3$&unb&hom&0.0&5.88 & 3.32 & \textbf{4.94} & 4.00 \\
$\chi^2_3$&unb&neg&0.0&8.12 & 3.88 & 6.92 & 7.08 \\
$\chi^2_3$&unb&pos&0.0&\textbf{4.48} & 2.90 & 3.82 & 4.10 \\
\bottomrule
\end{tabular}
\caption{Empirical FWER in \% of non-inferiority Tukey-type tests with interval-based estimator}
\end{table}\begin{table}[H]
\centering\footnotesize
\begin{tabular}{clllrrrr}
\toprule

\multicolumn{4}{c}{Setting} & \multicolumn{4}{c}{Method}\\

\cmidrule(lr){1-4}\cmidrule(lr){5-8}

Distribution & \multicolumn{2}{c}{Design} & $\delta$ & asymp. MCTP & boot. MCTP & B. asymp. & B. perm. \\
\midrule
$\chi^2_3$&bal&hom&0.0&6.02 & 1.22 & \textbf{4.92} & 4.14 \\
$\chi^2_3$&bal&neg&0.0&7.22 & 1.48 & 5.66 & 5.92 \\
$\chi^2_3$&bal&pos&0.0&6.94 & 1.38 & 5.84 & 5.66 \\
$\chi^2_3$&unb&hom&0.0&6.66 & 1.16 & \textbf{5.54} & 4.24 \\
$\chi^2_3$&unb&neg&0.0&8.18 & 1.54 & 6.96 & 6.22 \\
$\chi^2_3$&unb&pos&0.0&6.12 & 0.84 & \textbf{4.82} & 4.12 \\
\bottomrule
\end{tabular}
\caption{Empirical FWER in \% of two-sided Tukey-type tests with kernel estimator}
\end{table}\begin{table}[H]
\centering\footnotesize
\begin{tabular}{clllrrrr}
\toprule

\multicolumn{4}{c}{Setting} & \multicolumn{4}{c}{Method}\\

\cmidrule(lr){1-4}\cmidrule(lr){5-8}

Distribution & \multicolumn{2}{c}{Design} & $\delta$ & asymp. MCTP & boot. MCTP & B. asymp. & B. perm. \\
\midrule
$\chi^2_3$&bal&hom&0.0&5.80 & 2.48 & \textbf{4.96} & 4.20 \\
$\chi^2_3$&bal&neg&0.0&7.12 & 2.32 & 5.98 & 6.58 \\
$\chi^2_3$&bal&pos&0.0&\textbf{5.00} & 1.96 & 4.30 & \textbf{4.50} \\
$\chi^2_3$&unb&hom&0.0&6.46 & 3.12 & \textbf{5.54} & 4.38 \\
$\chi^2_3$&unb&neg&0.0&9.38 & 4.08 & 7.78 & 7.32 \\
$\chi^2_3$&unb&pos&0.0&\textbf{5.28} & 2.48 & \textbf{4.56} & 3.94 \\
\bottomrule
\end{tabular}
\caption{Empirical FWER in \% of non-inferiority Tukey-type tests with kernel estimator}
\end{table}\begin{table}[H]
\centering\footnotesize
\begin{tabular}{clllrrrr}
\toprule

\multicolumn{4}{c}{Setting} & \multicolumn{4}{c}{Method}\\

\cmidrule(lr){1-4}\cmidrule(lr){5-8}

Distribution & \multicolumn{2}{c}{Design} & $\delta$ & asymp. MCTP & boot. MCTP & B. asymp. & B. perm. \\
\midrule
$\chi^2_3$&bal&hom&0.0&4.24 & \textbf{5.24} & 3.96 & \textbf{5.14} \\
$\chi^2_3$&bal&neg&0.0&4.14 & \textbf{5.40} & 3.96 & \textbf{4.90} \\
$\chi^2_3$&bal&pos&0.0&4.08 & \textbf{5.60} & 3.84 & \textbf{5.02} \\
$\chi^2_3$&unb&hom&0.0&3.88 & \textbf{5.06} & 3.50 & 4.30 \\
$\chi^2_3$&unb&neg&0.0&\textbf{4.42} & \textbf{5.56} & 3.86 & \textbf{5.18} \\
$\chi^2_3$&unb&pos&0.0&4.24 & \textbf{5.28} & 3.72 & \textbf{5.02} \\
\bottomrule
\end{tabular}
\caption{Empirical FWER in \% of two-sided Grand-mean-type tests with bootstrap estimator}
\end{table}\begin{table}[H]
\centering\footnotesize
\begin{tabular}{clllrrrr}
\toprule

\multicolumn{4}{c}{Setting} & \multicolumn{4}{c}{Method}\\

\cmidrule(lr){1-4}\cmidrule(lr){5-8}

Distribution & \multicolumn{2}{c}{Design} & $\delta$ & asymp. MCTP & boot. MCTP & B. asymp. & B. perm. \\
\midrule
$\chi^2_3$&bal&hom&0.0&2.94 & \textbf{4.78} & 2.94 & \textbf{4.94} \\
$\chi^2_3$&bal&neg&0.0&2.70 & \textbf{5.34} & 2.70 & 6.20 \\
$\chi^2_3$&bal&pos&0.0&2.82 & \textbf{4.78} & 2.82 & \textbf{5.54} \\
$\chi^2_3$&unb&hom&0.0&2.76 & \textbf{4.60} & 2.72 & \textbf{4.54} \\
$\chi^2_3$&unb&neg&0.0&3.34 & \textbf{5.60} & 3.22 & 6.24 \\
$\chi^2_3$&unb&pos&0.0&2.98 & \textbf{4.78} & 2.94 & \textbf{5.32} \\
\bottomrule
\end{tabular}
\caption{Empirical FWER in \% of non-inferiority Grand-mean-type tests with bootstrap estimator}
\end{table}\begin{table}[H]
\centering\footnotesize
\begin{tabular}{clllrrrr}
\toprule

\multicolumn{4}{c}{Setting} & \multicolumn{4}{c}{Method}\\

\cmidrule(lr){1-4}\cmidrule(lr){5-8}

Distribution & \multicolumn{2}{c}{Design} & $\delta$ & asymp. MCTP & boot. MCTP & B. asymp. & B. perm. \\
\midrule
$\chi^2_3$&bal&hom&0.0&6.94 & 2.20 & 6.44 & \textbf{4.72} \\
$\chi^2_3$&bal&neg&0.0&7.66 & 2.48 & 7.02 & 5.86 \\
$\chi^2_3$&bal&pos&0.0&7.06 & 2.20 & 6.48 & \textbf{5.28} \\
$\chi^2_3$&unb&hom&0.0&\textbf{5.22} & 1.52 & \textbf{4.70} & 4.32 \\
$\chi^2_3$&unb&neg&0.0&\textbf{5.10} & 1.74 & \textbf{4.56} & 6.22 \\
$\chi^2_3$&unb&pos&0.0&\textbf{5.08} & 1.64 & \textbf{4.60} & 6.12 \\
\bottomrule
\end{tabular}
\caption{Empirical FWER in \% of two-sided Grand-mean-type tests with interval-based estimator}
\end{table}\begin{table}[H]
\centering\footnotesize
\begin{tabular}{clllrrrr}
\toprule

\multicolumn{4}{c}{Setting} & \multicolumn{4}{c}{Method}\\

\cmidrule(lr){1-4}\cmidrule(lr){5-8}

Distribution & \multicolumn{2}{c}{Design} & $\delta$ & asymp. MCTP & boot. MCTP & B. asymp. & B. perm. \\
\midrule
$\chi^2_3$&bal&hom&0.0&3.56 & 1.40 & 3.54 & \textbf{5.00} \\
$\chi^2_3$&bal&neg&0.0&4.12 & 1.80 & 4.10 & 6.28 \\
$\chi^2_3$&bal&pos&0.0&3.52 & 1.52 & 3.48 & \textbf{5.44} \\
$\chi^2_3$&unb&hom&0.0&2.78 & 1.20 & 2.72 & \textbf{5.02} \\
$\chi^2_3$&unb&neg&0.0&3.46 & 1.44 & 3.44 & 6.02 \\
$\chi^2_3$&unb&pos&0.0&2.94 & 1.14 & 2.92 & 6.16 \\
\bottomrule
\end{tabular}
\caption{Empirical FWER in \% of non-inferiority Grand-mean-type tests with interval-based estimator}
\end{table}\begin{table}[H]
\centering\footnotesize
\begin{tabular}{clllrrrr}
\toprule

\multicolumn{4}{c}{Setting} & \multicolumn{4}{c}{Method}\\

\cmidrule(lr){1-4}\cmidrule(lr){5-8}

Distribution & \multicolumn{2}{c}{Design} & $\delta$ & asymp. MCTP & boot. MCTP & B. asymp. & B. perm. \\
\midrule
$\chi^2_3$&bal&hom&0.0&6.74 & 1.56 & 6.28 & \textbf{4.78} \\
$\chi^2_3$&bal&neg&0.0&7.30 & 1.68 & 6.68 & 5.74 \\
$\chi^2_3$&bal&pos&0.0&7.04 & 1.64 & 6.40 & 5.72 \\
$\chi^2_3$&unb&hom&0.0&6.48 & 1.38 & 6.10 & 4.40 \\
$\chi^2_3$&unb&neg&0.0&6.98 & 1.80 & 6.30 & 5.84 \\
$\chi^2_3$&unb&pos&0.0&6.40 & 1.42 & 5.92 & \textbf{5.46} \\
\bottomrule
\end{tabular}
\caption{Empirical FWER in \% of two-sided Grand-mean-type tests with kernel estimator}
\end{table}\begin{table}[H]
\centering\footnotesize
\begin{tabular}{clllrrrr}
\toprule

\multicolumn{4}{c}{Setting} & \multicolumn{4}{c}{Method}\\

\cmidrule(lr){1-4}\cmidrule(lr){5-8}

Distribution & \multicolumn{2}{c}{Design} & $\delta$ & asymp. MCTP & boot. MCTP & B. asymp. & B. perm. \\
\midrule
$\chi^2_3$&bal&hom&0.0&3.94 & 1.50 & 3.92 & \textbf{4.78} \\
$\chi^2_3$&bal&neg&0.0&4.30 & 1.84 & 4.28 & 6.64 \\
$\chi^2_3$&bal&pos&0.0&3.96 & 1.58 & 3.90 & 5.78 \\
$\chi^2_3$&unb&hom&0.0&4.24 & 1.32 & 4.20 & \textbf{4.88} \\
$\chi^2_3$&unb&neg&0.0&\textbf{5.24} & 1.56 & \textbf{5.18} & 7.12 \\
$\chi^2_3$&unb&pos&0.0&4.04 & 1.60 & 4.00 & \textbf{4.98} \\
\bottomrule
\end{tabular}
\caption{Empirical FWER in \% of non-inferiority Grand-mean-type tests with kernel estimator}
\end{table}\begin{table}[H]
\centering\footnotesize
\begin{tabular}{clllrrrr}
\toprule

\multicolumn{4}{c}{Setting} & \multicolumn{4}{c}{Method}\\

\cmidrule(lr){1-4}\cmidrule(lr){5-8}

Distribution & \multicolumn{2}{c}{Design} & $\delta$ & asymp. MCTP & boot. MCTP & B. asymp. & B. perm. \\
\midrule
$\chi^2_3$&bal&hom&0.5&5.46 & 6.56 & 5.02 & 6.64 \\
$\chi^2_3$&bal&neg&0.5&5.22 & 6.98 & 4.78 & 6.42 \\
$\chi^2_3$&bal&pos&0.5&4.60 & 5.24 & 4.38 & 5.98 \\
$\chi^2_3$&unb&hom&0.5&5.94 & 7.50 & 5.18 & 6.74 \\
$\chi^2_3$&unb&neg&0.5&6.82 & 8.26 & 6.10 & 8.02 \\
$\chi^2_3$&unb&pos&0.5&4.18 & 5.02 & 3.98 & 5.80 \\
$\mathcal{LN}(0,1)$&bal&hom&0.5&7.48 & 10.58 & 7.00 & 12.72 \\
$\mathcal{LN}(0,1)$&bal&neg&0.5&5.68 & 8.62 & 5.20 & 10.22 \\
$\mathcal{LN}(0,1)$&bal&pos&0.5&4.18 & 5.60 & 4.00 & 8.70 \\
$\mathcal{LN}(0,1)$&unb&hom&0.5&10.26 & 14.02 & 9.28 & 15.60 \\
$\mathcal{LN}(0,1)$&unb&neg&0.5&9.52 & 12.06 & 8.58 & 13.24 \\
$\mathcal{LN}(0,1)$&unb&pos&0.5&4.66 & 6.54 & 4.38 & 8.46 \\
$\mathcal{N}(0,1)$&bal&hom&0.5&11.66 & 13.02 & 10.74 & 12.42 \\
$\mathcal{N}(0,1)$&bal&neg&0.5&7.50 & 9.32 & 6.68 & 7.84 \\
$\mathcal{N}(0,1)$&bal&pos&0.5&8.12 & 8.32 & 7.82 & 9.78 \\
$\mathcal{N}(0,1)$&unb&hom&0.5&12.32 & 13.48 & 11.08 & 11.74 \\
$\mathcal{N}(0,1)$&unb&neg&0.5&8.92 & 10.46 & 7.62 & 8.66 \\
$\mathcal{N}(0,1)$&unb&pos&0.5&8.40 & 9.10 & 7.84 & 9.42 \\
$t_2$&bal&hom&0.5&6.40 & 10.26 & 5.62 & 9.90 \\
$t_2$&bal&neg&0.5&4.48 & 8.60 & 3.64 & 7.90 \\
$t_2$&bal&pos&0.5&4.44 & 6.92 & 4.16 & 8.18 \\
$t_2$&unb&hom&0.5&6.72 & 10.04 & 5.92 & 10.08 \\
$t_2$&unb&neg&0.5&4.90 & 7.66 & 4.06 & 7.36 \\
$t_2$&unb&pos&0.5&3.66 & 6.48 & 3.44 & 7.78 \\
$t_3$&bal&hom&0.5&8.30 & 11.16 & 7.60 & 10.58 \\
$t_3$&bal&neg&0.5&5.90 & 8.86 & 5.24 & 7.56 \\
$t_3$&bal&pos&0.5&5.40 & 7.52 & 5.16 & 8.68 \\
$t_3$&unb&hom&0.5&8.06 & 10.86 & 7.18 & 10.42 \\
$t_3$&unb&neg&0.5&6.28 & 8.70 & 5.14 & 7.90 \\
$t_3$&unb&pos&0.5&5.62 & 7.78 & 5.08 & 8.32 \\
\bottomrule
\end{tabular}
\caption{Empirical Power for $\delta = 0.5$ in \% of two-sided Dunnett-type tests with bootstrap estimator}
\end{table}\begin{table}[H]
\centering\footnotesize
\begin{tabular}{clllrrrr}
\toprule

\multicolumn{4}{c}{Setting} & \multicolumn{4}{c}{Method}\\

\cmidrule(lr){1-4}\cmidrule(lr){5-8}

Distribution & \multicolumn{2}{c}{Design} & $\delta$ & asymp. MCTP & boot. MCTP & B. asymp. & B. perm. \\
\midrule
$\chi^2_3$&bal&hom&0.5&6.94 & 11.86 & 6.14 & 8.52 \\
$\chi^2_3$&bal&neg&0.5&5.96 & 11.34 & 5.28 & 7.54 \\
$\chi^2_3$&bal&pos&0.5&4.60 & 8.48 & 4.26 & 6.46 \\
$\chi^2_3$&unb&hom&0.5&8.16 & 13.66 & 7.06 & 8.14 \\
$\chi^2_3$&unb&neg&0.5&9.00 & 14.20 & 7.88 & 9.06 \\
$\chi^2_3$&unb&pos&0.5&5.24 & 9.70 & 4.96 & 5.78 \\
$\mathcal{LN}(0,1)$&bal&hom&0.5&10.56 & 18.98 & 9.72 & 16.48 \\
$\mathcal{LN}(0,1)$&bal&neg&0.5&8.26 & 15.48 & 7.32 & 13.10 \\
$\mathcal{LN}(0,1)$&bal&pos&0.5&5.60 & 11.98 & 5.34 & 10.94 \\
$\mathcal{LN}(0,1)$&unb&hom&0.5&15.20 & 23.94 & 14.00 & 17.98 \\
$\mathcal{LN}(0,1)$&unb&neg&0.5&12.86 & 19.62 & 11.54 & 14.92 \\
$\mathcal{LN}(0,1)$&unb&pos&0.5&6.84 & 14.10 & 6.36 & 10.28 \\
$\mathcal{N}(0,1)$&bal&hom&0.5&15.76 & 21.90 & 14.26 & 17.36 \\
$\mathcal{N}(0,1)$&bal&neg&0.5&9.82 & 16.20 & 8.20 & 10.12 \\
$\mathcal{N}(0,1)$&bal&pos&0.5&10.64 & 14.98 & 9.82 & 12.82 \\
$\mathcal{N}(0,1)$&unb&hom&0.5&17.00 & 23.78 & 14.52 & 14.76 \\
$\mathcal{N}(0,1)$&unb&neg&0.5&12.40 & 19.64 & 9.92 & 10.08 \\
$\mathcal{N}(0,1)$&unb&pos&0.5&11.04 & 15.94 & 10.22 & 10.76 \\
$t_2$&bal&hom&0.5&9.62 & 18.52 & 8.48 & 13.64 \\
$t_2$&bal&neg&0.5&6.48 & 14.48 & 5.28 & 9.20 \\
$t_2$&bal&pos&0.5&6.26 & 13.26 & 5.80 & 10.52 \\
$t_2$&unb&hom&0.5&9.68 & 19.94 & 8.32 & 11.72 \\
$t_2$&unb&neg&0.5&6.84 & 15.66 & 5.44 & 7.80 \\
$t_2$&unb&pos&0.5&5.82 & 13.42 & 5.14 & 9.84 \\
$t_3$&bal&hom&0.5&11.46 & 19.74 & 10.22 & 14.10 \\
$t_3$&bal&neg&0.5&8.42 & 15.80 & 6.84 & 10.10 \\
$t_3$&bal&pos&0.5&7.70 & 13.90 & 7.18 & 11.52 \\
$t_3$&unb&hom&0.5&11.36 & 19.62 & 9.68 & 11.94 \\
$t_3$&unb&neg&0.5&8.72 & 17.36 & 6.80 & 8.66 \\
$t_3$&unb&pos&0.5&7.58 & 14.60 & 6.68 & 9.46 \\
\bottomrule
\end{tabular}
\caption{Empirical Power for $\delta = 0.5$ in \% of non-inferiority Dunnett-type tests with bootstrap estimator}
\end{table}\begin{table}[H]
\centering\footnotesize
\begin{tabular}{clllrrrr}
\toprule

\multicolumn{4}{c}{Setting} & \multicolumn{4}{c}{Method}\\

\cmidrule(lr){1-4}\cmidrule(lr){5-8}

Distribution & \multicolumn{2}{c}{Design} & $\delta$ & asymp. MCTP & boot. MCTP & B. asymp. & B. perm. \\
\midrule
$\chi^2_3$&bal&hom&0.5&7.72 & 2.44 & 6.92 & 6.64 \\
$\chi^2_3$&bal&neg&0.5&7.78 & 3.34 & 7.04 & 7.78 \\
$\chi^2_3$&bal&pos&0.5&6.46 & 1.72 & 6.18 & 6.10 \\
$\chi^2_3$&unb&hom&0.5&6.56 & 2.46 & 5.78 & 6.92 \\
$\chi^2_3$&unb&neg&0.5&7.98 & 3.76 & 6.88 & 9.44 \\
$\chi^2_3$&unb&pos&0.5&5.38 & 1.68 & 4.96 & 6.48 \\
$\mathcal{LN}(0,1)$&bal&hom&0.5&14.32 & 4.50 & 13.20 & 14.90 \\
$\mathcal{LN}(0,1)$&bal&neg&0.5&12.14 & 4.74 & 10.62 & 13.34 \\
$\mathcal{LN}(0,1)$&bal&pos&0.5&8.22 & 1.68 & 7.92 & 8.46 \\
$\mathcal{LN}(0,1)$&unb&hom&0.5&14.52 & 6.70 & 13.44 & 17.10 \\
$\mathcal{LN}(0,1)$&unb&neg&0.5&13.58 & 6.86 & 12.20 & 16.38 \\
$\mathcal{LN}(0,1)$&unb&pos&0.5&6.64 & 2.08 & 6.22 & 8.82 \\
$\mathcal{N}(0,1)$&bal&hom&0.5&10.06 & 5.78 & 9.32 & 12.38 \\
$\mathcal{N}(0,1)$&bal&neg&0.5&6.62 & 3.78 & 5.78 & 8.64 \\
$\mathcal{N}(0,1)$&bal&pos&0.5&5.92 & 2.60 & 5.60 & 8.24 \\
$\mathcal{N}(0,1)$&unb&hom&0.5&7.88 & 4.40 & 6.82 & 11.34 \\
$\mathcal{N}(0,1)$&unb&neg&0.5&5.50 & 3.20 & 4.48 & 8.70 \\
$\mathcal{N}(0,1)$&unb&pos&0.5&5.40 & 2.78 & 5.04 & 10.28 \\
$t_2$&bal&hom&0.5&4.46 & 3.00 & 4.10 & 9.12 \\
$t_2$&bal&neg&0.5&3.10 & 2.12 & 2.78 & 7.64 \\
$t_2$&bal&pos&0.5&2.94 & 1.76 & 2.80 & 7.42 \\
$t_2$&unb&hom&0.5&3.02 & 2.06 & 2.54 & 8.64 \\
$t_2$&unb&neg&0.5&1.78 & 1.34 & 1.54 & 5.96 \\
$t_2$&unb&pos&0.5&1.58 & 1.04 & 1.52 & 7.26 \\
$t_3$&bal&hom&0.5&6.30 & 4.38 & 5.90 & 10.36 \\
$t_3$&bal&neg&0.5&4.70 & 3.12 & 4.00 & 8.52 \\
$t_3$&bal&pos&0.5&4.16 & 2.28 & 3.94 & 7.80 \\
$t_3$&unb&hom&0.5&3.64 & 2.24 & 3.22 & 8.94 \\
$t_3$&unb&neg&0.5&2.66 & 1.56 & 2.12 & 6.58 \\
$t_3$&unb&pos&0.5&2.56 & 1.52 & 2.30 & 8.30 \\
\bottomrule
\end{tabular}
\caption{Empirical Power for $\delta = 0.5$ in \% of two-sided Dunnett-type tests with interval-based estimator}
\end{table}\begin{table}[H]
\centering\footnotesize
\begin{tabular}{clllrrrr}
\toprule

\multicolumn{4}{c}{Setting} & \multicolumn{4}{c}{Method}\\

\cmidrule(lr){1-4}\cmidrule(lr){5-8}

Distribution & \multicolumn{2}{c}{Design} & $\delta$ & asymp. MCTP & boot. MCTP & B. asymp. & B. perm. \\
\midrule
$\chi^2_3$&bal&hom&0.5&9.38 & 8.48 & 8.82 & 8.58 \\
$\chi^2_3$&bal&neg&0.5&9.06 & 8.00 & 8.00 & 9.18 \\
$\chi^2_3$&bal&pos&0.5&6.32 & 5.56 & 6.22 & 5.96 \\
$\chi^2_3$&unb&hom&0.5&9.40 & 9.32 & 8.30 & 7.58 \\
$\chi^2_3$&unb&neg&0.5&10.82 & 10.96 & 9.30 & 10.46 \\
$\chi^2_3$&unb&pos&0.5&6.42 & 6.52 & 5.82 & 6.30 \\
$\mathcal{LN}(0,1)$&bal&hom&0.5&18.94 & 13.88 & 17.56 & 19.00 \\
$\mathcal{LN}(0,1)$&bal&neg&0.5&15.24 & 11.64 & 13.74 & 16.50 \\
$\mathcal{LN}(0,1)$&bal&pos&0.5&10.36 & 7.46 & 9.74 & 9.98 \\
$\mathcal{LN}(0,1)$&unb&hom&0.5&20.28 & 17.74 & 18.62 & 19.04 \\
$\mathcal{LN}(0,1)$&unb&neg&0.5&17.42 & 15.42 & 15.70 & 17.44 \\
$\mathcal{LN}(0,1)$&unb&pos&0.5&9.50 & 9.04 & 8.80 & 9.70 \\
$\mathcal{N}(0,1)$&bal&hom&0.5&14.08 & 15.42 & 12.38 & 16.62 \\
$\mathcal{N}(0,1)$&bal&neg&0.5&9.40 & 11.50 & 7.94 & 11.14 \\
$\mathcal{N}(0,1)$&bal&pos&0.5&8.06 & 9.82 & 7.60 & 10.90 \\
$\mathcal{N}(0,1)$&unb&hom&0.5&12.30 & 14.76 & 10.66 & 13.68 \\
$\mathcal{N}(0,1)$&unb&neg&0.5&8.56 & 11.80 & 6.60 & 10.04 \\
$\mathcal{N}(0,1)$&unb&pos&0.5&8.04 & 9.72 & 7.30 & 11.18 \\
$t_2$&bal&hom&0.5&7.22 & 11.46 & 6.40 & 12.36 \\
$t_2$&bal&neg&0.5&4.80 & 9.10 & 3.76 & 9.46 \\
$t_2$&bal&pos&0.5&4.74 & 7.86 & 4.38 & 10.26 \\
$t_2$&unb&hom&0.5&5.10 & 9.24 & 4.32 & 10.58 \\
$t_2$&unb&neg&0.5&3.26 & 6.38 & 2.50 & 6.86 \\
$t_2$&unb&pos&0.5&2.68 & 5.58 & 2.40 & 9.08 \\
$t_3$&bal&hom&0.5&8.96 & 12.94 & 7.98 & 13.66 \\
$t_3$&bal&neg&0.5&6.96 & 10.54 & 5.80 & 10.36 \\
$t_3$&bal&pos&0.5&5.72 & 9.02 & 5.24 & 10.24 \\
$t_3$&unb&hom&0.5&6.24 & 9.86 & 4.98 & 11.00 \\
$t_3$&unb&neg&0.5&4.38 & 8.84 & 3.44 & 7.38 \\
$t_3$&unb&pos&0.5&4.22 & 7.46 & 3.62 & 10.28 \\
\bottomrule
\end{tabular}
\caption{Empirical Power for $\delta = 0.5$ in \% of non-inferiority Dunnett-type tests with interval-based estimator}
\end{table}\begin{table}[H]
\centering\footnotesize
\begin{tabular}{clllrrrr}
\toprule

\multicolumn{4}{c}{Setting} & \multicolumn{4}{c}{Method}\\

\cmidrule(lr){1-4}\cmidrule(lr){5-8}

Distribution & \multicolumn{2}{c}{Design} & $\delta$ & asymp. MCTP & boot. MCTP & B. asymp. & B. perm. \\
\midrule
$\chi^2_3$&bal&hom&0.5&7.60 & 1.92 & 7.04 & 6.44 \\
$\chi^2_3$&bal&neg&0.5&7.80 & 2.68 & 6.94 & 7.36 \\
$\chi^2_3$&bal&pos&0.5&6.88 & 1.58 & 6.62 & 6.60 \\
$\chi^2_3$&unb&hom&0.5&8.30 & 1.98 & 7.56 & 6.56 \\
$\chi^2_3$&unb&neg&0.5&9.86 & 3.50 & 8.72 & 9.00 \\
$\chi^2_3$&unb&pos&0.5&5.96 & 1.54 & 5.72 & 5.48 \\
$\mathcal{LN}(0,1)$&bal&hom&0.5&12.08 & 4.02 & 11.22 & 15.22 \\
$\mathcal{LN}(0,1)$&bal&neg&0.5&9.56 & 4.16 & 8.76 & 13.16 \\
$\mathcal{LN}(0,1)$&bal&pos&0.5&6.22 & 1.36 & 5.86 & 8.88 \\
$\mathcal{LN}(0,1)$&unb&hom&0.5&15.14 & 5.56 & 14.26 & 17.62 \\
$\mathcal{LN}(0,1)$&unb&neg&0.5&14.74 & 6.54 & 13.38 & 16.40 \\
$\mathcal{LN}(0,1)$&unb&pos&0.5&6.28 & 1.62 & 5.90 & 7.66 \\
$\mathcal{N}(0,1)$&bal&hom&0.5&13.62 & 4.86 & 12.66 & 12.08 \\
$\mathcal{N}(0,1)$&bal&neg&0.5&9.16 & 3.36 & 7.76 & 7.84 \\
$\mathcal{N}(0,1)$&bal&pos&0.5&9.08 & 2.38 & 8.66 & 9.44 \\
$\mathcal{N}(0,1)$&unb&hom&0.5&15.08 & 4.68 & 13.58 & 11.60 \\
$\mathcal{N}(0,1)$&unb&neg&0.5&11.64 & 3.72 & 9.90 & 8.56 \\
$\mathcal{N}(0,1)$&unb&pos&0.5&10.72 & 2.70 & 10.12 & 8.70 \\
$t_2$&bal&hom&0.5&6.50 & 3.32 & 5.98 & 9.72 \\
$t_2$&bal&neg&0.5&4.82 & 2.38 & 3.94 & 7.52 \\
$t_2$&bal&pos&0.5&4.76 & 1.90 & 4.42 & 8.10 \\
$t_2$&unb&hom&0.5&8.74 & 3.38 & 7.72 & 10.04 \\
$t_2$&unb&neg&0.5&6.92 & 2.86 & 5.86 & 7.66 \\
$t_2$&unb&pos&0.5&5.18 & 1.98 & 4.92 & 6.80 \\
$t_3$&bal&hom&0.5&8.72 & 4.20 & 8.30 & 10.54 \\
$t_3$&bal&neg&0.5&6.58 & 3.26 & 5.72 & 7.60 \\
$t_3$&bal&pos&0.5&6.36 & 2.26 & 6.02 & 8.64 \\
$t_3$&unb&hom&0.5&10.20 & 3.34 & 9.18 & 10.06 \\
$t_3$&unb&neg&0.5&8.06 & 3.10 & 6.82 & 7.88 \\
$t_3$&unb&pos&0.5&6.68 & 2.08 & 6.20 & 7.40 \\
\bottomrule
\end{tabular}
\caption{Empirical Power for $\delta = 0.5$ in \% of two-sided Dunnett-type tests with kernel estimator}
\end{table}\begin{table}[H]
\centering\footnotesize
\begin{tabular}{clllrrrr}
\toprule

\multicolumn{4}{c}{Setting} & \multicolumn{4}{c}{Method}\\

\cmidrule(lr){1-4}\cmidrule(lr){5-8}

Distribution & \multicolumn{2}{c}{Design} & $\delta$ & asymp. MCTP & boot. MCTP & B. asymp. & B. perm. \\
\midrule
$\chi^2_3$&bal&hom&0.5&9.04 & 6.64 & 8.38 & 8.34 \\
$\chi^2_3$&bal&neg&0.5&8.94 & 7.20 & 7.84 & 9.06 \\
$\chi^2_3$&bal&pos&0.5&6.26 & 4.60 & 5.90 & 6.20 \\
$\chi^2_3$&unb&hom&0.5&10.24 & 8.02 & 9.30 & 8.18 \\
$\chi^2_3$&unb&neg&0.5&11.72 & 9.66 & 10.16 & 10.04 \\
$\chi^2_3$&unb&pos&0.5&6.62 & 5.62 & 6.30 & 5.60 \\
$\mathcal{LN}(0,1)$&bal&hom&0.5&16.42 & 13.44 & 15.28 & 19.82 \\
$\mathcal{LN}(0,1)$&bal&neg&0.5&13.64 & 11.30 & 11.82 & 16.64 \\
$\mathcal{LN}(0,1)$&bal&pos&0.5&7.56 & 6.66 & 7.28 & 10.24 \\
$\mathcal{LN}(0,1)$&unb&hom&0.5&20.66 & 17.10 & 18.94 & 20.94 \\
$\mathcal{LN}(0,1)$&unb&neg&0.5&18.42 & 15.34 & 16.78 & 18.60 \\
$\mathcal{LN}(0,1)$&unb&pos&0.5&8.38 & 7.98 & 7.88 & 9.40 \\
$\mathcal{N}(0,1)$&bal&hom&0.5&17.86 & 13.46 & 16.38 & 16.82 \\
$\mathcal{N}(0,1)$&bal&neg&0.5&11.48 & 10.58 & 9.60 & 10.10 \\
$\mathcal{N}(0,1)$&bal&pos&0.5&12.26 & 8.38 & 11.44 & 12.48 \\
$\mathcal{N}(0,1)$&unb&hom&0.5&19.90 & 13.64 & 17.40 & 14.68 \\
$\mathcal{N}(0,1)$&unb&neg&0.5&14.58 & 11.94 & 11.72 & 10.32 \\
$\mathcal{N}(0,1)$&unb&pos&0.5&12.72 & 8.58 & 11.82 & 10.88 \\
$t_2$&bal&hom&0.5&9.90 & 11.10 & 8.74 & 13.48 \\
$t_2$&bal&neg&0.5&6.60 & 8.80 & 5.52 & 9.02 \\
$t_2$&bal&pos&0.5&6.84 & 7.14 & 6.20 & 10.78 \\
$t_2$&unb&hom&0.5&12.64 & 11.84 & 10.78 & 12.36 \\
$t_2$&unb&neg&0.5&8.94 & 9.34 & 7.12 & 8.60 \\
$t_2$&unb&pos&0.5&6.72 & 7.06 & 6.12 & 8.78 \\
$t_3$&bal&hom&0.5&12.14 & 12.28 & 10.80 & 14.16 \\
$t_3$&bal&neg&0.5&8.86 & 9.84 & 7.38 & 10.20 \\
$t_3$&bal&pos&0.5&8.48 & 7.82 & 7.76 & 11.38 \\
$t_3$&unb&hom&0.5&13.72 & 11.48 & 11.84 & 12.30 \\
$t_3$&unb&neg&0.5&10.82 & 10.30 & 8.88 & 9.38 \\
$t_3$&unb&pos&0.5&8.88 & 7.46 & 7.90 & 9.08 \\
\bottomrule
\end{tabular}
\caption{Empirical Power for $\delta = 0.5$ in \% of non-inferiority Dunnett-type tests with kernel estimator}
\end{table}\begin{table}[H]
\centering\footnotesize
\begin{tabular}{clllrrrr}
\toprule

\multicolumn{4}{c}{Setting} & \multicolumn{4}{c}{Method}\\

\cmidrule(lr){1-4}\cmidrule(lr){5-8}

Distribution & \multicolumn{2}{c}{Design} & $\delta$ & asymp. MCTP & boot. MCTP & B. asymp. & B. perm. \\
\midrule
$\chi^2_3$&bal&hom&0.5&6.98 & 7.30 & 5.66 & 6.88 \\
$\chi^2_3$&bal&neg&0.5&7.32 & 7.24 & 6.06 & 8.00 \\
$\chi^2_3$&bal&pos&0.5&4.86 & 5.20 & 3.86 & 5.06 \\
$\chi^2_3$&unb&hom&0.5&8.44 & 8.64 & 7.24 & 8.28 \\
$\chi^2_3$&unb&neg&0.5&10.08 & 9.78 & 8.60 & 10.24 \\
$\chi^2_3$&unb&pos&0.5&5.10 & 5.72 & 4.26 & 5.08 \\
$\mathcal{LN}(0,1)$&bal&hom&0.5&12.18 & 14.28 & 10.30 & 16.90 \\
$\mathcal{LN}(0,1)$&bal&neg&0.5&10.62 & 12.58 & 9.22 & 16.56 \\
$\mathcal{LN}(0,1)$&bal&pos&0.5&4.94 & 6.04 & 4.12 & 8.08 \\
$\mathcal{LN}(0,1)$&unb&hom&0.5&17.84 & 19.96 & 15.60 & 23.54 \\
$\mathcal{LN}(0,1)$&unb&neg&0.5&17.80 & 19.30 & 15.54 & 23.64 \\
$\mathcal{LN}(0,1)$&unb&pos&0.5&6.64 & 8.04 & 5.50 & 9.38 \\
$\mathcal{N}(0,1)$&bal&hom&0.5&16.38 & 16.56 & 14.08 & 15.28 \\
$\mathcal{N}(0,1)$&bal&neg&0.5&12.80 & 11.68 & 10.94 & 11.98 \\
$\mathcal{N}(0,1)$&bal&pos&0.5&8.72 & 8.90 & 7.26 & 8.90 \\
$\mathcal{N}(0,1)$&unb&hom&0.5&19.40 & 17.62 & 16.90 & 17.16 \\
$\mathcal{N}(0,1)$&unb&neg&0.5&17.34 & 14.96 & 14.60 & 15.48 \\
$\mathcal{N}(0,1)$&unb&pos&0.5&10.06 & 9.82 & 8.72 & 9.26 \\
$t_2$&bal&hom&0.5&9.08 & 12.40 & 7.78 & 12.84 \\
$t_2$&bal&neg&0.5&7.12 & 10.14 & 5.74 & 10.14 \\
$t_2$&bal&pos&0.5&5.00 & 7.84 & 3.84 & 7.68 \\
$t_2$&unb&hom&0.5&10.94 & 13.24 & 9.30 & 13.56 \\
$t_2$&unb&neg&0.5&9.24 & 10.96 & 7.60 & 11.64 \\
$t_2$&unb&pos&0.5&5.16 & 7.60 & 4.18 & 7.62 \\
$t_3$&bal&hom&0.5&12.04 & 14.54 & 10.40 & 13.46 \\
$t_3$&bal&neg&0.5&9.16 & 11.04 & 8.04 & 10.66 \\
$t_3$&bal&pos&0.5&6.60 & 8.12 & 5.66 & 8.06 \\
$t_3$&unb&hom&0.5&13.20 & 14.88 & 11.04 & 14.42 \\
$t_3$&unb&neg&0.5&11.74 & 12.44 & 9.76 & 12.66 \\
$t_3$&unb&pos&0.5&7.78 & 9.60 & 6.48 & 9.06 \\
\bottomrule
\end{tabular}
\caption{Empirical Power for $\delta = 0.5$ in \% of two-sided Tukey-type tests with bootstrap estimator}
\end{table}\begin{table}[H]
\centering\footnotesize
\begin{tabular}{clllrrrr}
\toprule

\multicolumn{4}{c}{Setting} & \multicolumn{4}{c}{Method}\\

\cmidrule(lr){1-4}\cmidrule(lr){5-8}

Distribution & \multicolumn{2}{c}{Design} & $\delta$ & asymp. MCTP & boot. MCTP & B. asymp. & B. perm. \\
\midrule
$\chi^2_3$&bal&hom&0.5&8.84 & 11.86 & 7.70 & 9.80 \\
$\chi^2_3$&bal&neg&0.5&8.74 & 11.06 & 7.58 & 10.68 \\
$\chi^2_3$&bal&pos&0.5&5.20 & 7.68 & 4.32 & 5.98 \\
$\chi^2_3$&unb&hom&0.5&11.96 & 15.18 & 10.10 & 10.86 \\
$\chi^2_3$&unb&neg&0.5&13.54 & 15.30 & 11.78 & 13.34 \\
$\chi^2_3$&unb&pos&0.5&6.48 & 9.84 & 5.56 & 6.40 \\
$\mathcal{LN}(0,1)$&bal&hom&0.5&16.74 & 24.26 & 14.86 & 24.18 \\
$\mathcal{LN}(0,1)$&bal&neg&0.5&14.82 & 20.36 & 12.62 & 23.04 \\
$\mathcal{LN}(0,1)$&bal&pos&0.5&6.92 & 12.50 & 5.62 & 11.18 \\
$\mathcal{LN}(0,1)$&unb&hom&0.5&24.48 & 31.90 & 21.24 & 30.10 \\
$\mathcal{LN}(0,1)$&unb&neg&0.5&24.18 & 28.60 & 21.08 & 30.10 \\
$\mathcal{LN}(0,1)$&unb&pos&0.5&9.82 & 16.22 & 8.18 & 12.60 \\
$\mathcal{N}(0,1)$&bal&hom&0.5&22.32 & 25.24 & 20.16 & 22.82 \\
$\mathcal{N}(0,1)$&bal&neg&0.5&17.08 & 19.74 & 15.16 & 18.24 \\
$\mathcal{N}(0,1)$&bal&pos&0.5&11.86 & 14.54 & 10.44 & 12.90 \\
$\mathcal{N}(0,1)$&unb&hom&0.5&26.58 & 28.74 & 23.86 & 23.92 \\
$\mathcal{N}(0,1)$&unb&neg&0.5&23.58 & 25.32 & 20.76 & 21.42 \\
$\mathcal{N}(0,1)$&unb&pos&0.5&14.30 & 16.22 & 12.58 & 12.68 \\
$t_2$&bal&hom&0.5&12.92 & 20.88 & 11.30 & 18.46 \\
$t_2$&bal&neg&0.5&10.04 & 17.14 & 8.42 & 15.24 \\
$t_2$&bal&pos&0.5&7.20 & 12.56 & 5.96 & 11.20 \\
$t_2$&unb&hom&0.5&15.60 & 23.06 & 13.36 & 18.58 \\
$t_2$&unb&neg&0.5&13.32 & 19.78 & 11.00 & 15.98 \\
$t_2$&unb&pos&0.5&7.38 & 13.12 & 6.42 & 10.88 \\
$t_3$&bal&hom&0.5&16.58 & 22.82 & 14.60 & 19.22 \\
$t_3$&bal&neg&0.5&13.44 & 18.68 & 11.66 & 17.06 \\
$t_3$&bal&pos&0.5&9.20 & 14.22 & 7.94 & 12.42 \\
$t_3$&unb&hom&0.5&18.98 & 24.52 & 16.86 & 20.24 \\
$t_3$&unb&neg&0.5&16.48 & 21.74 & 14.10 & 17.56 \\
$t_3$&unb&pos&0.5&11.26 & 15.22 & 9.48 & 12.68 \\
\bottomrule
\end{tabular}
\caption{Empirical Power for $\delta = 0.5$ in \% of non-inferiority Tukey-type tests with bootstrap estimator}
\end{table}\begin{table}[H]
\centering\footnotesize
\begin{tabular}{clllrrrr}
\toprule

\multicolumn{4}{c}{Setting} & \multicolumn{4}{c}{Method}\\

\cmidrule(lr){1-4}\cmidrule(lr){5-8}

Distribution & \multicolumn{2}{c}{Design} & $\delta$ & asymp. MCTP & boot. MCTP & B. asymp. & B. perm. \\
\midrule
$\chi^2_3$&bal&hom&0.5&9.76 & 2.36 & 8.30 & 7.46 \\
$\chi^2_3$&bal&neg&0.5&11.14 & 2.76 & 9.64 & 9.70 \\
$\chi^2_3$&bal&pos&0.5&7.02 & 1.38 & 5.52 & 5.14 \\
$\chi^2_3$&unb&hom&0.5&9.12 & 3.02 & 7.48 & 7.74 \\
$\chi^2_3$&unb&neg&0.5&12.22 & 4.12 & 10.38 & 12.94 \\
$\chi^2_3$&unb&pos&0.5&6.34 & 1.52 & 5.14 & 6.00 \\
$\mathcal{LN}(0,1)$&bal&hom&0.5&22.72 & 6.62 & 20.14 & 20.84 \\
$\mathcal{LN}(0,1)$&bal&neg&0.5&21.38 & 7.28 & 18.76 & 21.46 \\
$\mathcal{LN}(0,1)$&bal&pos&0.5&10.06 & 1.76 & 8.28 & 8.42 \\
$\mathcal{LN}(0,1)$&unb&hom&0.5&25.60 & 9.96 & 22.68 & 26.72 \\
$\mathcal{LN}(0,1)$&unb&neg&0.5&24.86 & 10.44 & 22.18 & 28.20 \\
$\mathcal{LN}(0,1)$&unb&pos&0.5&9.46 & 2.44 & 7.78 & 9.78 \\
$\mathcal{N}(0,1)$&bal&hom&0.5&14.38 & 6.36 & 12.30 & 15.52 \\
$\mathcal{N}(0,1)$&bal&neg&0.5&11.92 & 4.96 & 10.14 & 13.94 \\
$\mathcal{N}(0,1)$&bal&pos&0.5&6.56 & 2.66 & 5.56 & 7.72 \\
$\mathcal{N}(0,1)$&unb&hom&0.5&15.70 & 7.36 & 13.52 & 17.22 \\
$\mathcal{N}(0,1)$&unb&neg&0.5&14.12 & 5.78 & 11.76 & 15.60 \\
$\mathcal{N}(0,1)$&unb&pos&0.5&7.78 & 3.00 & 6.34 & 10.16 \\
$t_2$&bal&hom&0.5&6.48 & 3.48 & 5.34 & 12.06 \\
$t_2$&bal&neg&0.5&5.18 & 2.72 & 4.26 & 10.20 \\
$t_2$&bal&pos&0.5&3.70 & 1.84 & 2.92 & 6.98 \\
$t_2$&unb&hom&0.5&7.16 & 4.12 & 5.48 & 12.28 \\
$t_2$&unb&neg&0.5&5.88 & 3.22 & 4.62 & 10.18 \\
$t_2$&unb&pos&0.5&3.20 & 1.90 & 2.64 & 7.54 \\
$t_3$&bal&hom&0.5&8.98 & 5.12 & 7.62 & 12.60 \\
$t_3$&bal&neg&0.5&7.42 & 3.94 & 5.90 & 11.42 \\
$t_3$&bal&pos&0.5&5.14 & 2.36 & 4.08 & 8.06 \\
$t_3$&unb&hom&0.5&9.56 & 4.90 & 7.92 & 13.36 \\
$t_3$&unb&neg&0.5&8.12 & 4.10 & 6.68 & 12.14 \\
$t_3$&unb&pos&0.5&4.78 & 2.52 & 4.08 & 9.14 \\
\bottomrule
\end{tabular}
\caption{Empirical Power for $\delta = 0.5$ in \% of two-sided Tukey-type tests with interval-based estimator}
\end{table}\begin{table}[H]
\centering\footnotesize
\begin{tabular}{clllrrrr}
\toprule

\multicolumn{4}{c}{Setting} & \multicolumn{4}{c}{Method}\\

\cmidrule(lr){1-4}\cmidrule(lr){5-8}

Distribution & \multicolumn{2}{c}{Design} & $\delta$ & asymp. MCTP & boot. MCTP & B. asymp. & B. perm. \\
\midrule
$\chi^2_3$&bal&hom&0.5&12.74 & 6.70 & 11.08 & 10.38 \\
$\chi^2_3$&bal&neg&0.5&14.52 & 7.02 & 12.62 & 13.66 \\
$\chi^2_3$&bal&pos&0.5&7.22 & 4.08 & 6.02 & 5.74 \\
$\chi^2_3$&unb&hom&0.5&13.88 & 7.84 & 11.72 & 10.18 \\
$\chi^2_3$&unb&neg&0.5&16.56 & 8.42 & 14.54 & 16.02 \\
$\chi^2_3$&unb&pos&0.5&8.14 & 4.90 & 6.94 & 6.42 \\
$\mathcal{LN}(0,1)$&bal&hom&0.5&29.72 & 16.62 & 26.76 & 28.76 \\
$\mathcal{LN}(0,1)$&bal&neg&0.5&27.06 & 13.92 & 24.64 & 28.94 \\
$\mathcal{LN}(0,1)$&bal&pos&0.5&12.82 & 6.98 & 10.90 & 11.40 \\
$\mathcal{LN}(0,1)$&unb&hom&0.5&33.60 & 20.14 & 30.32 & 32.84 \\
$\mathcal{LN}(0,1)$&unb&neg&0.5&32.56 & 18.98 & 29.02 & 34.26 \\
$\mathcal{LN}(0,1)$&unb&pos&0.5&13.54 & 8.54 & 11.70 & 12.84 \\
$\mathcal{N}(0,1)$&bal&hom&0.5&19.80 & 15.72 & 17.50 & 23.38 \\
$\mathcal{N}(0,1)$&bal&neg&0.5&16.70 & 12.04 & 15.00 & 20.90 \\
$\mathcal{N}(0,1)$&bal&pos&0.5&9.18 & 7.30 & 7.88 & 11.30 \\
$\mathcal{N}(0,1)$&unb&hom&0.5&22.50 & 15.44 & 19.72 & 22.90 \\
$\mathcal{N}(0,1)$&unb&neg&0.5&19.94 & 12.66 & 17.08 & 21.34 \\
$\mathcal{N}(0,1)$&unb&pos&0.5&10.80 & 7.92 & 9.64 & 13.12 \\
$t_2$&bal&hom&0.5&9.94 & 10.34 & 8.88 & 17.56 \\
$t_2$&bal&neg&0.5&7.94 & 7.62 & 6.54 & 15.98 \\
$t_2$&bal&pos&0.5&5.36 & 5.66 & 4.28 & 10.54 \\
$t_2$&unb&hom&0.5&11.06 & 10.18 & 9.30 & 16.92 \\
$t_2$&unb&neg&0.5&9.00 & 8.38 & 7.24 & 14.52 \\
$t_2$&unb&pos&0.5&4.88 & 4.82 & 4.08 & 10.84 \\
$t_3$&bal&hom&0.5&12.76 & 12.04 & 11.54 & 18.32 \\
$t_3$&bal&neg&0.5&10.48 & 9.38 & 9.16 & 17.24 \\
$t_3$&bal&pos&0.5&7.00 & 7.10 & 5.98 & 11.22 \\
$t_3$&unb&hom&0.5&13.44 & 11.06 & 11.66 & 19.02 \\
$t_3$&unb&neg&0.5&12.14 & 9.40 & 10.02 & 16.40 \\
$t_3$&unb&pos&0.5&7.34 & 6.52 & 6.04 & 12.52 \\
\bottomrule
\end{tabular}
\caption{Empirical Power for $\delta = 0.5$ in \% of non-inferiority Tukey-type tests with interval-based estimator}
\end{table}\begin{table}[H]
\centering\footnotesize
\begin{tabular}{clllrrrr}
\toprule

\multicolumn{4}{c}{Setting} & \multicolumn{4}{c}{Method}\\

\cmidrule(lr){1-4}\cmidrule(lr){5-8}

Distribution & \multicolumn{2}{c}{Design} & $\delta$ & asymp. MCTP & boot. MCTP & B. asymp. & B. perm. \\
\midrule
$\chi^2_3$&bal&hom&0.5&9.38 & 1.90 & 8.06 & 7.02 \\
$\chi^2_3$&bal&neg&0.5&11.22 & 2.36 & 9.24 & 9.60 \\
$\chi^2_3$&bal&pos&0.5&7.40 & 1.26 & 5.86 & 5.62 \\
$\chi^2_3$&unb&hom&0.5&10.94 & 1.92 & 9.26 & 7.88 \\
$\chi^2_3$&unb&neg&0.5&14.06 & 3.18 & 12.28 & 12.30 \\
$\chi^2_3$&unb&pos&0.5&7.12 & 1.36 & 6.00 & 4.94 \\
$\mathcal{LN}(0,1)$&bal&hom&0.5&18.64 & 5.56 & 15.96 & 20.68 \\
$\mathcal{LN}(0,1)$&bal&neg&0.5&17.76 & 5.42 & 15.76 & 21.58 \\
$\mathcal{LN}(0,1)$&bal&pos&0.5&7.18 & 1.50 & 5.84 & 8.20 \\
$\mathcal{LN}(0,1)$&unb&hom&0.5&25.14 & 6.94 & 21.96 & 28.56 \\
$\mathcal{LN}(0,1)$&unb&neg&0.5&25.88 & 8.12 & 22.48 & 30.82 \\
$\mathcal{LN}(0,1)$&unb&pos&0.5&8.62 & 1.88 & 7.40 & 9.24 \\
$\mathcal{N}(0,1)$&bal&hom&0.5&18.86 & 5.70 & 16.70 & 15.22 \\
$\mathcal{N}(0,1)$&bal&neg&0.5&15.18 & 4.34 & 13.20 & 12.06 \\
$\mathcal{N}(0,1)$&bal&pos&0.5&10.36 & 2.56 & 8.84 & 8.56 \\
$\mathcal{N}(0,1)$&unb&hom&0.5&22.74 & 5.46 & 20.10 & 17.08 \\
$\mathcal{N}(0,1)$&unb&neg&0.5&20.38 & 4.22 & 17.52 & 15.54 \\
$\mathcal{N}(0,1)$&unb&pos&0.5&12.62 & 2.62 & 11.10 & 8.94 \\
$t_2$&bal&hom&0.5&9.52 & 3.98 & 7.64 & 12.74 \\
$t_2$&bal&neg&0.5&7.50 & 2.84 & 6.14 & 10.34 \\
$t_2$&bal&pos&0.5&5.28 & 2.16 & 4.30 & 7.56 \\
$t_2$&unb&hom&0.5&12.58 & 3.56 & 10.50 & 13.36 \\
$t_2$&unb&neg&0.5&10.36 & 2.74 & 8.60 & 12.04 \\
$t_2$&unb&pos&0.5&6.20 & 2.02 & 5.22 & 7.08 \\
$t_3$&bal&hom&0.5&12.24 & 4.70 & 10.76 & 12.94 \\
$t_3$&bal&neg&0.5&9.96 & 3.64 & 8.30 & 10.80 \\
$t_3$&bal&pos&0.5&7.52 & 2.42 & 6.14 & 8.30 \\
$t_3$&unb&hom&0.5&14.52 & 4.16 & 12.24 & 14.34 \\
$t_3$&unb&neg&0.5&13.10 & 3.62 & 10.90 & 12.96 \\
$t_3$&unb&pos&0.5&8.54 & 2.10 & 7.08 & 8.12 \\
\bottomrule
\end{tabular}
\caption{Empirical Power for $\delta = 0.5$ in \% of two-sided Tukey-type tests with kernel estimator}
\end{table}\begin{table}[H]
\centering\footnotesize
\begin{tabular}{clllrrrr}
\toprule

\multicolumn{4}{c}{Setting} & \multicolumn{4}{c}{Method}\\

\cmidrule(lr){1-4}\cmidrule(lr){5-8}

Distribution & \multicolumn{2}{c}{Design} & $\delta$ & asymp. MCTP & boot. MCTP & B. asymp. & B. perm. \\
\midrule
$\chi^2_3$&bal&hom&0.5&12.04 & 5.02 & 10.34 & 9.78 \\
$\chi^2_3$&bal&neg&0.5&13.68 & 5.44 & 11.94 & 12.90 \\
$\chi^2_3$&bal&pos&0.5&7.26 & 3.18 & 6.04 & 5.86 \\
$\chi^2_3$&unb&hom&0.5&14.92 & 7.14 & 12.74 & 11.34 \\
$\chi^2_3$&unb&neg&0.5&17.78 & 7.84 & 15.64 & 15.96 \\
$\chi^2_3$&unb&pos&0.5&8.22 & 4.28 & 7.06 & 6.20 \\
$\mathcal{LN}(0,1)$&bal&hom&0.5&24.94 & 14.70 & 22.22 & 29.16 \\
$\mathcal{LN}(0,1)$&bal&neg&0.5&23.42 & 12.22 & 21.04 & 30.38 \\
$\mathcal{LN}(0,1)$&bal&pos&0.5&9.90 & 5.36 & 7.92 & 11.58 \\
$\mathcal{LN}(0,1)$&unb&hom&0.5&33.76 & 19.72 & 30.30 & 37.12 \\
$\mathcal{LN}(0,1)$&unb&neg&0.5&33.58 & 19.08 & 30.12 & 39.22 \\
$\mathcal{LN}(0,1)$&unb&pos&0.5&12.08 & 7.74 & 10.26 & 13.12 \\
$\mathcal{N}(0,1)$&bal&hom&0.5&25.00 & 12.50 & 22.40 & 21.76 \\
$\mathcal{N}(0,1)$&bal&neg&0.5&19.58 & 9.82 & 17.40 & 18.08 \\
$\mathcal{N}(0,1)$&bal&pos&0.5&13.54 & 6.12 & 11.80 & 12.84 \\
$\mathcal{N}(0,1)$&unb&hom&0.5&29.08 & 13.34 & 26.74 & 24.24 \\
$\mathcal{N}(0,1)$&unb&neg&0.5&26.66 & 11.74 & 23.28 & 22.02 \\
$\mathcal{N}(0,1)$&unb&pos&0.5&16.22 & 7.20 & 14.60 & 12.44 \\
$t_2$&bal&hom&0.5&13.90 & 9.80 & 12.12 & 18.84 \\
$t_2$&bal&neg&0.5&10.52 & 7.34 & 8.84 & 15.30 \\
$t_2$&bal&pos&0.5&7.28 & 5.22 & 6.20 & 10.98 \\
$t_2$&unb&hom&0.5&17.12 & 10.18 & 15.10 & 18.80 \\
$t_2$&unb&neg&0.5&14.68 & 8.46 & 12.34 & 16.20 \\
$t_2$&unb&pos&0.5&8.42 & 4.90 & 7.42 & 9.98 \\
$t_3$&bal&hom&0.5&16.84 & 11.24 & 15.26 & 18.82 \\
$t_3$&bal&neg&0.5&14.28 & 8.56 & 12.38 & 16.68 \\
$t_3$&bal&pos&0.5&10.10 & 6.26 & 8.92 & 12.38 \\
$t_3$&unb&hom&0.5&20.74 & 10.76 & 18.36 & 20.20 \\
$t_3$&unb&neg&0.5&17.86 & 9.60 & 15.36 & 17.84 \\
$t_3$&unb&pos&0.5&11.54 & 5.78 & 10.20 & 11.50 \\
\bottomrule
\end{tabular}
\caption{Empirical Power for $\delta = 0.5$ in \% of non-inferiority Tukey-type tests with kernel estimator}
\end{table}\begin{table}[H]
\centering\footnotesize
\begin{tabular}{clllrrrr}
\toprule

\multicolumn{4}{c}{Setting} & \multicolumn{4}{c}{Method}\\

\cmidrule(lr){1-4}\cmidrule(lr){5-8}

Distribution & \multicolumn{2}{c}{Design} & $\delta$ & asymp. MCTP & boot. MCTP & B. asymp. & B. perm. \\
\midrule
$\chi^2_3$&bal&hom&0.5&6.26 & 7.28 & 5.76 & 7.24 \\
$\chi^2_3$&bal&neg&0.5&6.10 & 7.14 & 5.34 & 7.16 \\
$\chi^2_3$&bal&pos&0.5&4.34 & 5.70 & 3.88 & 5.20 \\
$\chi^2_3$&unb&hom&0.5&6.72 & 8.34 & 6.30 & 8.42 \\
$\chi^2_3$&unb&neg&0.5&6.84 & 8.28 & 6.14 & 8.84 \\
$\chi^2_3$&unb&pos&0.5&5.38 & 6.54 & 4.94 & 6.46 \\
$\mathcal{LN}(0,1)$&bal&hom&0.5&9.56 & 14.68 & 8.96 & 16.00 \\
$\mathcal{LN}(0,1)$&bal&neg&0.5&7.12 & 10.76 & 6.54 & 13.20 \\
$\mathcal{LN}(0,1)$&bal&pos&0.5&5.28 & 7.80 & 4.72 & 9.32 \\
$\mathcal{LN}(0,1)$&unb&hom&0.5&12.40 & 18.18 & 11.46 & 21.46 \\
$\mathcal{LN}(0,1)$&unb&neg&0.5&9.94 & 14.80 & 9.18 & 18.46 \\
$\mathcal{LN}(0,1)$&unb&pos&0.5&5.60 & 8.88 & 5.28 & 10.86 \\
$\mathcal{N}(0,1)$&bal&hom&0.5&15.96 & 16.86 & 15.14 & 17.26 \\
$\mathcal{N}(0,1)$&bal&neg&0.5&10.80 & 11.72 & 9.94 & 12.30 \\
$\mathcal{N}(0,1)$&bal&pos&0.5&8.12 & 9.26 & 7.68 & 9.24 \\
$\mathcal{N}(0,1)$&unb&hom&0.5&18.30 & 18.26 & 17.14 & 20.02 \\
$\mathcal{N}(0,1)$&unb&neg&0.5&14.10 & 14.56 & 13.12 & 15.86 \\
$\mathcal{N}(0,1)$&unb&pos&0.5&9.46 & 10.18 & 9.00 & 11.32 \\
$t_2$&bal&hom&0.5&8.66 & 13.78 & 7.90 & 13.20 \\
$t_2$&bal&neg&0.5&6.00 & 10.66 & 5.30 & 10.18 \\
$t_2$&bal&pos&0.5&4.82 & 8.70 & 4.40 & 8.48 \\
$t_2$&unb&hom&0.5&9.40 & 14.58 & 8.58 & 15.54 \\
$t_2$&unb&neg&0.5&6.20 & 11.02 & 5.34 & 11.34 \\
$t_2$&unb&pos&0.5&4.42 & 8.22 & 4.12 & 9.30 \\
$t_3$&bal&hom&0.5&11.78 & 15.20 & 11.18 & 14.62 \\
$t_3$&bal&neg&0.5&8.28 & 11.44 & 7.74 & 11.58 \\
$t_3$&bal&pos&0.5&6.36 & 9.12 & 5.80 & 9.42 \\
$t_3$&unb&hom&0.5&11.98 & 15.20 & 11.02 & 16.14 \\
$t_3$&unb&neg&0.5&8.64 & 11.88 & 7.72 & 11.94 \\
$t_3$&unb&pos&0.5&7.32 & 10.14 & 6.76 & 11.00 \\
\bottomrule
\end{tabular}
\caption{Empirical Power for $\delta = 0.5$ in \% of two-sided Grand-mean-type tests with bootstrap estimator}
\end{table}\begin{table}[H]
\centering\footnotesize
\begin{tabular}{clllrrrr}
\toprule

\multicolumn{4}{c}{Setting} & \multicolumn{4}{c}{Method}\\

\cmidrule(lr){1-4}\cmidrule(lr){5-8}

Distribution & \multicolumn{2}{c}{Design} & $\delta$ & asymp. MCTP & boot. MCTP & B. asymp. & B. perm. \\
\midrule
$\chi^2_3$&bal&hom&0.5&4.58 & 6.94 & 4.58 & 7.74 \\
$\chi^2_3$&bal&neg&0.5&4.54 & 7.32 & 4.50 & 8.14 \\
$\chi^2_3$&bal&pos&0.5&3.36 & 5.66 & 3.36 & 6.26 \\
$\chi^2_3$&unb&hom&0.5&5.52 & 8.86 & 5.48 & 8.98 \\
$\chi^2_3$&unb&neg&0.5&5.78 & 8.76 & 5.72 & 8.92 \\
$\chi^2_3$&unb&pos&0.5&3.74 & 5.90 & 3.72 & 6.78 \\
$\mathcal{LN}(0,1)$&bal&hom&0.5&8.32 & 16.08 & 8.30 & 18.58 \\
$\mathcal{LN}(0,1)$&bal&neg&0.5&5.86 & 12.90 & 5.80 & 16.36 \\
$\mathcal{LN}(0,1)$&bal&pos&0.5&3.46 & 6.92 & 3.36 & 9.36 \\
$\mathcal{LN}(0,1)$&unb&hom&0.5&13.02 & 21.72 & 12.94 & 24.64 \\
$\mathcal{LN}(0,1)$&unb&neg&0.5&9.96 & 17.48 & 9.82 & 20.78 \\
$\mathcal{LN}(0,1)$&unb&pos&0.5&4.76 & 9.52 & 4.68 & 11.82 \\
$\mathcal{N}(0,1)$&bal&hom&0.5&16.62 & 18.86 & 16.56 & 19.74 \\
$\mathcal{N}(0,1)$&bal&neg&0.5&11.32 & 13.12 & 11.22 & 13.44 \\
$\mathcal{N}(0,1)$&bal&pos&0.5&8.50 & 10.12 & 8.50 & 11.00 \\
$\mathcal{N}(0,1)$&unb&hom&0.5&20.04 & 21.02 & 19.94 & 21.38 \\
$\mathcal{N}(0,1)$&unb&neg&0.5&16.14 & 17.46 & 15.90 & 17.12 \\
$\mathcal{N}(0,1)$&unb&pos&0.5&10.04 & 10.92 & 10.00 & 11.40 \\
$t_2$&bal&hom&0.5&9.32 & 14.96 & 9.32 & 15.62 \\
$t_2$&bal&neg&0.5&6.72 & 11.38 & 6.66 & 11.86 \\
$t_2$&bal&pos&0.5&5.42 & 9.24 & 5.34 & 10.14 \\
$t_2$&unb&hom&0.5&11.38 & 16.58 & 11.28 & 16.58 \\
$t_2$&unb&neg&0.5&7.50 & 12.70 & 7.32 & 12.06 \\
$t_2$&unb&pos&0.5&5.16 & 9.08 & 5.14 & 10.30 \\
$t_3$&bal&hom&0.5&11.80 & 16.00 & 11.76 & 16.24 \\
$t_3$&bal&neg&0.5&9.48 & 13.12 & 9.44 & 13.44 \\
$t_3$&bal&pos&0.5&6.38 & 9.34 & 6.34 & 10.32 \\
$t_3$&unb&hom&0.5&13.34 & 17.70 & 13.20 & 17.26 \\
$t_3$&unb&neg&0.5&9.92 & 14.70 & 9.76 & 13.26 \\
$t_3$&unb&pos&0.5&7.92 & 11.10 & 7.90 & 11.16 \\
\bottomrule
\end{tabular}
\caption{Empirical Power for $\delta = 0.5$ in \% of non-inferiority Grand-mean-type tests with bootstrap estimator}
\end{table}\begin{table}[H]
\centering\footnotesize
\begin{tabular}{clllrrrr}
\toprule

\multicolumn{4}{c}{Setting} & \multicolumn{4}{c}{Method}\\

\cmidrule(lr){1-4}\cmidrule(lr){5-8}

Distribution & \multicolumn{2}{c}{Design} & $\delta$ & asymp. MCTP & boot. MCTP & B. asymp. & B. perm. \\
\midrule
$\chi^2_3$&bal&hom&0.5&10.30 & 3.24 & 9.68 & 7.22 \\
$\chi^2_3$&bal&neg&0.5&9.68 & 3.14 & 8.98 & 8.90 \\
$\chi^2_3$&bal&pos&0.5&7.96 & 1.98 & 7.24 & 5.66 \\
$\chi^2_3$&unb&hom&0.5&8.20 & 3.14 & 7.36 & 8.16 \\
$\chi^2_3$&unb&neg&0.5&8.94 & 3.36 & 8.02 & 11.76 \\
$\chi^2_3$&unb&pos&0.5&7.02 & 2.60 & 6.54 & 8.00 \\
$\mathcal{LN}(0,1)$&bal&hom&0.5&20.46 & 7.14 & 19.46 & 17.84 \\
$\mathcal{LN}(0,1)$&bal&neg&0.5&16.68 & 5.82 & 15.32 & 16.92 \\
$\mathcal{LN}(0,1)$&bal&pos&0.5&12.50 & 3.72 & 11.26 & 10.02 \\
$\mathcal{LN}(0,1)$&unb&hom&0.5&18.94 & 8.58 & 17.66 & 24.74 \\
$\mathcal{LN}(0,1)$&unb&neg&0.5&15.48 & 7.26 & 14.36 & 22.72 \\
$\mathcal{LN}(0,1)$&unb&pos&0.5&9.46 & 4.10 & 8.80 & 11.76 \\
$\mathcal{N}(0,1)$&bal&hom&0.5&13.92 & 7.22 & 12.88 & 16.84 \\
$\mathcal{N}(0,1)$&bal&neg&0.5&10.24 & 5.48 & 9.76 & 14.16 \\
$\mathcal{N}(0,1)$&bal&pos&0.5&6.42 & 3.26 & 5.92 & 8.52 \\
$\mathcal{N}(0,1)$&unb&hom&0.5&13.58 & 7.32 & 12.68 & 19.94 \\
$\mathcal{N}(0,1)$&unb&neg&0.5&9.70 & 4.78 & 8.58 & 16.00 \\
$\mathcal{N}(0,1)$&unb&pos&0.5&7.14 & 3.68 & 6.64 & 12.76 \\
$t_2$&bal&hom&0.5&5.28 & 4.32 & 4.88 & 12.32 \\
$t_2$&bal&neg&0.5&4.16 & 3.30 & 3.74 & 11.02 \\
$t_2$&bal&pos&0.5&3.06 & 2.50 & 2.76 & 8.18 \\
$t_2$&unb&hom&0.5&4.06 & 3.26 & 3.70 & 13.16 \\
$t_2$&unb&neg&0.5&1.90 & 1.42 & 1.64 & 8.74 \\
$t_2$&unb&pos&0.5&1.78 & 1.54 & 1.70 & 9.84 \\
$t_3$&bal&hom&0.5&8.50 & 6.58 & 7.98 & 13.56 \\
$t_3$&bal&neg&0.5&6.06 & 4.14 & 5.48 & 11.34 \\
$t_3$&bal&pos&0.5&4.50 & 3.08 & 4.04 & 8.78 \\
$t_3$&unb&hom&0.5&5.52 & 3.92 & 4.94 & 14.86 \\
$t_3$&unb&neg&0.5&3.46 & 2.34 & 3.06 & 10.78 \\
$t_3$&unb&pos&0.5&3.84 & 2.64 & 3.50 & 11.74 \\
\bottomrule
\end{tabular}
\caption{Empirical Power for $\delta = 0.5$ in \% of two-sided Grand-mean-type tests with interval-based estimator}
\end{table}\begin{table}[H]
\centering\footnotesize
\begin{tabular}{clllrrrr}
\toprule

\multicolumn{4}{c}{Setting} & \multicolumn{4}{c}{Method}\\

\cmidrule(lr){1-4}\cmidrule(lr){5-8}

Distribution & \multicolumn{2}{c}{Design} & $\delta$ & asymp. MCTP & boot. MCTP & B. asymp. & B. perm. \\
\midrule
$\chi^2_3$&bal&hom&0.5&5.84 & 2.82 & 5.84 & 7.96 \\
$\chi^2_3$&bal&neg&0.5&6.44 & 3.00 & 6.42 & 9.52 \\
$\chi^2_3$&bal&pos&0.5&4.26 & 1.74 & 4.18 & 6.06 \\
$\chi^2_3$&unb&hom&0.5&5.86 & 3.44 & 5.86 & 9.30 \\
$\chi^2_3$&unb&neg&0.5&6.66 & 3.64 & 6.58 & 10.58 \\
$\chi^2_3$&unb&pos&0.5&4.56 & 2.00 & 4.50 & 7.12 \\
$\mathcal{LN}(0,1)$&bal&hom&0.5&15.50 & 8.56 & 15.44 & 21.92 \\
$\mathcal{LN}(0,1)$&bal&neg&0.5&11.96 & 6.90 & 11.86 & 20.66 \\
$\mathcal{LN}(0,1)$&bal&pos&0.5&6.50 & 3.00 & 6.40 & 9.56 \\
$\mathcal{LN}(0,1)$&unb&hom&0.5&18.42 & 11.96 & 18.28 & 28.14 \\
$\mathcal{LN}(0,1)$&unb&neg&0.5&14.08 & 9.74 & 13.86 & 24.44 \\
$\mathcal{LN}(0,1)$&unb&pos&0.5&6.76 & 3.92 & 6.72 & 11.74 \\
$\mathcal{N}(0,1)$&bal&hom&0.5&13.84 & 10.98 & 13.76 & 20.80 \\
$\mathcal{N}(0,1)$&bal&neg&0.5&9.76 & 7.16 & 9.72 & 16.38 \\
$\mathcal{N}(0,1)$&bal&pos&0.5&5.74 & 4.24 & 5.74 & 9.82 \\
$\mathcal{N}(0,1)$&unb&hom&0.5&16.06 & 12.30 & 16.02 & 21.40 \\
$\mathcal{N}(0,1)$&unb&neg&0.5&11.62 & 8.98 & 11.58 & 17.82 \\
$\mathcal{N}(0,1)$&unb&pos&0.5&7.50 & 5.66 & 7.42 & 12.14 \\
$t_2$&bal&hom&0.5&6.70 & 7.38 & 6.62 & 14.80 \\
$t_2$&bal&neg&0.5&4.18 & 4.82 & 4.16 & 12.72 \\
$t_2$&bal&pos&0.5&3.40 & 3.60 & 3.40 & 9.18 \\
$t_2$&unb&hom&0.5&5.18 & 5.46 & 5.06 & 15.00 \\
$t_2$&unb&neg&0.5&2.70 & 2.70 & 2.64 & 10.16 \\
$t_2$&unb&pos&0.5&2.38 & 2.82 & 2.30 & 11.32 \\
$t_3$&bal&hom&0.5&8.70 & 8.34 & 8.66 & 15.94 \\
$t_3$&bal&neg&0.5&6.12 & 5.74 & 6.04 & 14.02 \\
$t_3$&bal&pos&0.5&4.28 & 4.02 & 4.26 & 10.02 \\
$t_3$&unb&hom&0.5&7.44 & 7.22 & 7.38 & 16.30 \\
$t_3$&unb&neg&0.5&4.42 & 4.22 & 4.32 & 12.00 \\
$t_3$&unb&pos&0.5&4.36 & 4.36 & 4.32 & 12.74 \\
\bottomrule
\end{tabular}
\caption{Empirical Power for $\delta = 0.5$ in \% of non-inferiority Grand-mean-type tests with interval-based estimator}
\end{table}\begin{table}[H]
\centering\footnotesize
\begin{tabular}{clllrrrr}
\toprule

\multicolumn{4}{c}{Setting} & \multicolumn{4}{c}{Method}\\

\cmidrule(lr){1-4}\cmidrule(lr){5-8}

Distribution & \multicolumn{2}{c}{Design} & $\delta$ & asymp. MCTP & boot. MCTP & B. asymp. & B. perm. \\
\midrule
$\chi^2_3$&bal&hom&0.5&9.78 & 2.52 & 9.36 & 7.26 \\
$\chi^2_3$&bal&neg&0.5&9.76 & 2.26 & 8.92 & 8.38 \\
$\chi^2_3$&bal&pos&0.5&7.70 & 1.58 & 7.18 & 5.94 \\
$\chi^2_3$&unb&hom&0.5&9.62 & 2.62 & 8.88 & 7.82 \\
$\chi^2_3$&unb&neg&0.5&10.68 & 2.86 & 9.68 & 10.50 \\
$\chi^2_3$&unb&pos&0.5&7.96 & 2.32 & 7.42 & 6.52 \\
$\mathcal{LN}(0,1)$&bal&hom&0.5&18.00 & 5.80 & 16.90 & 18.44 \\
$\mathcal{LN}(0,1)$&bal&neg&0.5&13.30 & 4.28 & 12.32 & 16.62 \\
$\mathcal{LN}(0,1)$&bal&pos&0.5&9.84 & 2.62 & 9.06 & 9.88 \\
$\mathcal{LN}(0,1)$&unb&hom&0.5&19.34 & 6.60 & 18.04 & 25.10 \\
$\mathcal{LN}(0,1)$&unb&neg&0.5&16.00 & 6.38 & 14.64 & 23.74 \\
$\mathcal{LN}(0,1)$&unb&pos&0.5&9.10 & 3.08 & 8.54 & 10.98 \\
$\mathcal{N}(0,1)$&bal&hom&0.5&19.04 & 6.38 & 17.98 & 17.10 \\
$\mathcal{N}(0,1)$&bal&neg&0.5&13.54 & 4.50 & 12.56 & 12.50 \\
$\mathcal{N}(0,1)$&bal&pos&0.5&10.18 & 2.70 & 9.52 & 9.42 \\
$\mathcal{N}(0,1)$&unb&hom&0.5&21.56 & 5.74 & 20.24 & 19.28 \\
$\mathcal{N}(0,1)$&unb&neg&0.5&18.14 & 4.48 & 16.88 & 16.16 \\
$\mathcal{N}(0,1)$&unb&pos&0.5&11.34 & 3.08 & 10.88 & 10.60 \\
$t_2$&bal&hom&0.5&8.78 & 4.48 & 8.10 & 13.10 \\
$t_2$&bal&neg&0.5&6.38 & 3.18 & 5.94 & 9.90 \\
$t_2$&bal&pos&0.5&5.02 & 2.54 & 4.64 & 8.68 \\
$t_2$&unb&hom&0.5&11.48 & 4.58 & 10.60 & 15.48 \\
$t_2$&unb&neg&0.5&8.50 & 3.12 & 7.82 & 12.22 \\
$t_2$&unb&pos&0.5&5.32 & 2.22 & 5.00 & 8.64 \\
$t_3$&bal&hom&0.5&11.96 & 5.76 & 11.46 & 13.78 \\
$t_3$&bal&neg&0.5&8.70 & 3.90 & 8.18 & 11.40 \\
$t_3$&bal&pos&0.5&7.02 & 2.68 & 6.56 & 9.38 \\
$t_3$&unb&hom&0.5&13.52 & 5.00 & 12.70 & 16.72 \\
$t_3$&unb&neg&0.5&10.96 & 3.46 & 9.90 & 12.88 \\
$t_3$&unb&pos&0.5&8.14 & 2.66 & 7.62 & 10.30 \\
\bottomrule
\end{tabular}
\caption{Empirical Power for $\delta = 0.5$ in \% of two-sided Grand-mean-type tests with kernel estimator}
\end{table}\begin{table}[H]
\centering\footnotesize
\begin{tabular}{clllrrrr}
\toprule

\multicolumn{4}{c}{Setting} & \multicolumn{4}{c}{Method}\\

\cmidrule(lr){1-4}\cmidrule(lr){5-8}

Distribution & \multicolumn{2}{c}{Design} & $\delta$ & asymp. MCTP & boot. MCTP & B. asymp. & B. perm. \\
\midrule
$\chi^2_3$&bal&hom&0.5&6.10 & 2.54 & 6.08 & 7.80 \\
$\chi^2_3$&bal&neg&0.5&6.68 & 3.10 & 6.70 & 9.42 \\
$\chi^2_3$&bal&pos&0.5&4.62 & 1.70 & 4.54 & 6.20 \\
$\chi^2_3$&unb&hom&0.5&7.08 & 2.70 & 7.04 & 9.44 \\
$\chi^2_3$&unb&neg&0.5&8.10 & 3.08 & 8.02 & 10.86 \\
$\chi^2_3$&unb&pos&0.5&5.24 & 1.82 & 5.18 & 6.52 \\
$\mathcal{LN}(0,1)$&bal&hom&0.5&12.66 & 7.52 & 12.60 & 22.22 \\
$\mathcal{LN}(0,1)$&bal&neg&0.5&9.72 & 6.50 & 9.66 & 20.84 \\
$\mathcal{LN}(0,1)$&bal&pos&0.5&4.72 & 2.66 & 4.66 & 9.68 \\
$\mathcal{LN}(0,1)$&unb&hom&0.5&17.80 & 9.30 & 17.62 & 30.66 \\
$\mathcal{LN}(0,1)$&unb&neg&0.5&14.92 & 8.24 & 14.84 & 27.52 \\
$\mathcal{LN}(0,1)$&unb&pos&0.5&5.64 & 2.90 & 5.56 & 11.80 \\
$\mathcal{N}(0,1)$&bal&hom&0.5&19.16 & 9.02 & 19.10 & 19.54 \\
$\mathcal{N}(0,1)$&bal&neg&0.5&13.34 & 5.58 & 13.32 & 14.04 \\
$\mathcal{N}(0,1)$&bal&pos&0.5&10.12 & 4.36 & 10.08 & 10.80 \\
$\mathcal{N}(0,1)$&unb&hom&0.5&22.66 & 7.80 & 22.64 & 21.60 \\
$\mathcal{N}(0,1)$&unb&neg&0.5&20.20 & 5.70 & 20.00 & 18.32 \\
$\mathcal{N}(0,1)$&unb&pos&0.5&11.28 & 3.70 & 11.26 & 11.30 \\
$t_2$&bal&hom&0.5&10.06 & 6.64 & 10.04 & 15.78 \\
$t_2$&bal&neg&0.5&7.12 & 4.40 & 7.10 & 12.20 \\
$t_2$&bal&pos&0.5&5.74 & 3.72 & 5.68 & 10.06 \\
$t_2$&unb&hom&0.5&13.08 & 6.06 & 13.02 & 16.52 \\
$t_2$&unb&neg&0.5&9.98 & 3.82 & 9.80 & 13.46 \\
$t_2$&unb&pos&0.5&5.72 & 2.78 & 5.70 & 9.36 \\
$t_3$&bal&hom&0.5&12.12 & 7.68 & 12.12 & 15.86 \\
$t_3$&bal&neg&0.5&9.88 & 5.36 & 9.88 & 13.50 \\
$t_3$&bal&pos&0.5&6.96 & 4.06 & 6.96 & 10.62 \\
$t_3$&unb&hom&0.5&15.40 & 6.22 & 15.36 & 17.90 \\
$t_3$&unb&neg&0.5&13.00 & 4.48 & 12.80 & 14.16 \\
$t_3$&unb&pos&0.5&8.72 & 3.34 & 8.64 & 10.70 \\
\bottomrule
\end{tabular}
\caption{Empirical Power for $\delta = 0.5$ in \% of non-inferiority Grand-mean-type tests with kernel estimator}
\end{table}\begin{table}[H]
\centering\footnotesize
\begin{tabular}{clllrrrr}
\toprule

\multicolumn{4}{c}{Setting} & \multicolumn{4}{c}{Method}\\

\cmidrule(lr){1-4}\cmidrule(lr){5-8}

Distribution & \multicolumn{2}{c}{Design} & $\delta$ & asymp. MCTP & boot. MCTP & B. asymp. & B. perm. \\
\midrule
$\chi^2_3$&bal&hom&1.0&9.90 & 11.32 & 9.14 & 12.22 \\
$\chi^2_3$&bal&neg&1.0&8.36 & 10.10 & 7.62 & 9.92 \\
$\chi^2_3$&bal&pos&1.0&5.94 & 6.66 & 5.58 & 8.18 \\
$\chi^2_3$&unb&hom&1.0&12.08 & 13.46 & 11.08 & 13.06 \\
$\chi^2_3$&unb&neg&1.0&10.38 & 12.12 & 9.16 & 11.58 \\
$\chi^2_3$&unb&pos&1.0&5.98 & 6.48 & 5.66 & 7.48 \\
$\mathcal{LN}(0,1)$&bal&hom&1.0&30.82 & 36.26 & 29.38 & 38.16 \\
$\mathcal{LN}(0,1)$&bal&neg&1.0&17.84 & 23.16 & 16.44 & 23.22 \\
$\mathcal{LN}(0,1)$&bal&pos&1.0&12.96 & 15.22 & 12.52 & 21.64 \\
$\mathcal{LN}(0,1)$&unb&hom&1.0&37.30 & 42.82 & 35.30 & 40.60 \\
$\mathcal{LN}(0,1)$&unb&neg&1.0&22.98 & 27.50 & 20.76 & 24.98 \\
$\mathcal{LN}(0,1)$&unb&pos&1.0&15.88 & 19.80 & 15.12 & 23.80 \\
$\mathcal{N}(0,1)$&bal&hom&1.0&37.26 & 37.20 & 35.42 & 39.30 \\
$\mathcal{N}(0,1)$&bal&neg&1.0&20.64 & 22.56 & 18.46 & 20.76 \\
$\mathcal{N}(0,1)$&bal&pos&1.0&20.00 & 19.94 & 19.32 & 23.18 \\
$\mathcal{N}(0,1)$&unb&hom&1.0&38.94 & 38.58 & 35.90 & 38.38 \\
$\mathcal{N}(0,1)$&unb&neg&1.0&22.30 & 23.76 & 19.80 & 20.74 \\
$\mathcal{N}(0,1)$&unb&pos&1.0&21.60 & 21.70 & 20.32 & 22.72 \\
$t_2$&bal&hom&1.0&23.54 & 30.64 & 22.26 & 29.96 \\
$t_2$&bal&neg&1.0&11.98 & 17.80 & 10.48 & 15.68 \\
$t_2$&bal&pos&1.0&11.62 & 16.20 & 11.14 & 19.08 \\
$t_2$&unb&hom&1.0&20.32 & 26.80 & 18.38 & 25.72 \\
$t_2$&unb&neg&1.0&11.30 & 15.66 & 9.86 & 14.10 \\
$t_2$&unb&pos&1.0&11.50 & 16.24 & 10.64 & 17.94 \\
$t_3$&bal&hom&1.0&29.68 & 34.14 & 27.76 & 34.18 \\
$t_3$&bal&neg&1.0&15.04 & 19.42 & 13.32 & 17.48 \\
$t_3$&bal&pos&1.0&14.74 & 17.64 & 14.04 & 20.42 \\
$t_3$&unb&hom&1.0&27.70 & 31.72 & 25.42 & 30.72 \\
$t_3$&unb&neg&1.0&14.90 & 18.52 & 13.04 & 16.12 \\
$t_3$&unb&pos&1.0&14.68 & 18.38 & 13.74 & 19.78 \\
\bottomrule
\end{tabular}
\caption{Empirical Power for $\delta = 1.0$ in \% of two-sided Dunnett-type tests with bootstrap estimator}
\end{table}\begin{table}[H]
\centering\footnotesize
%
\caption{Empirical Power for $\delta = 1.0$ in \% of non-inferiority Dunnett-type tests with bootstrap estimator}
\end{table}\begin{table}[H]
\centering\footnotesize
%
\caption{Empirical Power for $\delta = 1.0$ in \% of two-sided Dunnett-type tests with interval-based estimator}
\end{table}\begin{table}[H]
\centering\footnotesize
%
\caption{Empirical Power for $\delta = 1.0$ in \% of non-inferiority Dunnett-type tests with interval-based estimator}
\end{table}\begin{table}[H]
\centering\footnotesize
%
\caption{Empirical Power for $\delta = 1.0$ in \% of two-sided Dunnett-type tests with kernel estimator}
\end{table}\begin{table}[H]
\centering\footnotesize
%
\caption{Empirical Power for $\delta = 1.0$ in \% of non-inferiority Dunnett-type tests with kernel estimator}
\end{table}\begin{table}[H]
\centering\footnotesize
%
\caption{Empirical Power for $\delta = 1.0$ in \% of two-sided Tukey-type tests with bootstrap estimator}
\end{table}\begin{table}[H]
\centering\footnotesize
%
\caption{Empirical Power for $\delta = 1.0$ in \% of non-inferiority Tukey-type tests with bootstrap estimator}
\end{table}\begin{table}[H]
\centering\footnotesize
%
\caption{Empirical Power for $\delta = 1.0$ in \% of two-sided Tukey-type tests with interval-based estimator}
\end{table}\begin{table}[H]
\centering\footnotesize
%
\caption{Empirical Power for $\delta = 1.0$ in \% of non-inferiority Tukey-type tests with interval-based estimator}
\end{table}\begin{table}[H]
\centering\footnotesize
%
\caption{Empirical Power for $\delta = 1.0$ in \% of two-sided Tukey-type tests with kernel estimator}
\end{table}\begin{table}[H]
\centering\footnotesize
%
\caption{Empirical Power for $\delta = 1.0$ in \% of non-inferiority Tukey-type tests with kernel estimator}
\end{table}\begin{table}[H]
\centering\footnotesize
%
\caption{Empirical Power for $\delta = 1.0$ in \% of two-sided Grand-mean-type tests with bootstrap estimator}
\end{table}\begin{table}[H]
\centering\footnotesize
%
\caption{Empirical Power for $\delta = 1.0$ in \% of non-inferiority Grand-mean-type tests with bootstrap estimator}
\end{table}\begin{table}[H]
\centering\footnotesize
%
\caption{Empirical Power for $\delta = 1.0$ in \% of two-sided Grand-mean-type tests with interval-based estimator}
\end{table}\begin{table}[H]
\centering\footnotesize
%
\caption{Empirical Power for $\delta = 1.0$ in \% of non-inferiority Grand-mean-type tests with interval-based estimator}
\end{table}\begin{table}[H]
\centering\footnotesize
%
\caption{Empirical Power for $\delta = 1.0$ in \% of two-sided Grand-mean-type tests with kernel estimator}
\end{table}\begin{table}[H]
\centering\footnotesize
%
\caption{Empirical Power for $\delta = 1.0$ in \% of non-inferiority Grand-mean-type tests with kernel estimator}
\end{table}\begin{table}[H]
\centering\footnotesize
%
\caption{Empirical Power for $\delta = 1.5$ in \% of two-sided Dunnett-type tests with bootstrap estimator}
\end{table}\begin{table}[H]
\centering\footnotesize
%
\caption{Empirical Power for $\delta = 1.5$ in \% of non-inferiority Dunnett-type tests with bootstrap estimator}
\end{table}\begin{table}[H]
\centering\footnotesize
%
\caption{Empirical Power for $\delta = 1.5$ in \% of two-sided Dunnett-type tests with interval-based estimator}
\end{table}\begin{table}[H]
\centering\footnotesize
%
\caption{Empirical Power for $\delta = 1.5$ in \% of non-inferiority Dunnett-type tests with interval-based estimator}
\end{table}\begin{table}[H]
\centering\footnotesize
%
\caption{Empirical Power for $\delta = 1.5$ in \% of two-sided Dunnett-type tests with kernel estimator}
\end{table}\begin{table}[H]
\centering\footnotesize
%
\caption{Empirical Power for $\delta = 1.5$ in \% of non-inferiority Dunnett-type tests with kernel estimator}
\end{table}\begin{table}[H]
\centering\footnotesize
%
\caption{Empirical Power for $\delta = 1.5$ in \% of two-sided Tukey-type tests with bootstrap estimator}
\end{table}\begin{table}[H]
\centering\footnotesize
%
\caption{Empirical Power for $\delta = 1.5$ in \% of non-inferiority Tukey-type tests with bootstrap estimator}
\end{table}\begin{table}[H]
\centering\footnotesize
%
\caption{Empirical Power for $\delta = 1.5$ in \% of two-sided Tukey-type tests with interval-based estimator}
\end{table}\begin{table}[H]
\centering\footnotesize
%
\caption{Empirical Power for $\delta = 1.5$ in \% of non-inferiority Tukey-type tests with interval-based estimator}
\end{table}\begin{table}[H]
\centering\footnotesize
%
\caption{Empirical Power for $\delta = 1.5$ in \% of two-sided Tukey-type tests with kernel estimator}
\end{table}\begin{table}[H]
\centering\footnotesize
%
\caption{Empirical Power for $\delta = 1.5$ in \% of non-inferiority Tukey-type tests with kernel estimator}
\end{table}\begin{table}[H]
\centering\footnotesize
%
\caption{Empirical Power for $\delta = 1.5$ in \% of two-sided Grand-mean-type tests with bootstrap estimator}
\end{table}\begin{table}[H]
\centering\footnotesize
%
\caption{Empirical Power for $\delta = 1.5$ in \% of non-inferiority Grand-mean-type tests with bootstrap estimator}
\end{table}\begin{table}[H]
\centering\footnotesize
%
\caption{Empirical Power for $\delta = 1.5$ in \% of two-sided Grand-mean-type tests with interval-based estimator}
\end{table}\begin{table}[H]
\centering\footnotesize
%
\caption{Empirical Power for $\delta = 1.5$ in \% of non-inferiority Grand-mean-type tests with interval-based estimator}
\end{table}\begin{table}[H]
\centering\footnotesize
%
\caption{Empirical Power for $\delta = 1.5$ in \% of two-sided Grand-mean-type tests with kernel estimator}
\end{table}\begin{table}[H]
\centering\footnotesize
%
\caption{Empirical Power for $\delta = 1.5$ in \% of non-inferiority Grand-mean-type tests with kernel estimator}\label{tab:2}
\end{table}

\section{Plots for Empirical Local Power}
\begin{figure}[H]\centering
\includegraphics[scale=0.4]{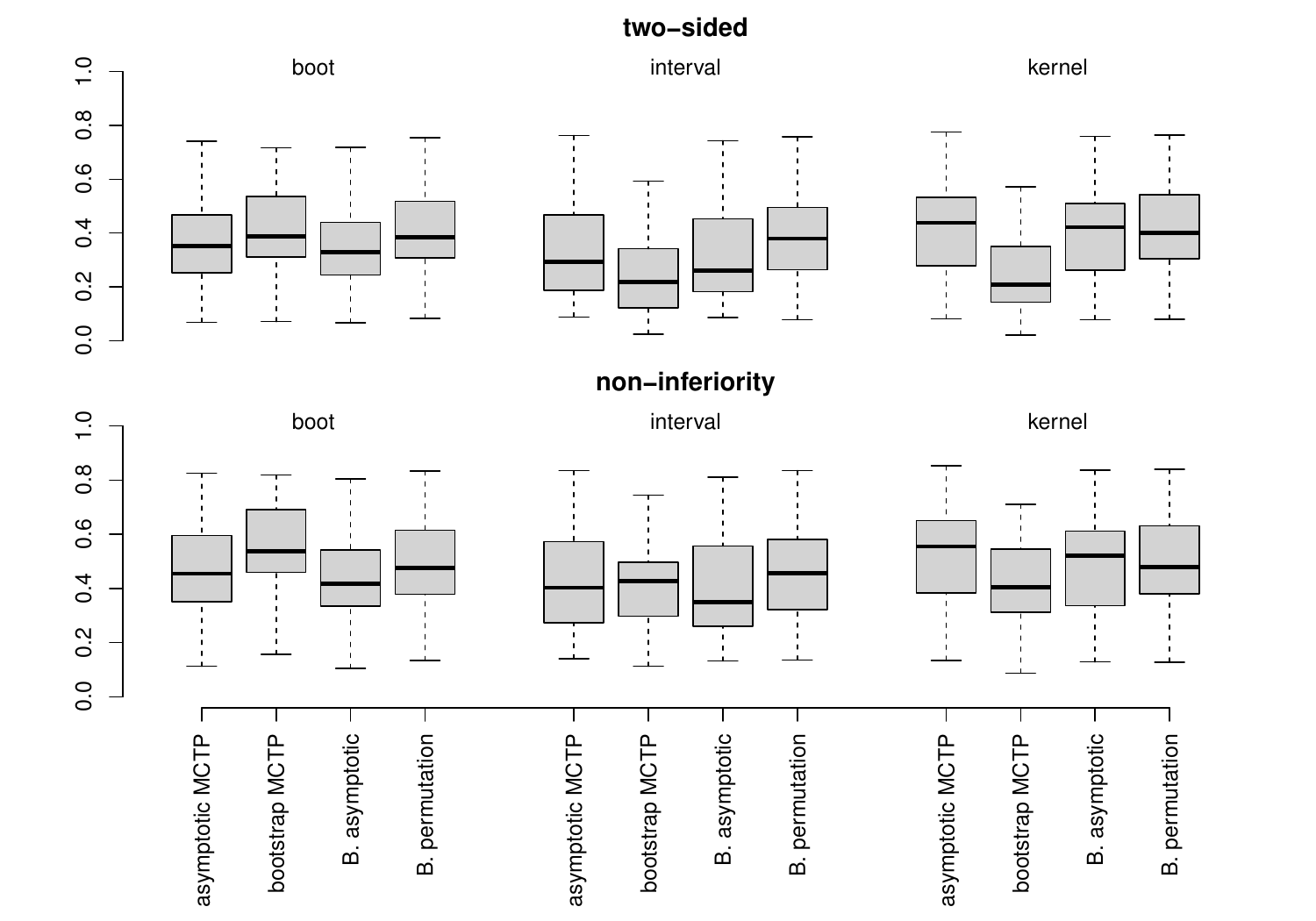} 
\caption{Empirical  \new{local} power of $\mathcal{H}_{0,3}: m_4 - m_1 = 0$ (top) and $\mathcal{H}^I_{0,3}: m_4 - m_1 \leq 0$ (bottom) with $\delta = 1.5$ for Dunnett-type contrasts with different variance estimators.}
\label{fig:localDunnettStart}
\end{figure}

\begin{figure}[H]\centering
\includegraphics[scale=0.4]{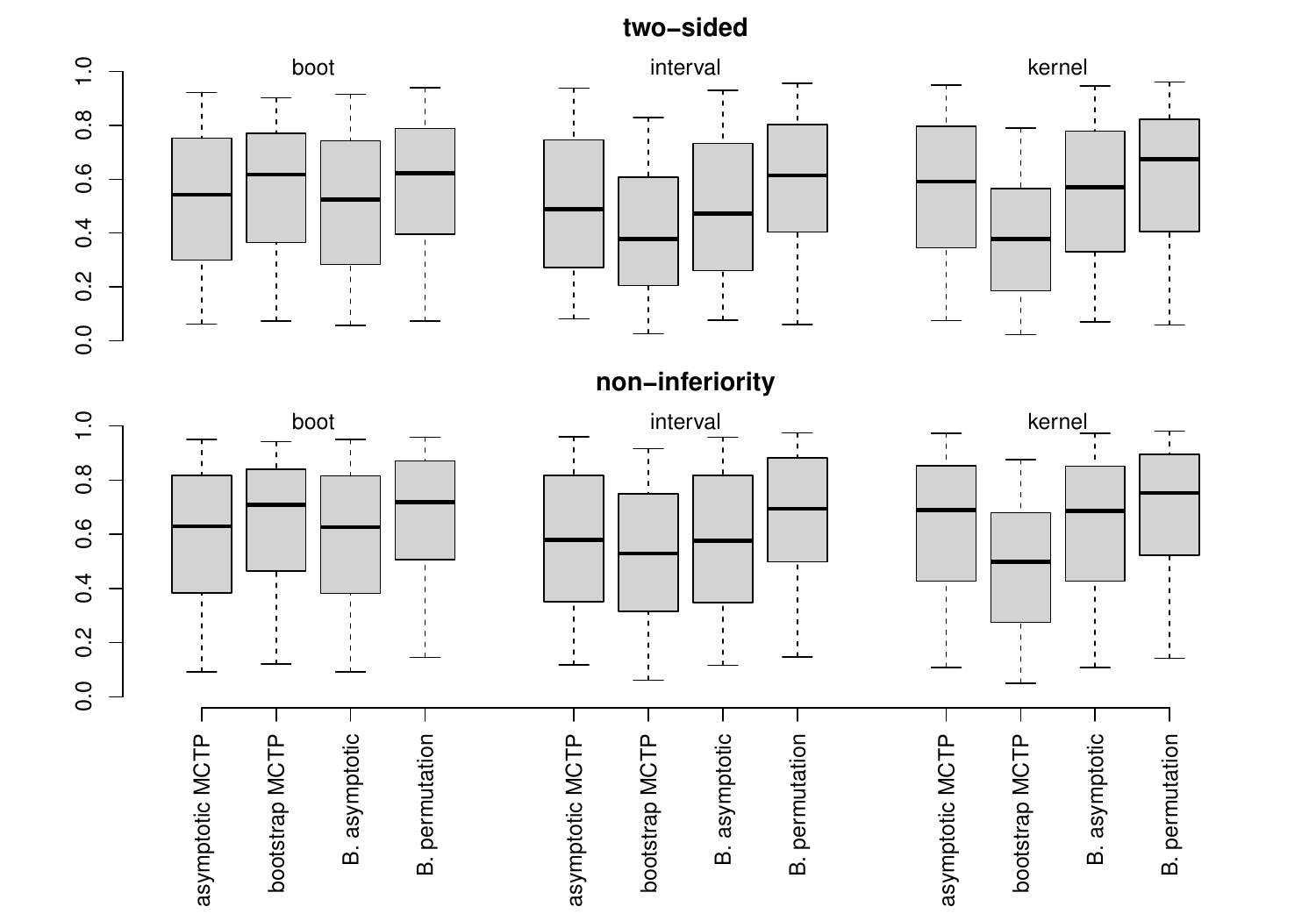} 
\caption{Empirical local power of $\mathcal{H}_{0,4}: m_4 - \bar{m} = 0$ (top) and  of $\mathcal{H}_{0,4}^I: m_4 - \bar{m} \leq 0$ (bottom) with $\delta = 1.5$ for Grand-mean-type contrasts with different variance estimators.}
\label{fig:localGMEnd}
\end{figure}

In this section, boxplots to analyze the local power of the different methods are provided in Figures~\ref{fig:localDunnettStart}--\ref{fig:localGMEnd}.
In detail, the rejection rates for the false local hypotheses are plotted separately.
Here, we only show the results with median difference $\delta = 1.5$ for the sake of clarity.
Note that for the Grand-mean-type contrast matrix, the two-sided hypotheses $\mathcal{H}_{0,1}: m_1 - \bar{m} = 0, \mathcal{H}_{0,2}: m_2 - \bar{m} = 0$ and $\mathcal{H}_{0,3}: m_2 - \bar{m} = 0$ are false under the considered simulation scenario with $m_1 = m_2 = m_3 = 0$ and $m_4 = 1.5$ but the non-inferiority hypotheses $\mathcal{H}_{0,1}^I: m_1 - \bar{m} \leq 0, \mathcal{H}_{0,2}^I: m_2 - \bar{m} = 0$ and $\mathcal{H}_{0,3}^I: m_3 - \bar{m} = 0$ are true, which is why only the rejection rates of the two-sided hypotheses are shown in the corresponding figures.
Moreover, the empirical local power is relatively small for these hypotheses, which can be explained by the smaller effect parameter of $m_\ell - \bar{m} = -\delta/4$ for $\ell\in\{1,2,3\}$.

\begin{figure}[H]\centering
\includegraphics[scale=0.4]{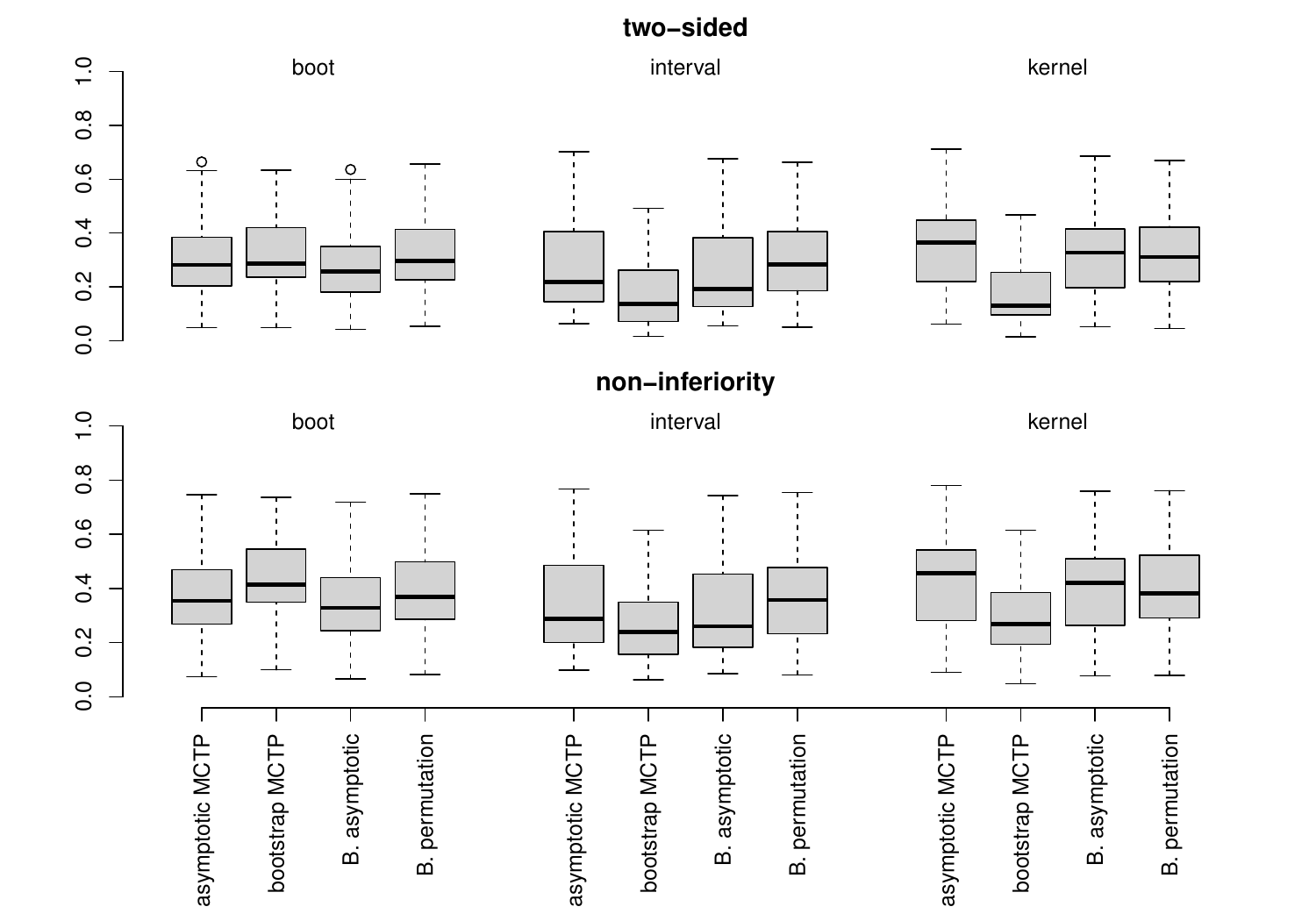} 
\caption{Empirical local power of $\mathcal{H}_{0,3}: m_4 - m_1 = 0$ (top) and $\mathcal{H}_{0,3}^I: m_4 - m_1 \leq 0$ (bottom) with $\delta = 1.5$ for Tukey-type contrasts with different variance estimators.}
\end{figure}

\begin{figure}[H]\centering
\includegraphics[scale=0.4]{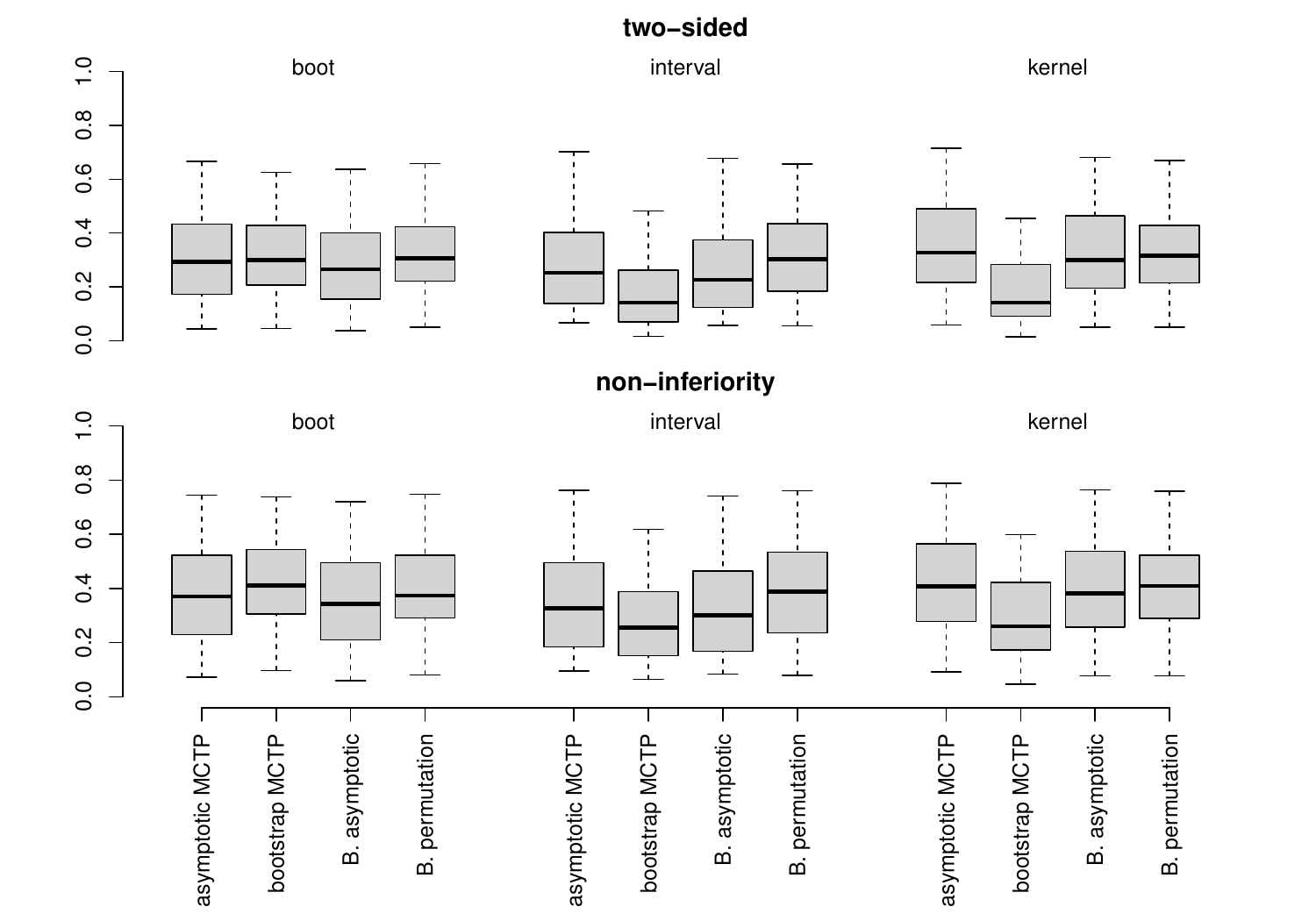} 
\caption{Empirical local power of $\mathcal{H}_{0,5}: m_4 - m_2 = 0$ (top) and $\mathcal{H}_{0,5}^I: m_4 - m_2 \leq 0$ (bottom) with $\delta = 1.5$ for Tukey-type contrasts with different variance estimators.}
\end{figure}

\begin{figure}[H]\centering
\includegraphics[scale=0.4]{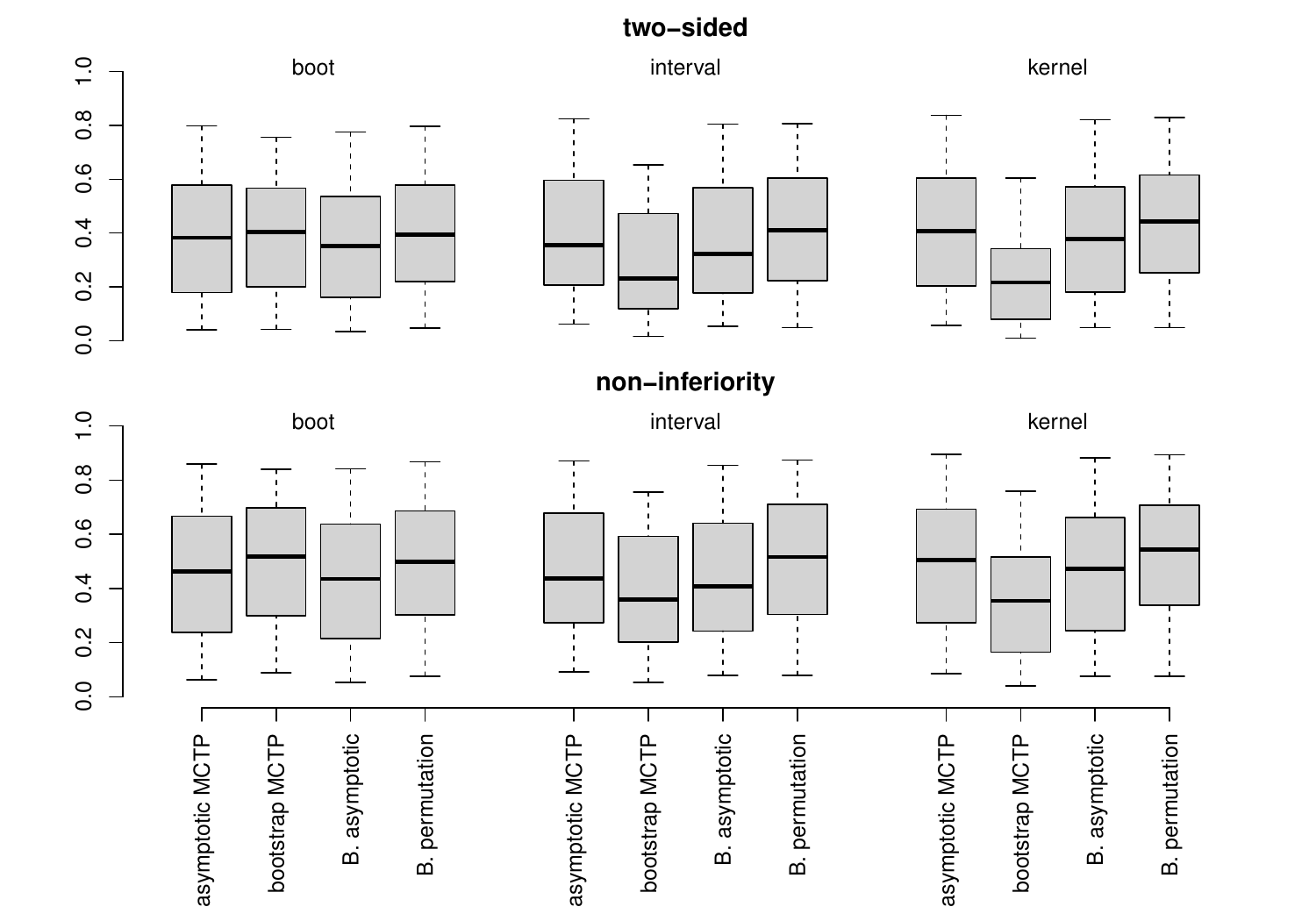} 
\caption{Empirical local power of $\mathcal{H}_{0,6}: m_4 - m_3 = 0$ (top) and $\mathcal{H}_{0,6}^I: m_4 - m_3 \leq 0$ (bottom) with $\delta = 1.5$ for Tukey-type contrasts with different variance estimators.}
\end{figure}

\begin{figure}[H]\centering
\includegraphics[scale=0.6]{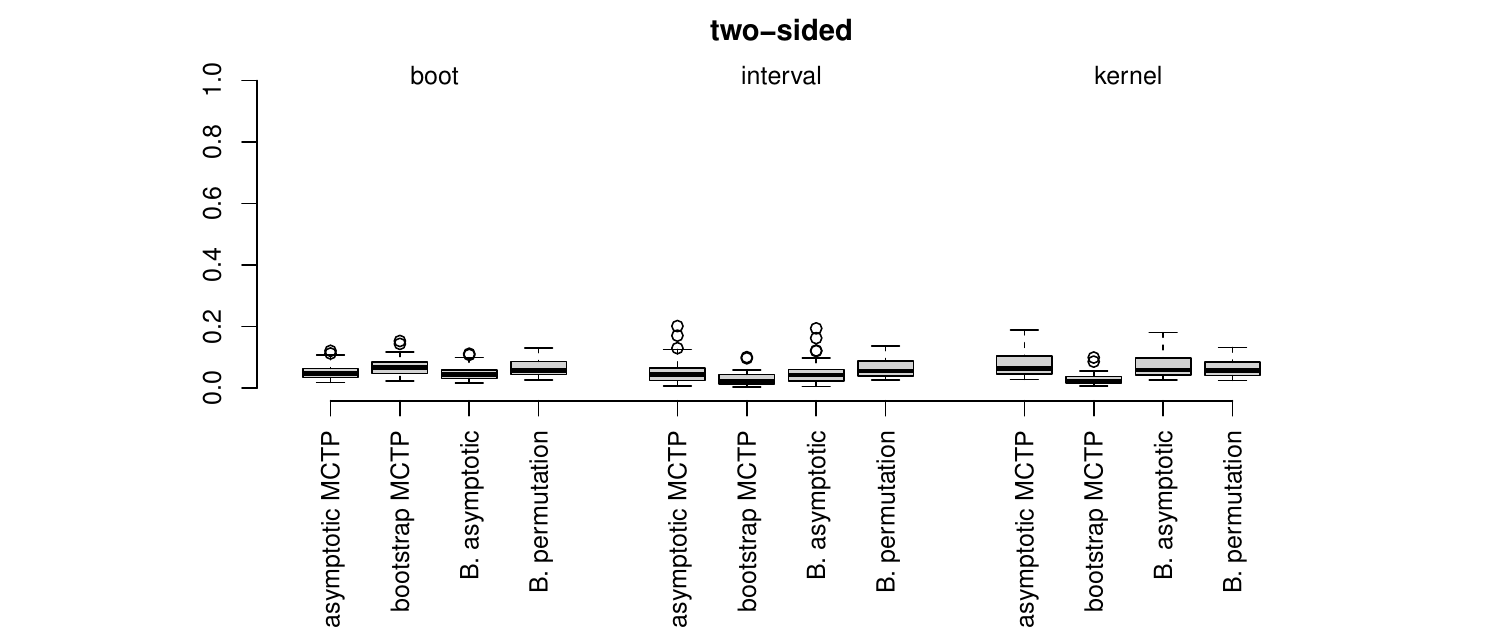} 
\caption{Empirical local power of $\mathcal{H}_{0,1}: m_1 - \bar{m} = 0$ with $\delta = 1.5$ for Grand-mean-type contrasts with different variance estimators.}
\end{figure}

\begin{figure}[H]\centering
\includegraphics[scale=0.6]{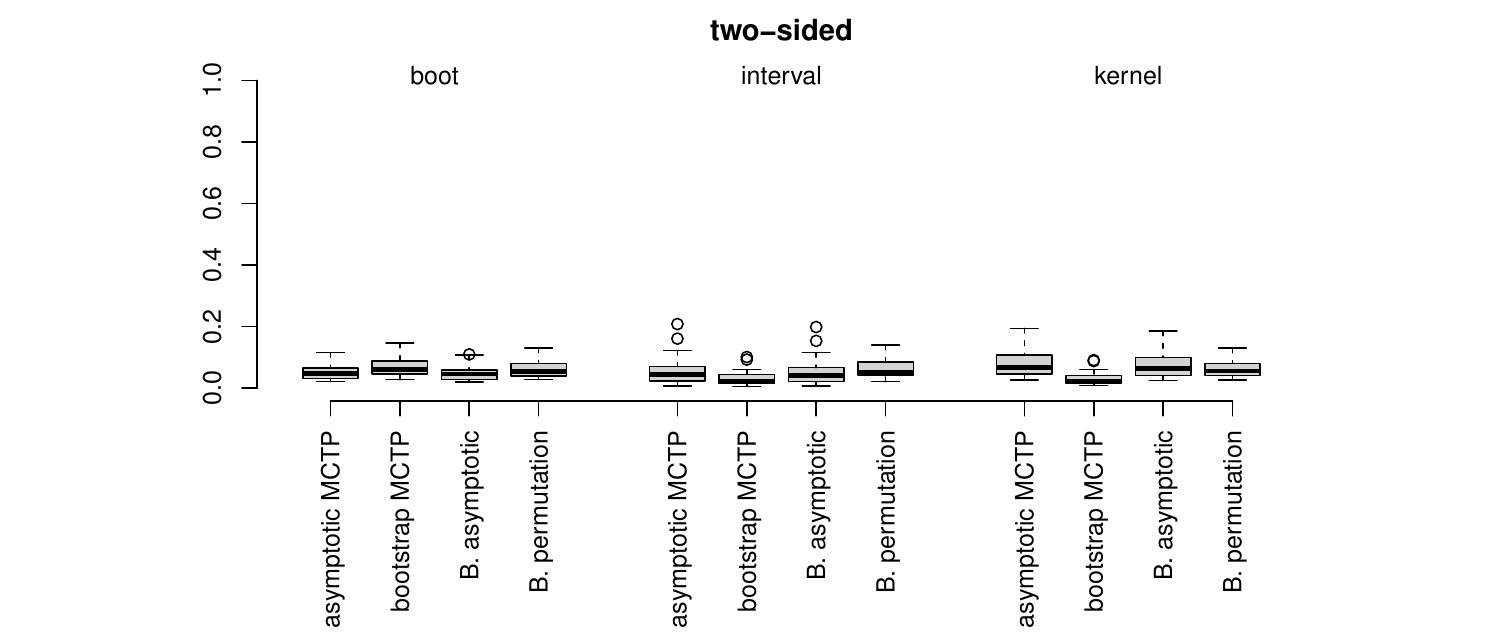} 
\caption{Empirical local power of $\mathcal{H}_{0,2}: m_2 - \bar{m} = 0$ with $\delta = 1.5$ for Grand-mean-type contrasts with different variance estimators.}
\end{figure}

\begin{figure}[H]\centering
\includegraphics[scale=0.6]{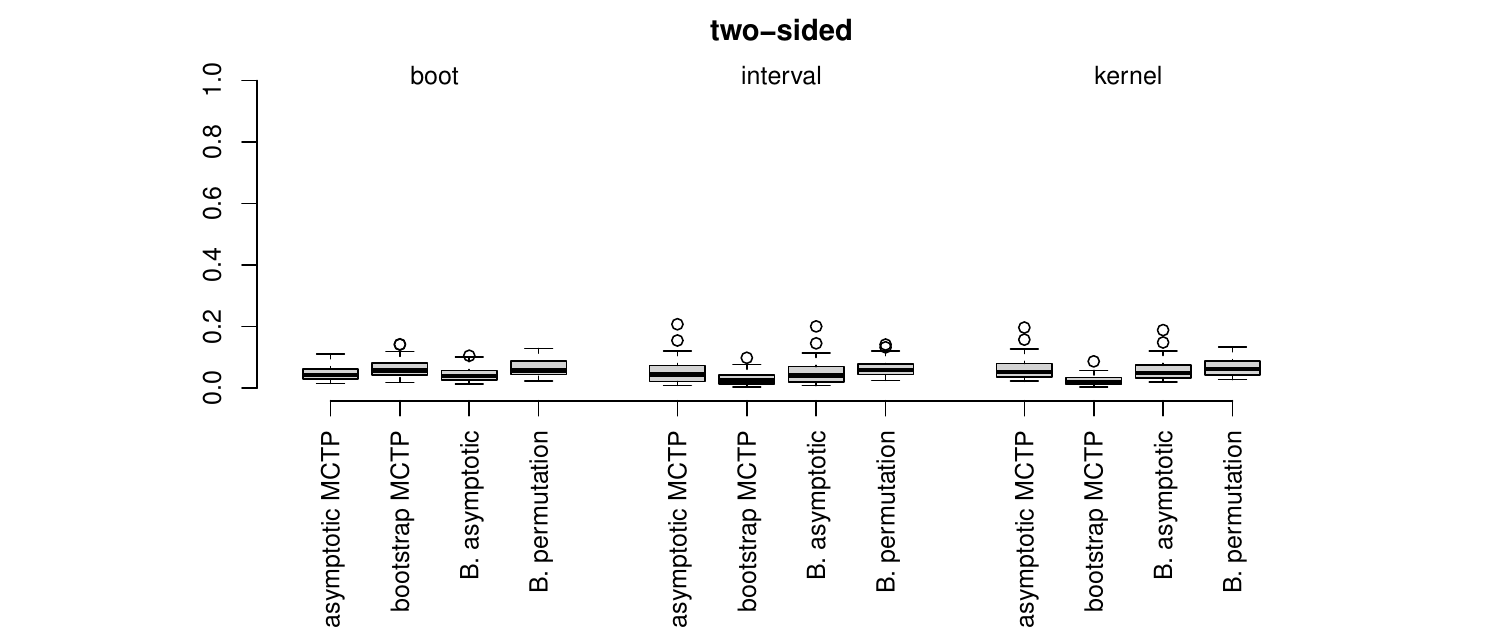} 
\caption{Empirical local power of $\mathcal{H}_{0,3}: m_3 - \bar{m} = 0$ with $\delta = 1.5$ for Grand-mean-type contrasts with different variance estimators.}
\end{figure}

\section[Additional Simulation Study]{Additional Simulation Study \new{for medians and interquartile ranges}}	
We conducted additional simulations to consider the type I error and power for simultaneous comparisons of medians and interquartile ranges (IQRs) across $k=4$ different groups. Hence, we choose the probabilities $p_1 = 0.25, p_2 = 0.5, p_3 = 0.75$ with $m=3$. As contrast matrices, the proposed matrices in the end of Section 2 of the paper are used, i.e. Kronecker products of Dunnett-, Tukey-, and Grand-mean-type matrices with $$\begin{bmatrix}
0 & 1 & 0\\ -1 & 0 & 1
\end{bmatrix}.$$ All other parameters as, e.g., the data generation is as described in Section \new{4\newnew{.1}} of the paper. 
Now, also the IQRs are compared and, thus, the settings with positive and negative pairing ($\boldsymbol{\sigma}_2, \boldsymbol{\sigma}_3$) become settings under the alternative. Hence, the model implicates that the data is exchangeable under the null hypothesis and, thus, the global permutation test should be exact. 

\begin{figure}[H]\centering
\includegraphics[width=\textwidth]{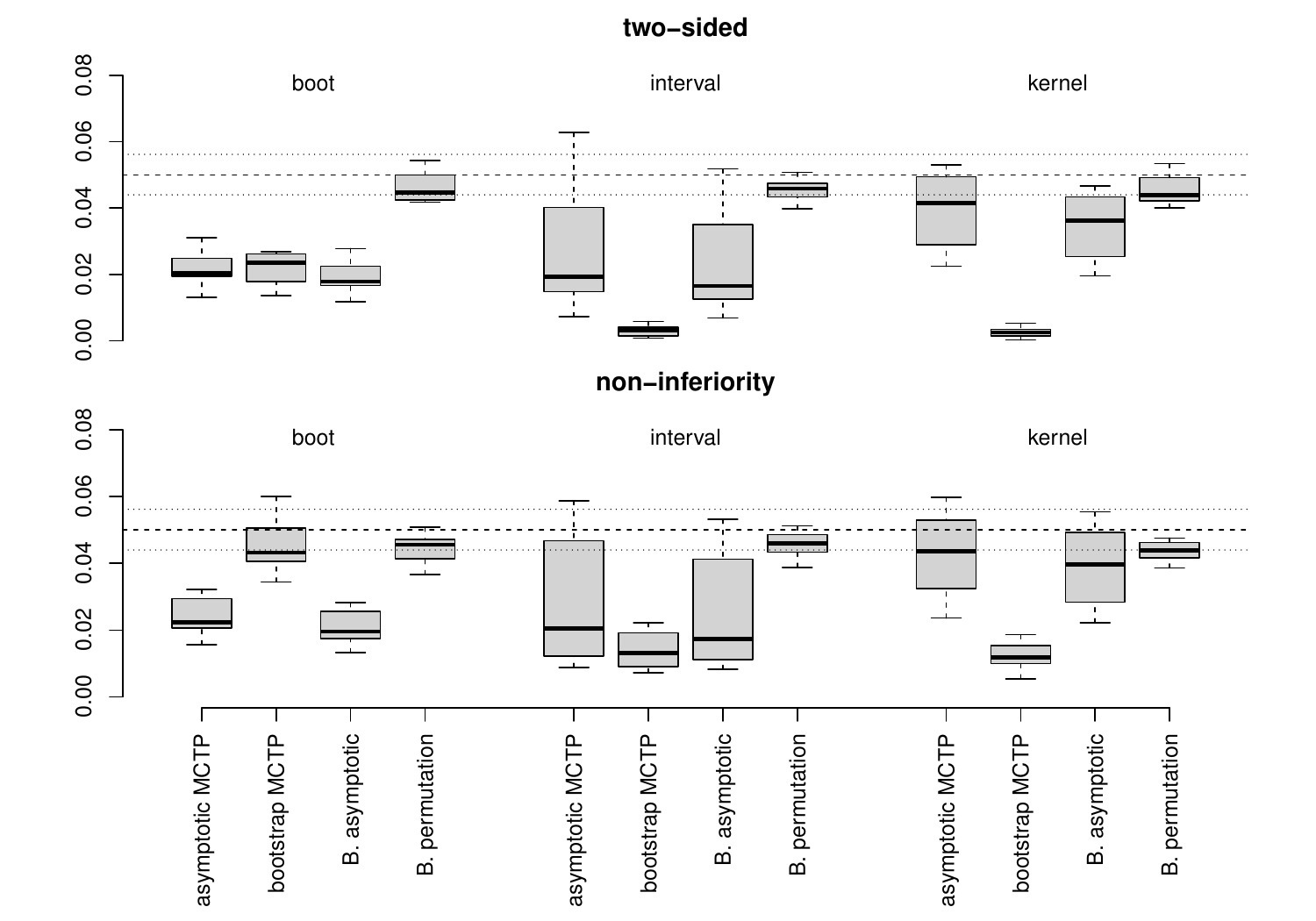} 
\caption{Empirical FWERs for Dunnett-type contrasts with different hypotheses (top: two-sided and bottom: non-inferiority) and variance estimators (from left to right: bootstrap, interval-based or kernel). The dashed line represents the desired level of significance of $\alpha=5\%$ and the dotted lines represent the Binomial interval $[0.044,0.0562]$ for $N_{sim} = 5000$ repetitions. }
\label{fig:AddDunnett}
\end{figure}

\begin{figure}[H]\centering
\includegraphics[width=\textwidth]{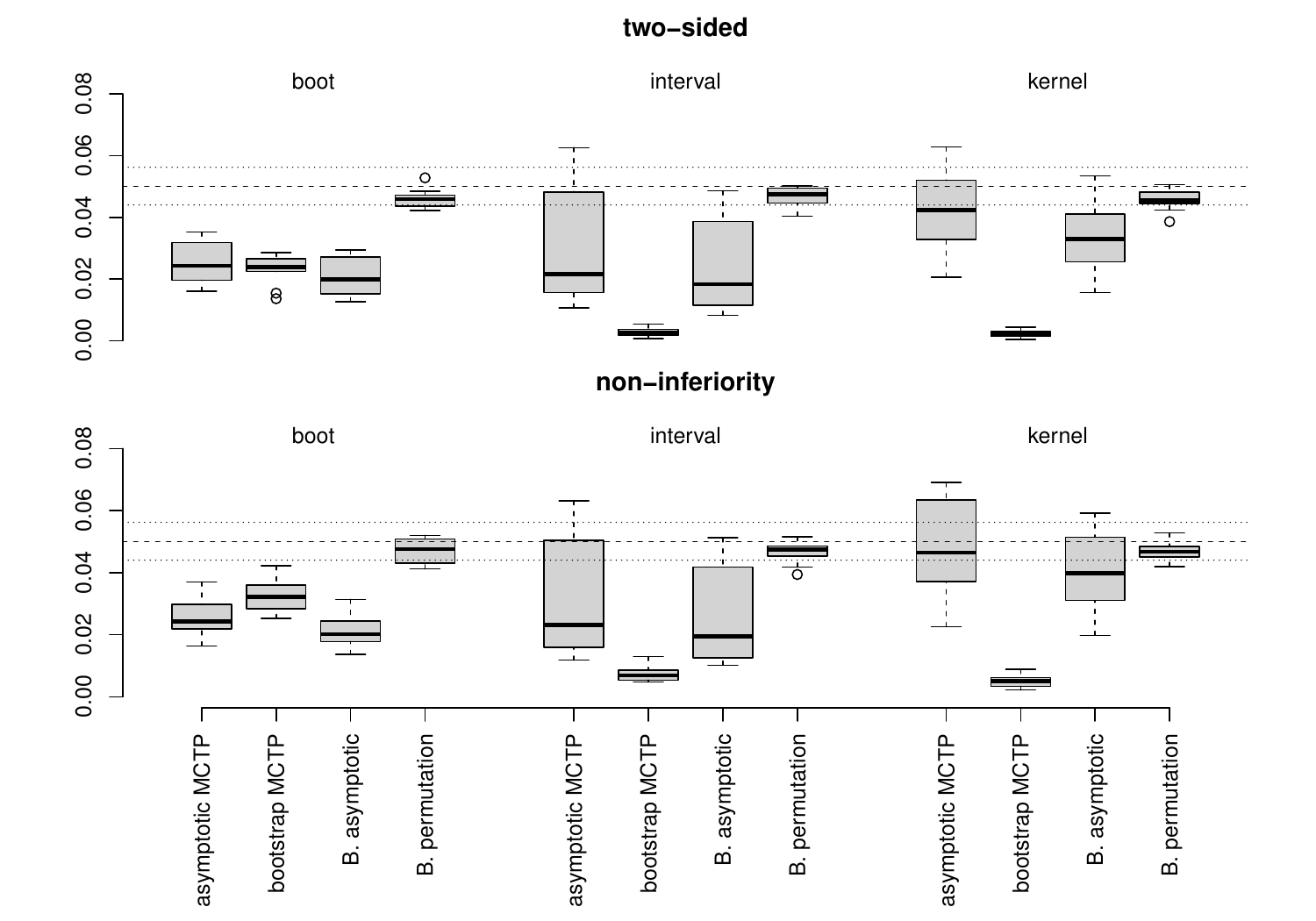} 
\caption{Empirical FWERs for Tukey-type contrasts with different hypotheses (top: two-sided and bottom: non-inferiority) and variance estimators (from left to right: bootstrap, interval-based or kernel). The dashed line represents the desired level of significance of $\alpha=5\%$ and the dotted lines represent the Binomial interval $[0.044,0.0562]$ for $N_{sim} = 5000$ repetitions. }
\label{fig:AddTukey}
\end{figure}

\begin{figure}[H]\centering
\includegraphics[width=\textwidth]{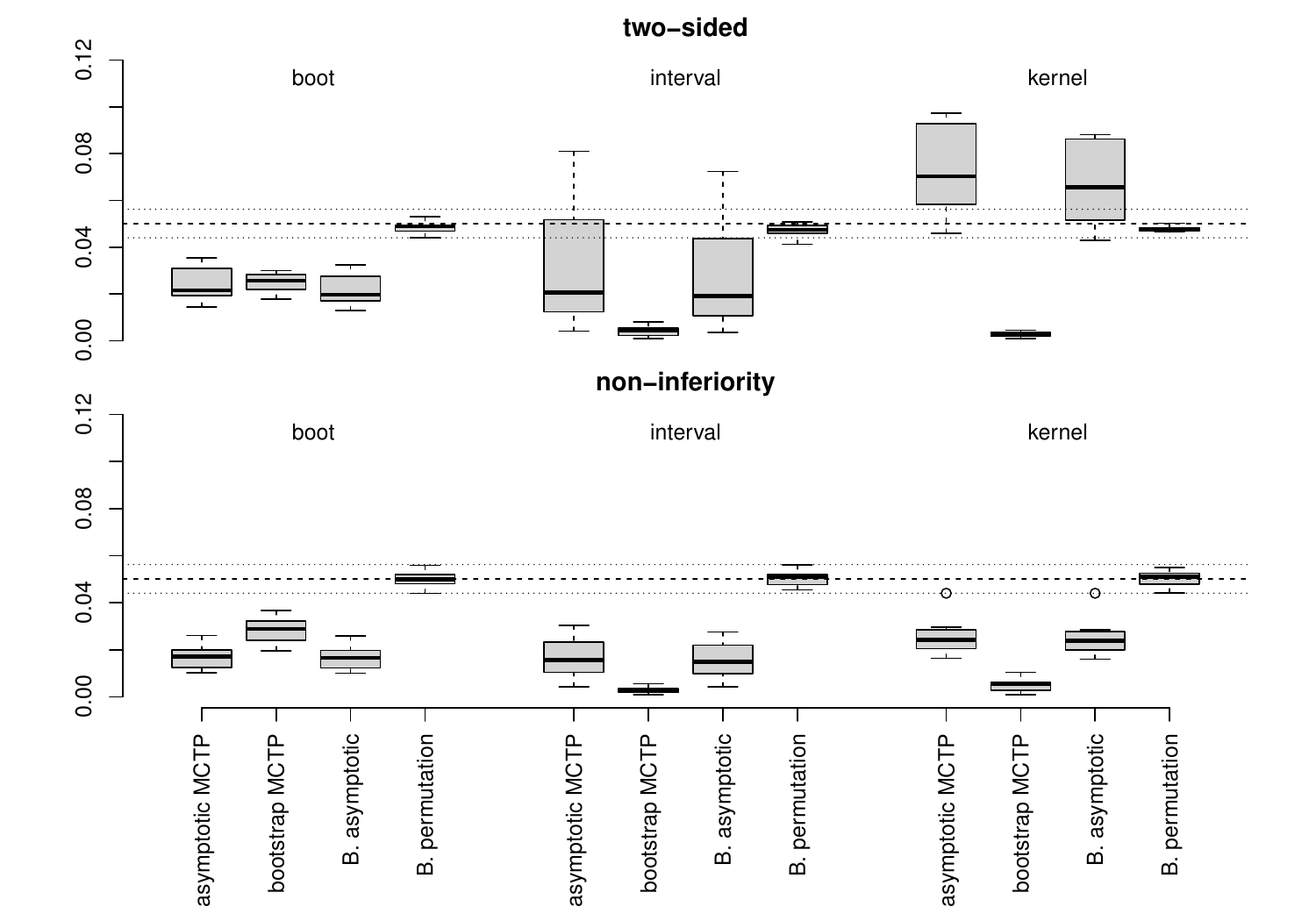} 
\caption{Empirical FWERs for Grand-mean-type contrasts with different hypotheses (top: two-sided and bottom: non-inferiority) and variance estimators (from left to right: bootstrap, interval-based or kernel). The dashed line represents the desired level of significance of $\alpha=5\%$ and the dotted lines represent the Binomial interval $[0.044,0.0562]$ for $N_{sim} = 5000$ repetitions. }
\label{fig:AddGM}
\end{figure}

\begin{figure}[H]\centering
\includegraphics[height=0.4\textheight]{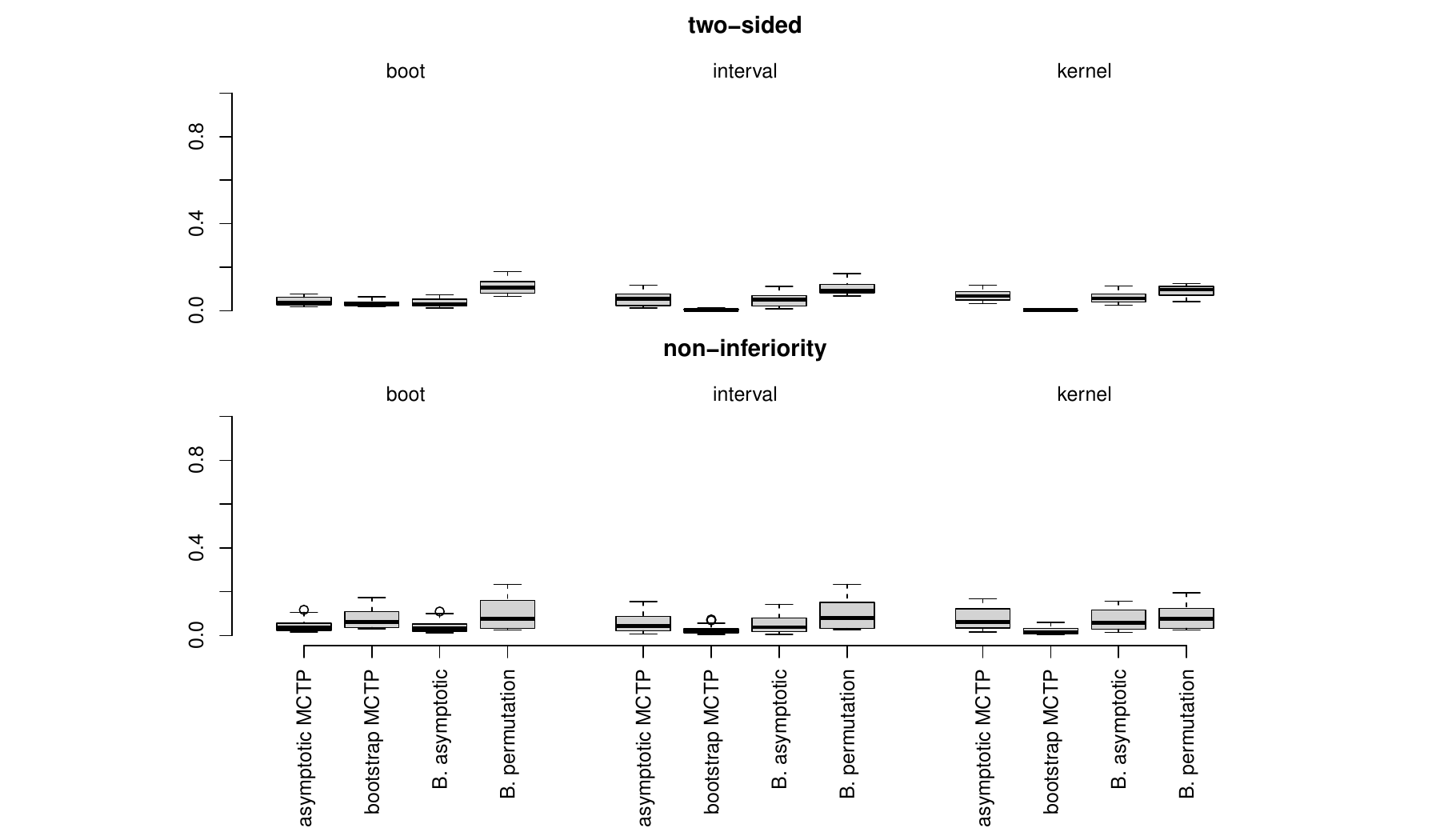} 
\caption{Empirical power with $\delta = 0.0$ for Dunnett-type contrasts with different hypotheses (top: two-sided and bottom: non-inferiority) and variance estimators.}
\label{fig:AddDunnettStart}
\end{figure}

\begin{figure}[H]\centering
\includegraphics[height=0.4\textheight]{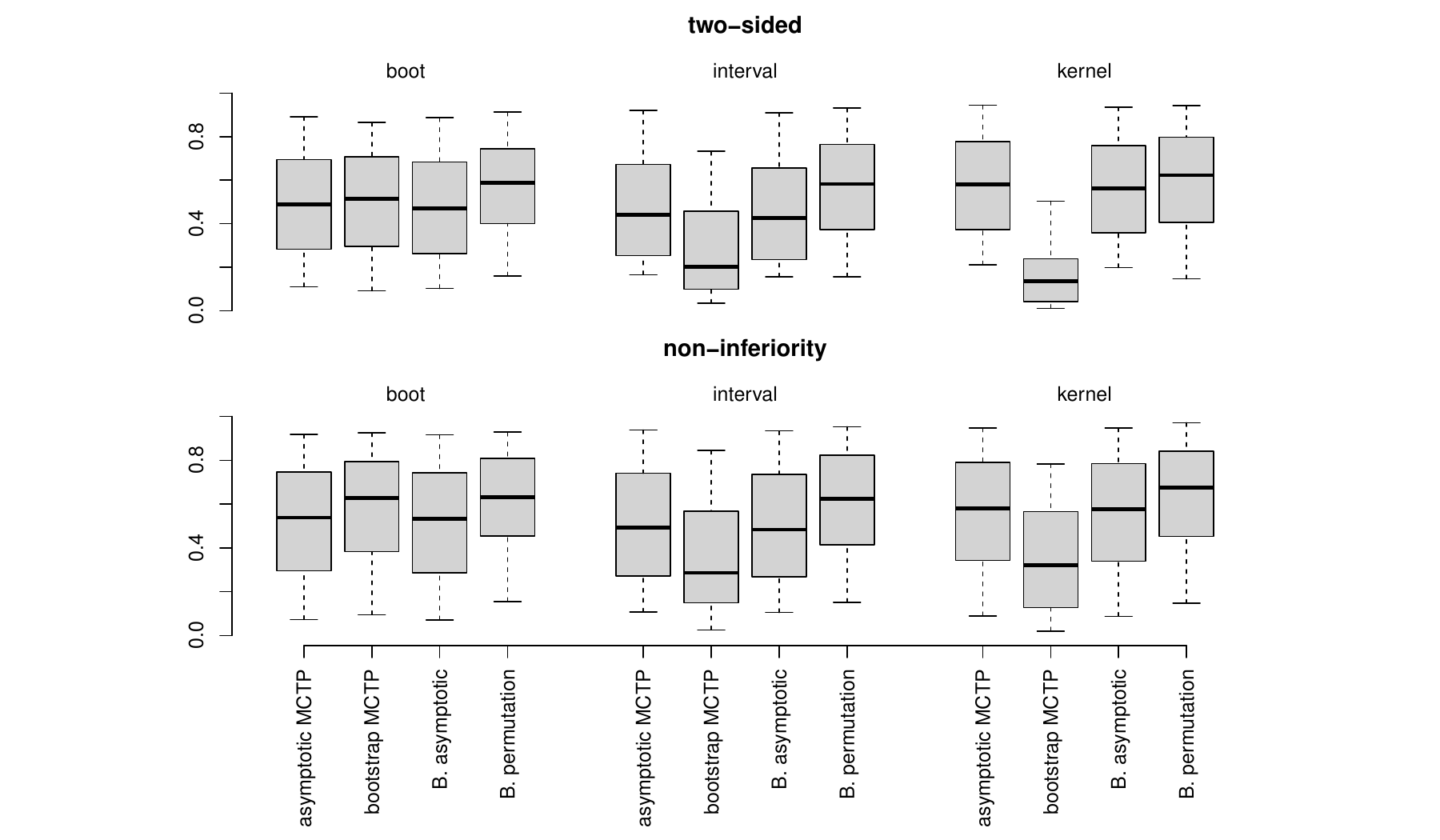} 
\caption{Empirical power with $\delta = 1.5$ for Grand-mean-type contrasts with different hypotheses (top: two-sided and bottom: non-inferiority) and variance estimators.}
\label{fig:AddGMEnd}
\end{figure}

In Figures~\ref{fig:AddDunnett}--\ref{fig:AddGM}, it is observable that the Bonferroni-adjusted permutation test is accurate or slightly conservative (due to the Bonferroni-correction) regarding the type I error control. The asymptotic approaches tend to be conservative except for the two-sided Gand-mean-type hypotheses in combination with the kernel estimator, where it occurs a liberal behaviour of the asymptotic tests. The groupwise bootstrap tests perform too conservative in nearly all scenarios. Only for Dunnett-type contrasts in combination with the bootstrap variance estimator, the groupwise bootstrap tests seem to perform accurate regarding the empirical FWER. The permutation approach can outperform the other methods regarding the power in nearly all scenarios, see Figures~\ref{fig:AddDunnettStart}--\ref{fig:AddGMEnd}. All in all, the results of this additional simulation study are similar to the results in Section \new{4} of the paper.

\begin{figure}[H]\centering
\includegraphics[height=0.4\textheight]{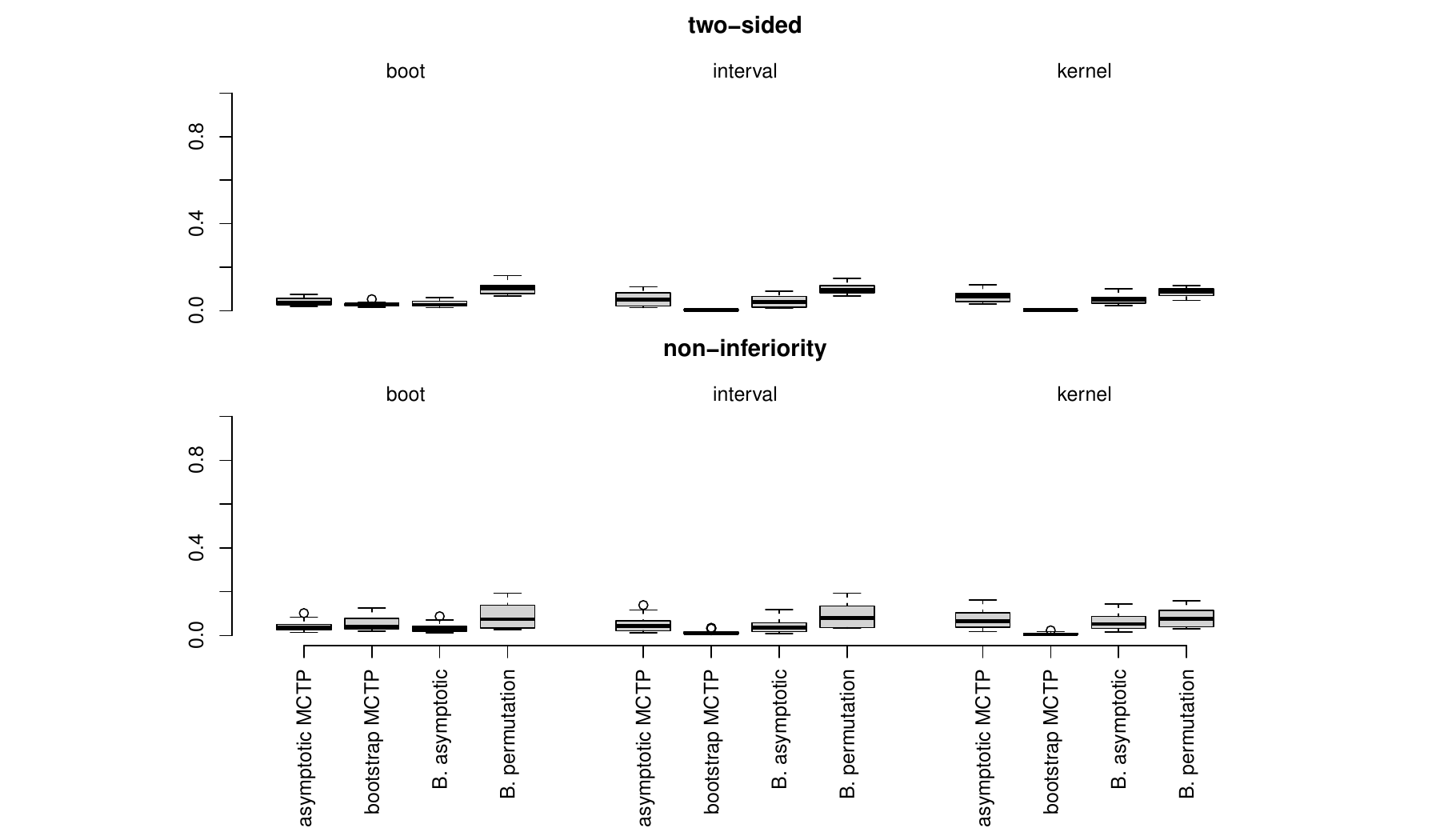} 
\caption{Empirical power with $\delta = 0.0$ for Tukey-type contrasts with different hypotheses (top: two-sided and bottom: non-inferiority) and variance estimators.}
\end{figure}

\begin{figure}[H]\centering
\includegraphics[height=0.4\textheight]{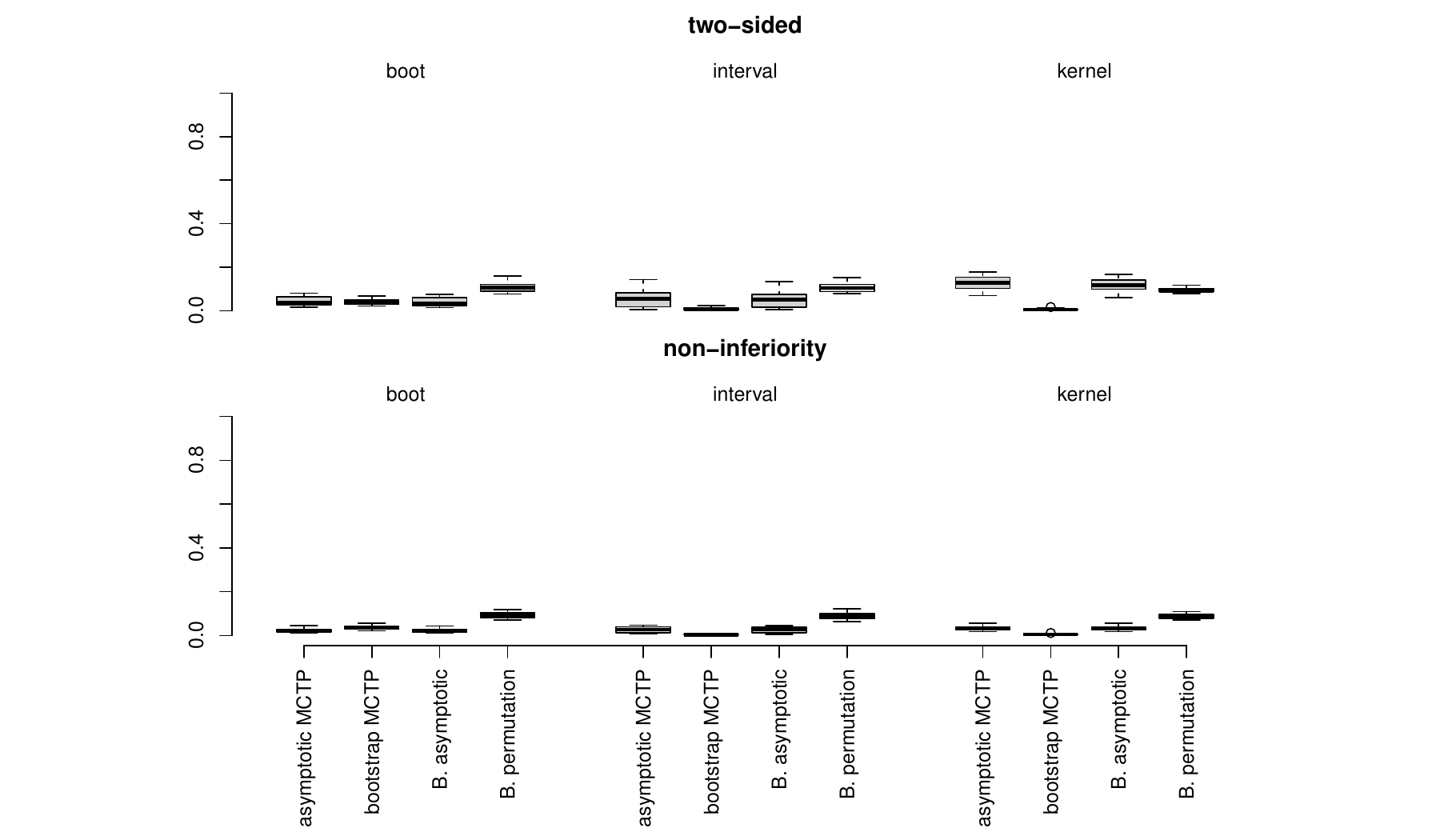} 
\caption{Empirical power with $\delta = 0.0$ for Grand-mean-type contrasts with different hypotheses (top: two-sided and bottom: non-inferiority) and variance estimators.}
\end{figure}

\begin{figure}[H]\centering
\includegraphics[height=0.4\textheight]{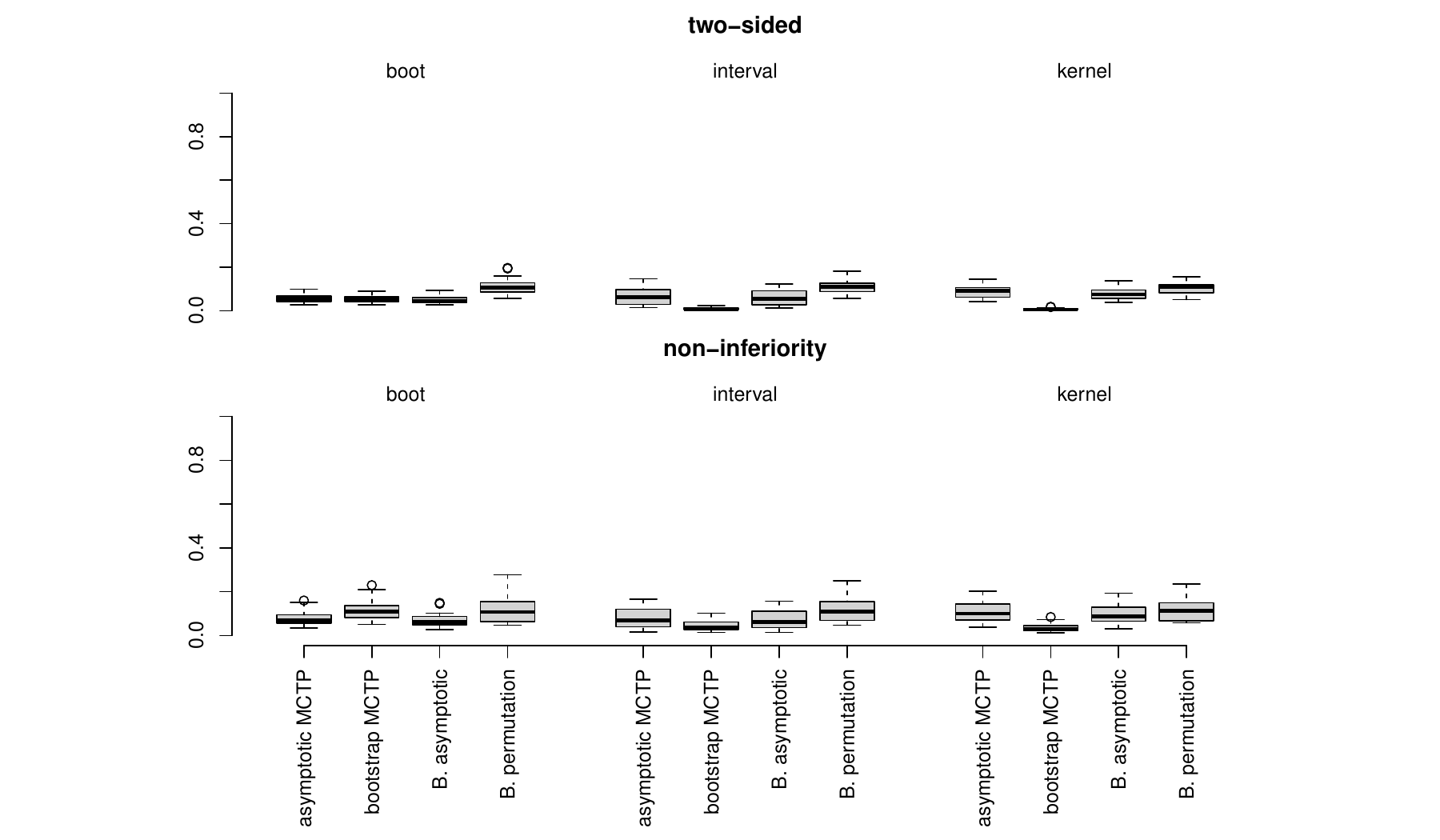} 
\caption{Empirical power with $\delta = 0.5$ for Dunnett-type contrasts with different hypotheses (top: two-sided and bottom: non-inferiority) and variance estimators.}
\end{figure}

\begin{figure}[H]\centering
\includegraphics[height=0.4\textheight]{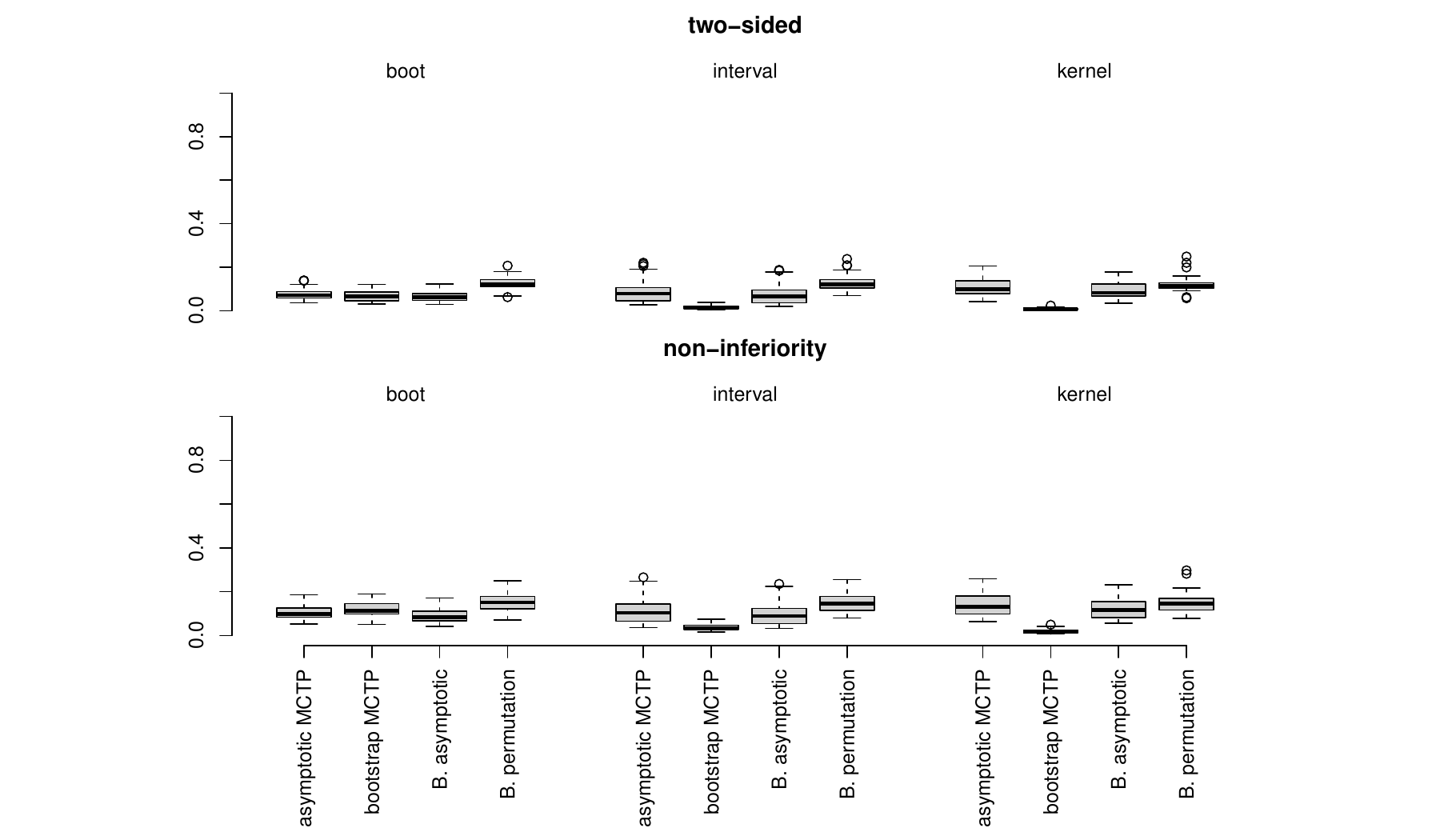} 
\caption{Empirical power with $\delta = 0.5$ for Tukey-type contrasts with different hypotheses (top: two-sided and bottom: non-inferiority) and variance estimators.}
\end{figure}

\begin{figure}[H]\centering
\includegraphics[height=0.4\textheight]{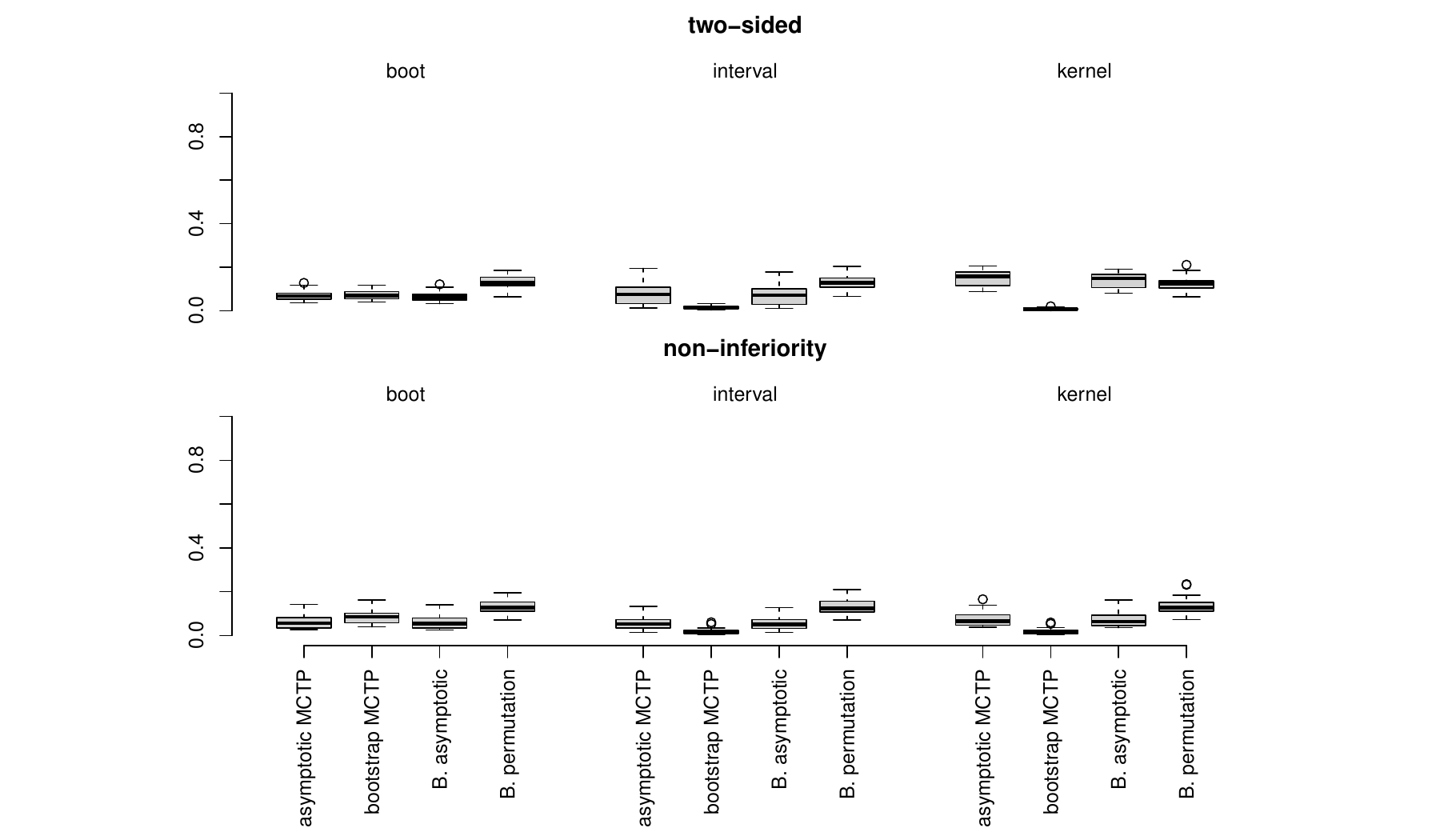} 
\caption{Empirical power with $\delta = 0.5$ for Grand-mean-type contrasts with different hypotheses (top: two-sided and bottom: non-inferiority) and variance estimators.}
\end{figure}

\begin{figure}[H]\centering
\includegraphics[height=0.4\textheight]{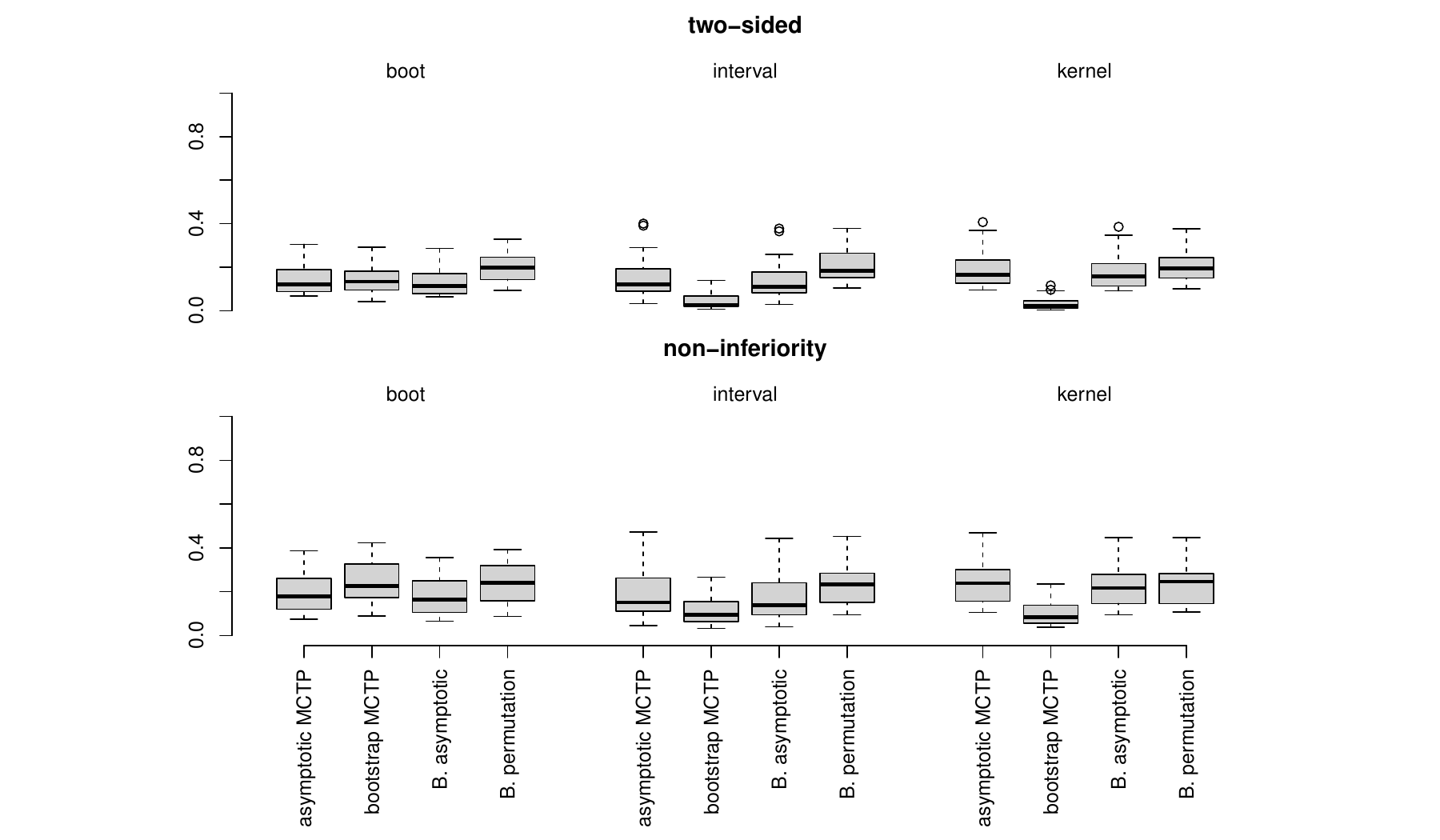} 
\caption{Empirical power with $\delta = 1.0$ for Dunnett-type contrasts with different hypotheses (top: two-sided and bottom: non-inferiority) and variance estimators.}
\end{figure}

\begin{figure}[H]\centering
\includegraphics[height=0.4\textheight]{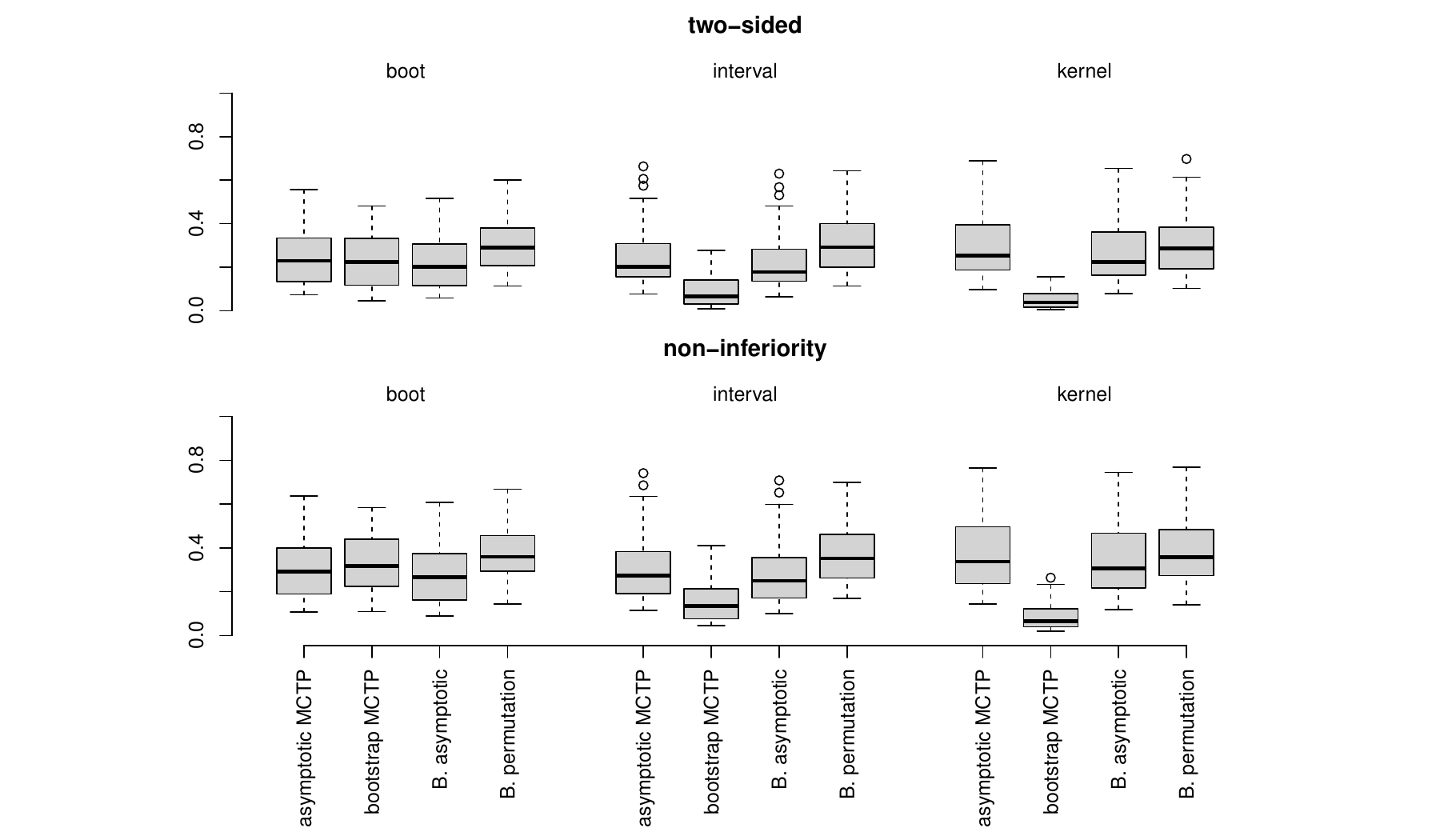} 
\caption{Empirical power with $\delta = 1.0$ for Tukey-type contrasts with different hypotheses (top: two-sided and bottom: non-inferiority) and variance estimators.}
\end{figure}

\begin{figure}[H]\centering
\includegraphics[height=0.4\textheight]{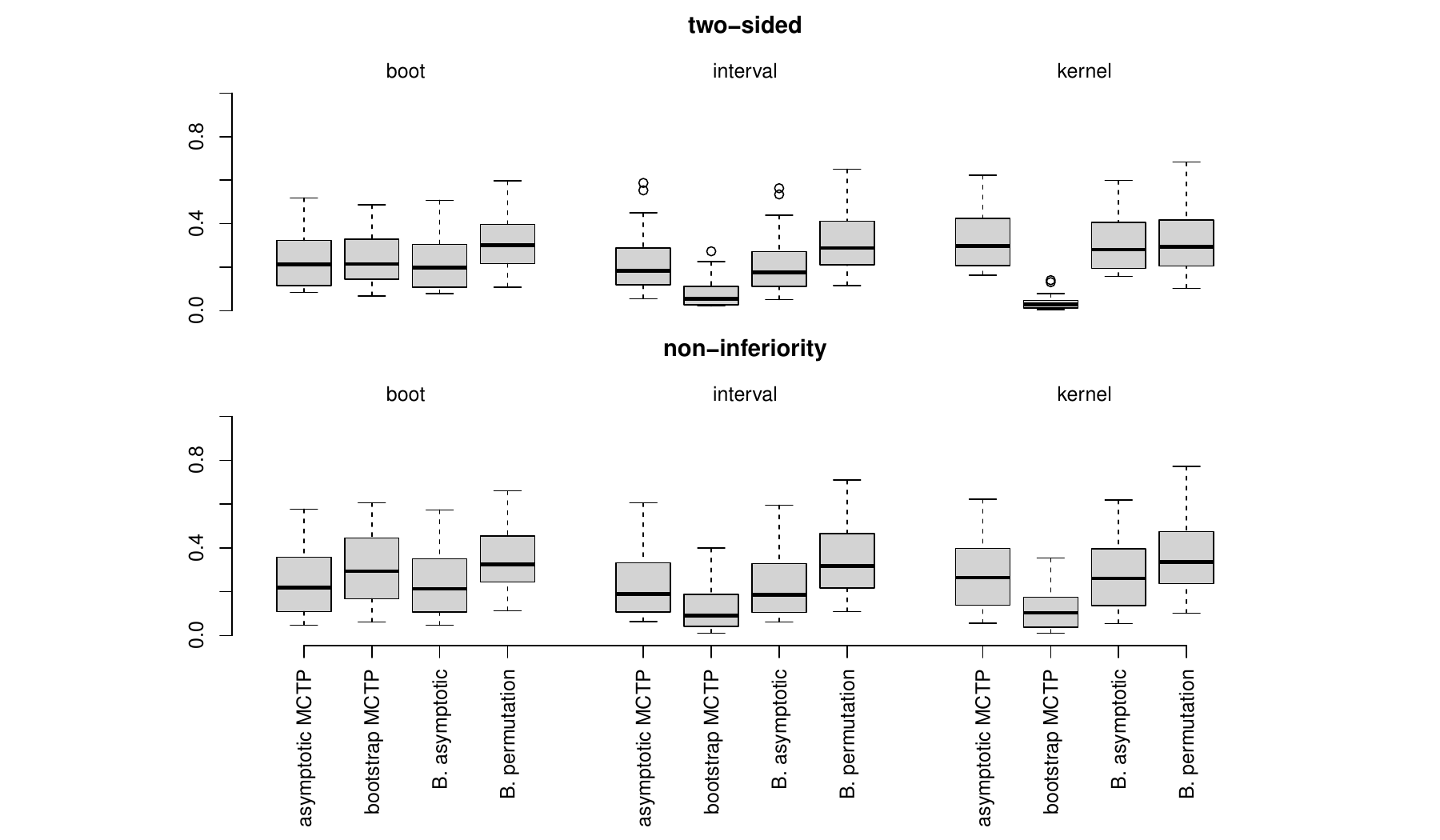} 
\caption{Empirical power with $\delta = 1.0$ for Grand-mean-type contrasts with different hypotheses (top: two-sided and bottom: non-inferiority) and variance estimators.}
\end{figure}

\begin{figure}[H]\centering
\includegraphics[height=0.4\textheight]{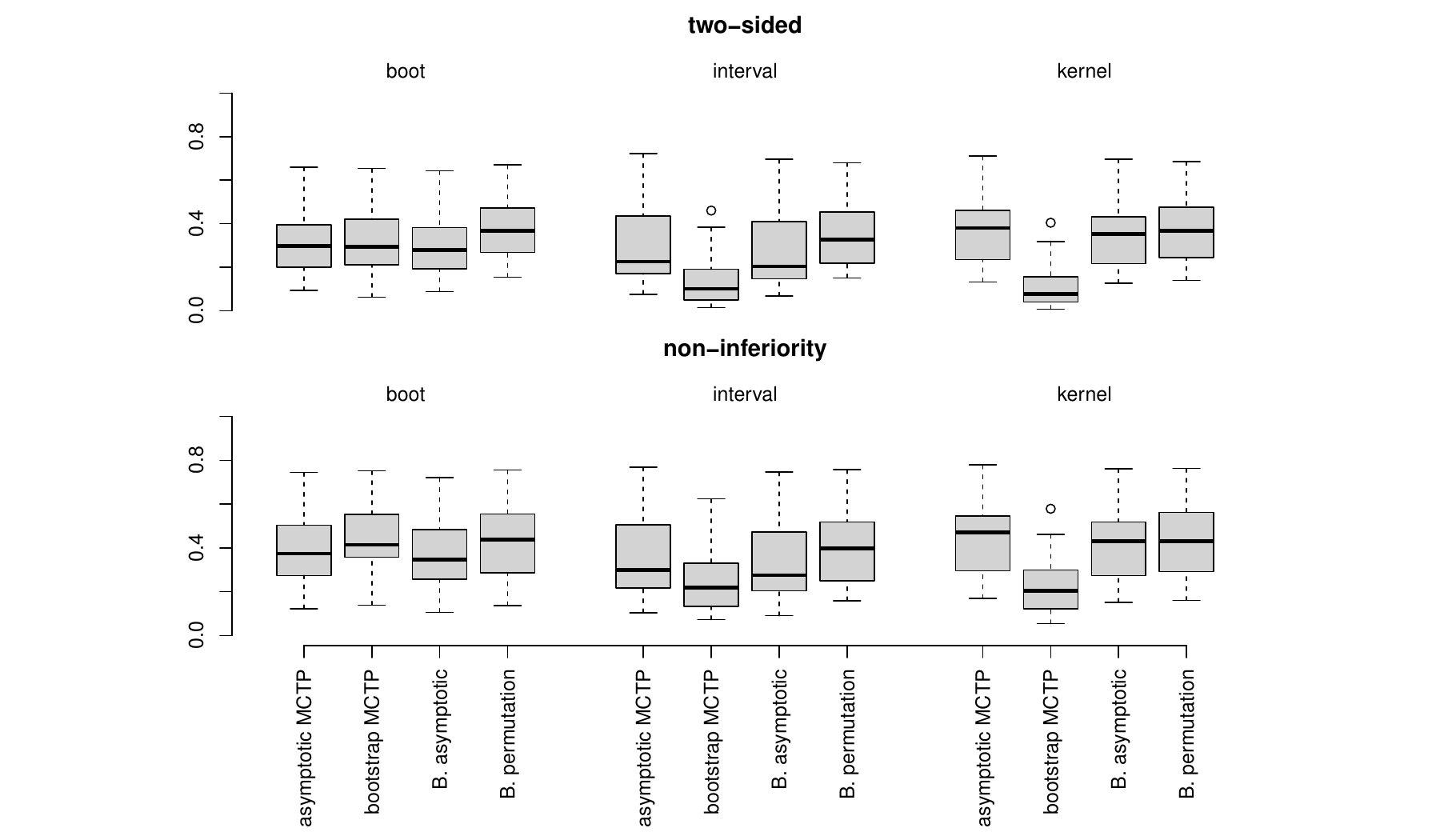} 
\caption{Empirical power with $\delta = 1.5$ for Dunnett-type contrasts with different hypotheses (top: two-sided and bottom: non-inferiority) and variance estimators.}
\end{figure}

\begin{figure}[H]\centering
\includegraphics[height=0.4\textheight]{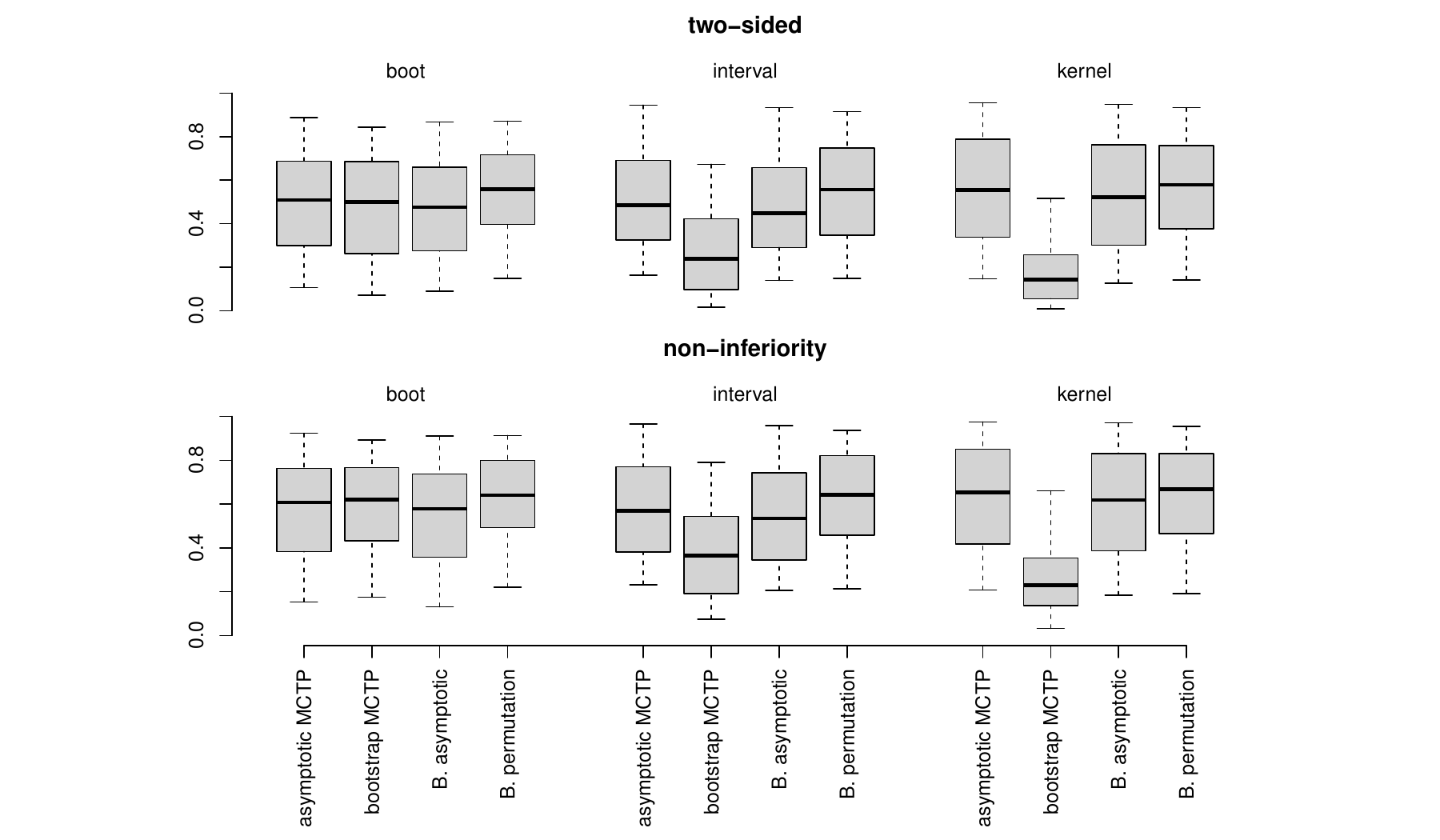} 
\caption{Empirical power with $\delta = 1.5$ for Tukey-type contrasts with different hypotheses (top: two-sided and bottom: non-inferiority) and variance estimators.}
\end{figure}

